\documentclass[preprint,12pt,authoryear]{elsarticle}

\usepackage{amssymb}
\usepackage{mathtools}
\usepackage{booktabs}
\usepackage{circledsteps}
\usepackage{graphics}
\usepackage[group-separator={,}]{siunitx}
\usepackage{acronym}
\usepackage{microtype}
\usepackage{multirow}
\usepackage{gensymb}
\usepackage{hyperref}
\usepackage{xcolor}
\hypersetup{
    colorlinks,
    linkcolor={red!50!black},
    citecolor={blue!50!black},
    urlcolor={blue!80!black}
}

\journal{International Journal of Multiphase Flow}

\newcommand\Fr{\mbox{\textit{Fr}}}
\newcommand\FrD{(\mbox{\textit{Fr}})_{\scriptscriptstyle D}}
\newcommand\We{\mbox{\textit{We}}}

\newcommand\muL{\mu_{\scriptscriptstyle L}}
\newcommand\muLN{{\mu}^\circ_{\scriptscriptstyle L}}
\newcommand\muG{\mu_{\scriptscriptstyle G}}
\newcommand\rhoL{\rho_{\scriptscriptstyle L}}

\newcommand\hf{h_{\scriptscriptstyle f}}
\newcommand\hL{h_{\scriptscriptstyle L}}
\newcommand\hLc{(h_{\scriptscriptstyle L})_c}
\newcommand\HLs{H_{Ls}}

\newcommand\HGs{H_{Gs}}
\newcommand\HGsc{(H_{Gs})_c}
\newcommand\HLAc{(H_{LA})_c}

\newcommand\bu{\mathbf{u}}
\newcommand\bn{\boldsymbol{\nabla}}
\newcommand\tauf{\tau_{\scriptscriptstyle f}}

\newcommand\uGS{u_{\scriptscriptstyle G}^{\scriptscriptstyle S}}
\newcommand\uLS{u_{\scriptscriptstyle L}^{\scriptscriptstyle S}}

\newcommand\uf{u_{\scriptscriptstyle f}}
\newcommand\vf{v_{\scriptscriptstyle f}}
\newcommand\ut{u_{\scriptstyle t}}
\newcommand\ud{u_{\scriptstyle d}}
\newcommand\uL{u_{\scriptscriptstyle L}}
\newcommand\um{u_{\scriptstyle m}}

\newcommand\lD{l_{\scriptscriptstyle D}}
\newcommand\lC{l_{\scriptscriptstyle St}}
\newcommand\lf{l_{\scriptstyle f}}
\newcommand\ls{l_{\scriptstyle s}}
\newcommand\lm{l_{\scriptscriptstyle M}}
\newcommand\lB{l_{\scriptscriptstyle B}}
\newcommand\luf{l_{\scriptstyle uf}}

\newcommand\Rey{\mbox{\textit{Re}}}
\newcommand\Reyf{\Rey_{\scriptstyle f}}
\newcommand\Reyfc{(\Rey_{\scriptstyle f})_c}
\newcommand\Reychi{\Rey_\chi}
\newcommand\Reychic{(\Rey_\chi)_c}

\newcommand\tB{t_{\scriptscriptstyle B}}
\newcommand\tm{t_{\scriptscriptstyle M}}

\newcommand\ts{t_{\scriptstyle s}}

\newcommand\yOD{\lambda_{\scriptscriptstyle 1D}}
\newcommand\yTD{\lambda_{\scriptscriptstyle 2D}}
\newcommand\yN{\lambda_{\scriptscriptstyle N}}
\newcommand\yB{\lambda_{\scriptscriptstyle B}}
\newcommand\yT{\lambda_{\scriptscriptstyle T}}
\newcommand\yk{\lambda_{\scriptscriptstyle K}}

\newcommand\oyOD{{\lambda}^\circ_{\scriptscriptstyle 1D}}
\newcommand\oyTD{{\lambda}^\circ_{\scriptscriptstyle 2D}}
\newcommand\oyN{{\lambda}^\circ_{\scriptscriptstyle N}}
\newcommand\oyB{{\lambda}^\circ_{\scriptscriptstyle B}}
\newcommand\oyT{{\lambda}^\circ_{\scriptscriptstyle T}}
\newcommand\oyk{{\lambda}^\circ_{\scriptscriptstyle K}}
\newcommand\oy{{\lambda}^\circ}

\newcommand\mydots{\makebox[1em][c]{.\hfil.\hfil.}}

\begin{document}

\begin{frontmatter}

\title{Elongated bubble centring and high-viscosity liquids in horizontal gas-liquid slug flow: Empirical analyses and novel theory}

\author{Sean J. Perkins}

\affiliation{organization={University of Alberta, Department of Civil \& Environmental Engineering and the School of Mining \& Petroleum Engineering},
            addressline={116 St and 85 Ave}, 
            city={Edmonton},
            postcode={T6G 2R3}, 
            state={Alberta},
            country={Canada}}

\begin{abstract}

Elongated bubble centring\textemdash an obscure counter-buoyant phenomenon encountered in horizontal gas-liquid slug flow\textemdash is correlated with liquid viscosity and their connection is theorized. Extracting from three sets of high-viscosity liquid (HVL) photographic data with $\muL\in[1,960]\unit{mPa.s}$ and $D\in[20,50.8]\unit{mm}$, the degree of incurred centring is found to increase, generally, in proportion to $\muL$ for a wide range of operational rates as evidenced through measurements at bubble nose, body and tail. It is demonstrated that full and nearly-symmetric centring can occur in HVL-containing flows\textemdash the former at relatively low inertial supply in contradiction to water-based dynamics. Qualitative advancements regarding the mechanistic nature of bubble centring and its plausible function within flow pattern transition theory are presented. Elaborating on recent modelling efforts, four distinct hypotheses are formulated: 1) film region laminarity as a modulator for centring; 2) boundary layer theory in slug flow to differentiate an outer-layer, relative motion-dominated film flow necessary for the initiation of centring; 3) wedge theory\textemdash a plausible alternative mechanism for partial-centring; and 4) a novel framework for the slug-annular transition composed of two unique mechanisms\textemdash centring and coalescence. The postulated boundary layer theory is investigated using a calibrated case of HVL slug flow and a dynamical environment conducive to centring mechanism proliferation is calculated.

\end{abstract}

\begin{keyword}
Bubble centring \sep High-viscosity liquids \sep Slug flow \sep Horizontal pipe \sep Slug-annular transition \sep Boundary layer theory 
\end{keyword}

\end{frontmatter}

%% \linenumbers

%% main text
\tableofcontents
\section{Introduction}
\label{sec:1}
Concurrent flow of gas and liquid in a circular pipe yields a complex system featuring multilayered dynamic behaviour. Depending on pipe, fluid and operational parameters, various discrete geometric configurations\textemdash known as flow patterns\textemdash may be observed. In a horizontally-oriented system, primary flow patterns include: stratified (smooth or wavy), intermittent (plug or slug), annular and dispersed bubble flows \citep{TaitelEA78}. Accurate prediction and modelling of flow patterns is supremely important in the preservation of engineering feasibility. Multiphase pipe flow phenomena such as heat transfer, pressure distribution and mechanical fatigue are captured using flow pattern-specific methodologies; therefore, a robust, validated conceptual paradigm is necessary to ensure competent design protocol in petroleum, processing and nuclear industries where such flow-types are commonplace. Transitional boundaries which differentiate flow patterns are of tantamount theoretic interest to the intrinsic physicality that defines them; as such, considerable efforts have been dedicated to understanding the nature of flow pattern conversion. Historic empirical works include \citet{LockhartMartinelli49}, \citet{Baker54} and \citet{MandhaneEA74} while the analytic endeavours of \citet{TaitelDukler76} represent a baseline reference of mechanistic inquiry. 

To foster economic sustainability in the global energy sector, efficient production and transportation of heavy oil resources is critical. Due to intricate rheology and chemical constituency, hydrocarbon liquids pose a unique challenge to both researchers and operators. Notably, they are characterized by a remarkable range of potential viscosity values; for example, under in situ conditions, certain crude oils may have $\muL<1\unit{mPa.s}$ whereas others are found with up to $\muL=\num{100000}\unit{mPa.s}$ while remaining flowable \citep{Islam23}.\footnote{With respect to typical oilfield units, $1\unit{mPa.s}=1\unit{cp}$.} Viscosity\textemdash a fluid's internal resistance to deformation\textemdash is proportional to the structural complexity of composing molecules; subsequently, $\muL$ is a function of temperature, pressure and solution-gas since all three affect positioning of hydrocarbon chains \citep{McCain90}.

Compared to otherwise equivalent water-based systems, gas-liquid pipe flows involving a high-viscosity liquid (HVL) are known to exhibit a variety of macroscopic peculiarities \citep{ZhangEA12}.\footnote{HVL: high-viscosity liquid; defined here broadly as a liquid with $\muL>1\unit{mPa.s}$. This is not a universal definition\textemdash in this context, an HVL is any liquid more viscous than water. However, owing to parameterization constraints, empirical conclusions derived here apply only for HVL-systems with $\muL\ge 5.5\unit{mPa.s}$.} Furthermore, these flow-types are ubiquitous in the realm of practical application; for instance, pipelined flow of heavy oil is often accompanied by hydrocarbon gases, stemming either organically from the reservoir or flashed from a miscible solution below a pressure threshold known as the bubble point. Gas-HVL pipe flow can also emerge during enhanced oil recovery schemes which utilize injected steam to mobilize bitumen\textemdash a naturally existing semi-solid with $\muL>\num{1000000}\unit{mPa.s}$ \citep{Islam23}\textemdash into a flowing commodity, particularly when live steam enters producer wells after breaking through highly-permeable reservoir trajectories \citep{DongEA21}. Consequently, relevant theoretical explorations have matured into a popularized area of research which lends itself to the facilitation of heavy oil resource extraction.

\begin{figure}[!ht]
\centering
\includegraphics[width=3.3in]{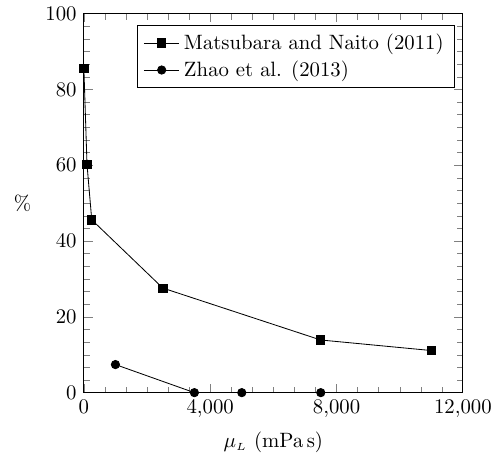}
\caption{Predictive capability of the \citet{TaitelDukler76} (TD76) flow pattern transition model applied to HVL data from \citet{MatsubaraNaito11} and \citet{ZhaoEA13}. \%-values represent portion of total flow pattern points correctly determined. }
\label{fig:01}
\end{figure}

Over the last two decades, the study of gas-HVL flow patterns in horizontal pipes has benefited from a marked influx of experimental publication \citep{GokcalEA08, MatsubaraNaito11, AlSafranEA13, ZhaoEA13, AlSafranEA15, ZhaoEA15, AlSafranAlQenae18}. Evidently, the subfield's contemporary status is primarily one of empiricism. Lacking is a base of purely theoretical works which query fundamental phenomenology corresponding to the introduced presence of an HVL.\footnote{This is not to say that such works do not exist; for example, a paper by \citet{TaitelDukler87} studies the impact of pipe length on flow pattern transition boundaries for horizontal, HVL-containing flow-systems. However, their model is 1) established only for $\muL\le 150\unit{mPa.s}$ and 2) limited to a singular outlet configuration.} Invoking the aforementioned model from \citet{TaitelDukler76} (TD76)\textemdash which provides a mechanistic framework for steady-state flow pattern formation in horizontal gas-liquid pipe flows\textemdash the situation is exemplified.\footnote{TD76: \citet{TaitelDukler76} steady-state flow pattern transition model for horizontal gas-liquid pipe flows.} TD76 is widely accepted and thoroughly validated for water-based flows \citep{Shoham06}; further, because it dissects individual boundaries, seemingly, into their most rudimentary components, it is considered to be a foundational explanation for flow pattern transition phenomena. However, a problem is encountered when implementing TD76 for the prediction of HVL flow pattern data; for example, inputting laboratory results from \citet{MatsubaraNaito11} and \citet{ZhaoEA13} indicates that modelling efficacy diminishes exponentially with increasing $\muL$, in certain cases demonstrating null predictability\textemdash shown here in figure \ref{fig:01}.

As is discussed in \S\ref{sec:4.4}, other papers\textemdash such as \citet{GokcalEA08}\textemdash have reported wide margins of predictive discrepancy for HVL data, even when utilizing mechanistic models alternative to TD76. Since these models are non-empirical, it is clear that imposing an HVL in a horizontal gas-liquid pipe flow induces a profound shift in underlying physics\textemdash particularly for $\muL\gg 1$. This divergence is precisely motivation for the present study. Ideally, existing data-based efforts will be transmuted into a cohesive, comprehensive narrative which describes exactly the implications of HVL-inclusion. A similar evolution happened in the 1970s when researchers harnessed a culmination of air-water data to craft a working framework of governing mechanics\textemdash a period designated the ``awakening years'' by \citet{ShippenBailey12}. I hypothesize here that an analogous era is unfolding in the pursuit of knowledge for HVL-based flow dynamics\textemdash within which the genesis phase has arrived\textemdash and that an obscure flow event known as elongated bubble centring will prove itself an invaluable keystone in the course of this transformation.

Bubble centring is a counterintuitive element of horizontal gas-liquid slug flow\textemdash a heterogeneous flow pattern in which elongated bubbles and liquid slugs flow in alternating sequence \citep{TaitelBarnea90, DuklerHubbard75}. Typically, long bubbles flow flush to the upper pipe wall while a slower-moving liquid film flows beneath; however, under certain conditions, at least part of the bubble\textemdash starting from its nose or most downstream point\textemdash detaches from the upper pipe wall in defiance of natural buoyant proclivities. This effect was first discussed by \citet{Bendiksen84} who discovered that, in air-water systems, centring initiates when the flowing phases reach a critical value of Froude number: a metric which measures the balance of inertial and gravitational forces given by
\begin{equation} \label{eqn:01} 
\Fr=\frac{\um}{\sqrt{gD}} 
\end{equation}
where $\um=\uGS+\uLS$ is mixture velocity or the summation of gas and liquid superficial velocities and $D$ is pipe diameter.\footnote{In analyzing data derived from a flow-system featuring constant pipe diameter $D$, relational trends based on variable $\Fr$ represent changes due to mixture velocity $\um$. As will be seen, this is the case within datasets studied here.} In the first strictly mechanistic study of its kind, \citet{PerkinsLi20} (PL20) provided a thorough theoretical evaluation of long bubble centring, including a simplistic model which calculates its operational manifestation, validated using externally generated visual data from air-water experiments.\footnote{PL20: \citet{PerkinsLi20} mechanistic bubble centring study. The present study is a continuation of PL20.} In doing so, it was determined that centring is not only a function of $\Fr$ but of the ratio between gas and liquid superficial velocities, defined as 
\begin{equation} \label{eqn:02} 
\gamma \equiv \frac{\uGS}{\uLS} = \frac{Q_{\scriptscriptstyle G}}{Q_{\scriptscriptstyle L}}
\end{equation}
where $Q_{\scriptscriptstyle P}$ is input volumetric flow rate of phase-${\scriptstyle P}$. PL20 postulates that bubble centring is crucially important in the conceptual architecture of HVL slug flow, reasoned in part due to prior researchers' observation of a thin liquid film above the elongated bubble in HVL systems \citep{ZhaoEA15, ZhangEA12} and from the obvious applicability of a singular utilized modelling simplification; namely, the existence of negligible turbulence (i.e., laminarity) throughout the underlying liquid film such that unbroken streamlines of relative motion can form\textemdash adjacent to the gas-liquid interface\textemdash long enough for a downward force to be transmitted due to a pressure differential and Bernoulli's principle.\footnote{Accurate prediction of the centring phenomenon is important from a design perspective in that film region pressure differential will be under-valuated without consideration of detached bubble regions, as discussed in PL20. Since centring increases liquid-wall contact area, this effect is more pronounced in HVL systems because frictional forces and thus pressure losses are proportional to $\muL$.} In amplifying liquid viscousness, there is, in general, a decreasing probability of turbulence production; thus, for HVL flow-systems, this vital assumption is more likely to be valid in contrast to air-water flows of otherwise same design.

Fortuitously, high-resolution photographic data which elucidate the correlative role of liquid viscosity in the dynamics of bubble centring have recently been published. Documented in this paper, these images were extracted from their original sources and analyzed thoroughly. The resultant illustration shows that the degree of centring incurred during horizontal slug flow is indeed a positively increasing function of $\muL$, thereby supporting theoretical musings put forth in PL20. Further, it provides a perfect segue for the creation of a mechanistic framework which qualitatively describes causality, implications and theoretical extensions relevant to the centring phenomenon and HVLs. 

In light of this opportunity, the present paper is dual-purpose: \S\S\ref{sec:2} and \ref{sec:3} present implemented methodology, garnered relationality and established results based on experimentalism while \S\ref{sec:4} is devoted to expounding novel theories related to bubble centring and its phenomenological placement in the greater sphere of flow pattern transition philosophy. Subsequently, \S\S\ref{sec:5} and \ref{sec:6} discuss applicability and critical conclusions/future work, respectively. To summarize, the objectives of this study are threefold: 

\begin{enumerate}
\item To provide a comprehensive analytical breakdown of raw elongated bubble centring data implicitly provided by \citet{NaidekEA23}, \citet{ShinEA24} and \citet{KimEA20}, including potential anomalies and operational dependencies on liquid viscosity, mixture Froude number and, where available, superficial velocity ratio;
\item To develop a structure of logical, falsifiable hypotheses that\textemdash in combination with integrated known theory\textemdash attempt to characterize the mechanistic nature of bubble centring as it relates to gas-HVL flow physics in horizontal pipes; and 
\item To offer guidance and suggest specific, feasible required future work designed to reliably expand and advance the field of HVL multiphase pipe flow theory, consequently assisting in the optimization of heavy oil production and distribution.
\end{enumerate}

\section{Methodology}
\label{sec:2}
\subsection{Data curation}\label{sec:2.1}

Emerging as a research trend is the production and publication of high-resolution still-images which showcase the internal workings of multiphase pipe flows. Undoubtedly, the topological complexity inherent in such flows poses a significant challenge, necessitating improved capacities in photography and visual data processing; for example, a study by \citet{WidyatamaEA18} demonstrates a growing joint interest between multiphase fluid dynamicists and signal processing engineers. 

Here, three peer-reviewed sources of pictorial data are studied, all of which offer compelling snapshots of the bubble centring phenomenon in horizontal gas-liquid slug flow generated from systems with liquid dynamic viscosity ($\muL$) values larger than that of water (i.e., HVLs). For the sake of clarity, each dataset is assigned a short-form label and simplified set notation is used to differentiate isolated flow-cases.\footnote{See \ref{sec:A4} for an overview of basic set notation used here.} Table \ref{tab:01} outlines $\muL$-range, $\Fr$-range, number of cases $N$, set notation, pipe diameter $D$ and abbreviation used for the three datasets, along with liquid density $\rhoL$ and surface tension $\sigma$.\footnote{K20 $\rhoL$-values are obtained using figure 10 of their paper (temperature-dependent correlations for $\muL$ and $\rhoL$); no such correlation is available for $\sigma$ and thus a singular value is utilized.}$^{,}$\footnote{Although diameter $D$ varies between datasets, the impact of changing $D$ on centring extent is not explicitly studied here. That is, other than continuity analysis provided in \S\ref{sec:3.1}, centring relationality is investigated within datasets which have constant values of $D$ (see also footnote 5).} Included are a total of $N_{\scriptscriptstyle \Sigma}=38$ cases across a liquid viscosity spectrum of $\muL\in[1,960]\unit{mPa.s}$, all displaying variable degrees of observable long bubble centring. Other than N23's $\{1.\mathrm{j}\mid\forall \mathrm{j}\}$\textemdash the referential water baseline\textemdash all cases considered are herein classified as HVL-containing.

\begin{table} [t!]
\begin{center}
\begin{tabular}{c|ccc} 
  & \citet{NaidekEA23} & \citet{ShinEA24} & \citet{KimEA20}   \\  \toprule 
Abbrev.  & \textbf{N23} & \textbf{S24} & \textbf{K20}  \\
$N$  & $30$ & $5$ & $3$  \\
\midrule
 \multirow{4}{*}{\shortstack{Set \\ notation}} & \multirow{4}{*}{\shortstack{$\mathbf{i.j}$ \\ $\mathrm{i}\in[1,5]\in\mathbb{Z}$ \\ $\mathrm{j}\in[1,6]\in\mathbb{Z}$}} & \multirow{4}{*}{\shortstack{$\mathbf{m.k}$ \\ $\mathrm{m}\in\{\mathrm{A},\mathrm{B},\mathrm{C}\}$ \\ $\mathrm{k}_{\scriptscriptstyle \mathrm{A},\mathrm{B}}\in\{1,2\}$ \\ $\mathrm{k}_{\scriptscriptstyle \mathrm{C}}=1$}} & \multirow{4}{*}{\shortstack{$\mathbf{\Omega.q}$ \\ $\mathrm{q}\in[1,3]\in\mathbb{Z}$}} \\
  & & & \\
    & & & \\
      & & & \\
\midrule
$\muL$ (\unit{mPa.s})  & $1\rightarrow30.4$ & $37.7\rightarrow352$ & $510\rightarrow960$  \\
$\Fr$  & $1\rightarrow4$ & $0.68\rightarrow3.23$ & $0.57$  \\
$D$ (\unit{mm})  & $26$ & $20$ & $50.8$  \\
$\rhoL$ (\unit{kg/m^3}) & $997.1\rightarrow1188$ & $878\rightarrow970$ & $846.6\rightarrow852.8$  \\
$\sigma$ (\unit{mN/m})  & $64.9\rightarrow73.2$ & $19.4\rightarrow26.8$ & $33$  \\
 \bottomrule
\end{tabular}
\caption{Relevant parameterization and notation for data sources utilized in this study.}
\label{tab:01}
\end{center}
\end{table}

To retrieve centring data from supplied photos, rigorous analyses were performed using open-source, vectorized graphics software Inkscape (v.1.3.2) which offers exceptional resolution and, by proxy, measurement capabilities. Although the data have, technically, already been published in raw plainness, the incurred extent of bubble centring was not formally investigated, nor was it measured and correlated to metrics such as $\muL$ and $\gamma$. Observations made through eyesight are inadequate; typically, high levels of magnification are required to properly locate gas-liquid and liquid-solid interfaces, as was done for this study. 

For all three datasets, an equivalent, generalized data collection methodology was utilized to measure, using Inkscape, liquid separation between upper pipe wall and elongated bubble-top at strategically predefined spatial coordinates.\footnote{Doing so requires an important assumption; namely, that long bubble photographs, as supplied, represent the entire vertical span of the inner pipe region. This assertion is maintained despite, for example, the authors of S24 stating that their editing involved ``removing an outer region of the flow and adjusting the resolution from the original records''\textemdash the specifics of this are unknown; thus, a potential source of error is acknowledged in S24 centring analyses.} Measurement specifics and data availability vary uniquely between sets, however, as is discussed later in this section. A complete list of obtained metrics is given below wherein ``separation'' indiscriminately refers to the aforementioned distance between top-of-bubble and upper pipe wall:

\begin{figure}[!t]
\centering
\includegraphics[width=\textwidth]{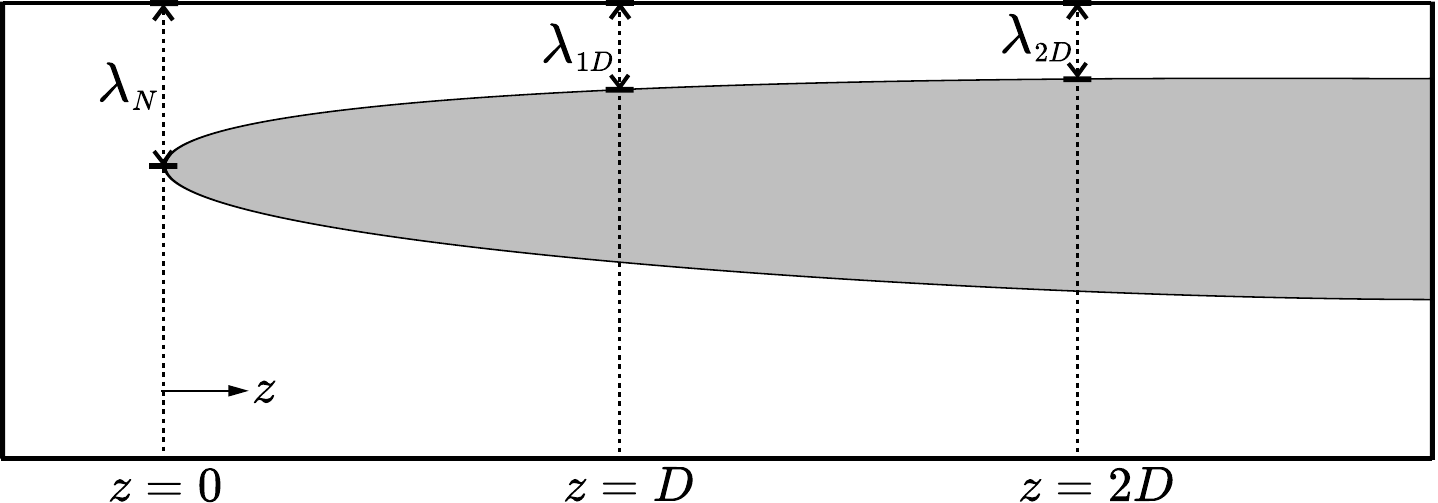}
\caption{Simplistic depiction of a centred elongated bubble (nose region) in horizontal gas-liquid slug flow overlaid with $\yN$, $\yOD$ and $\yTD$ metrics.}
\label{fig:02}
\end{figure}

\begin{enumerate}
\item[]\begin{enumerate}
\item[$\mathbf{\yOD}$:] separation one diameter upstream of bubble nose-tip
\item[$\mathbf{\yTD}$:] separation two diameters upstream of bubble nose-tip
\item[$\mathbf{\yN}$:] separation at bubble nose-tip
\item[$\mathbf{\yB}$:] separation at an arbitrary location within bubble body, averaged using three points each spaced apart by one diameter
\item[$\mathbf{\yT}$:] separation at bubble tail\textemdash an upstream location at which average bubble body shape deteriorates
\end{enumerate}
\end{enumerate}

Figure \ref{fig:02} provides a visual depiction of $\yOD$, $\yTD$ and $\yN$ whereas $\yB$ and $\yT$ are shown in figure \ref{fig:03}. Centring- or $\lambda$-metrics\textemdash meaningless in absolute terms\textemdash must be described relative to pipe diameter. Throughout this study they are given in fractional or normalized forms such as
\begin{equation} \label{eqn:03} 
\oyk=\frac{\yk}{D}
\end{equation}
where the index ${\scriptstyle \mathrm{K}}$ represents any of the earlier described subscripts and $\oyk$ is a percentage. Using the outlined metrics, four types of long bubble centring are formally defined: 

\begin{figure}[!t]
\centering
\includegraphics[width=\textwidth]{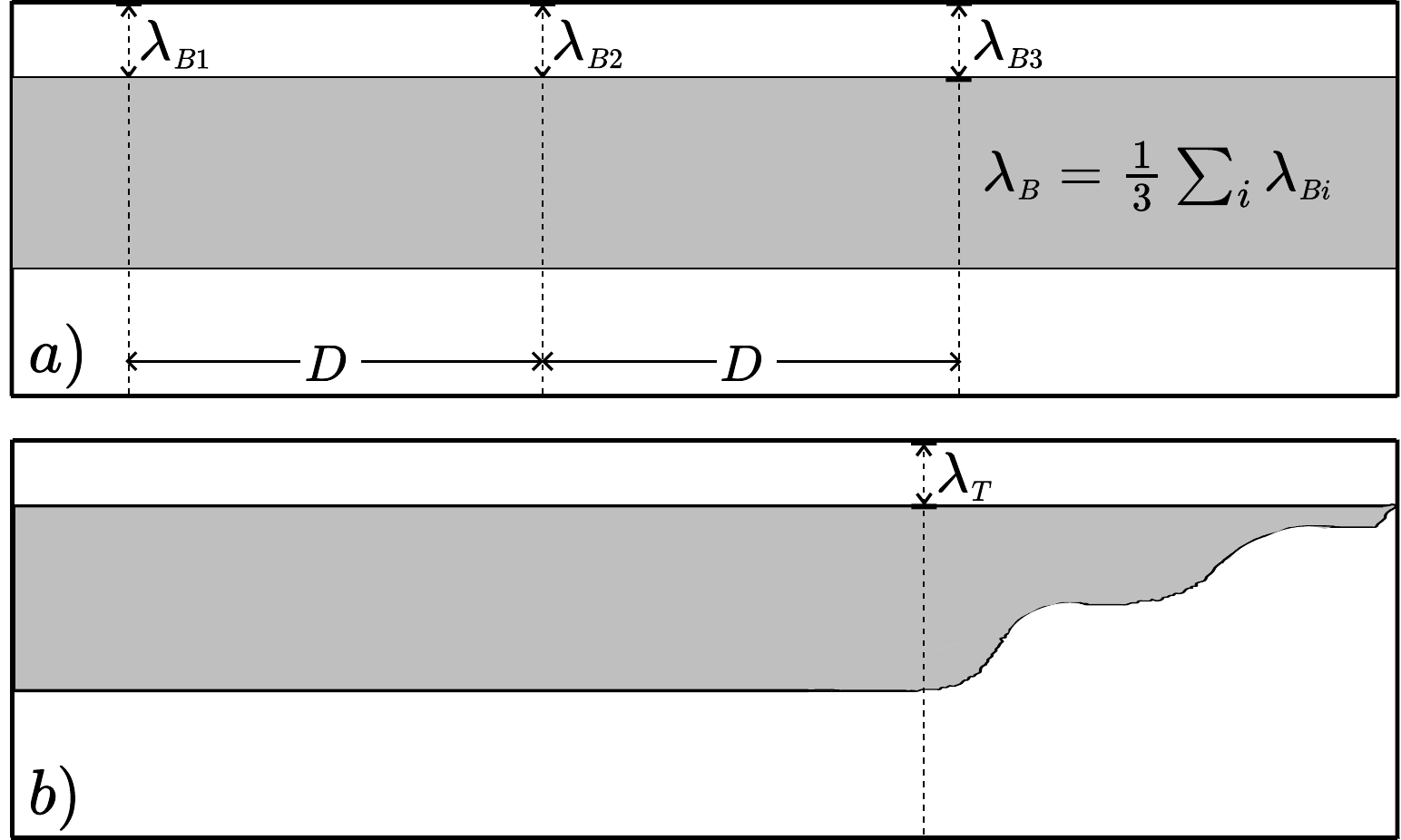}
\caption{Bubble centring metrics a) in the body $\yB=\sum_i \lambda_{\scriptscriptstyle \mathrm{B}i}/3$ and b) at the tail $\yT$. }
\label{fig:03}
\end{figure}

\begin{enumerate}
\item[]\textbf{No-centring}: null or negligible $\oyOD$
\item[]\textbf{Partial-centring}: non-negligible $\oyOD$ and null or negligible $\oyT$
\item[]\textbf{Full-centring}: non-negligible $\oyOD$, $\oyB$ and $\oyT$
\item[]\textbf{Perfect-centring}: non-negligible $\oyOD$, $\oyB$ and $\oyT$ with equivalent spacing beneath the bubble\textemdash radial symmetry\footnote{The variable $\oyOD$ alone defines no-centring because centring extent begins at long bubble nose; thus, a flow-case wherein $\oyOD$ is null or negligible while farther upstream locations are detached is non-physical. Partial-centring is classified using $\oyOD$ and $\oyT$ since it can manifest in a wide range of configurations. Full-centring requires that the entire bubble is centred; however, only when $\oyT$ is near negligibility must $\oyB$ also be checked.}
\end{enumerate} 
The downstream delimiter is selected to be $\oyOD$, rather than $\oyN$, because the bubble nose-tip is subject to a high level of variability \citep{Diaz16} and, as such, $\oyN$ could be large while the rest of the bubble remains attached to the upper pipe wall. The concept of perfect-centring is an idealization originally discussed by \citet{Bendiksen84} wherein the entire bubble flows in complete alignment with the pipe centreline. Partial- and full-centring terminologies were put forth in the PL20 study; the former is commonly observed in air-water slug flow wherein approximately $1-3D$ of the bubble's nose region is separated from the upper pipe wall.\footnote{The term ``fully centred'' is used in \citet{Bendiksen84} but formally defined in PL20.} Also defined for convenience is a normalized form of liquid viscosity
\begin{equation} \label{eqn:04} 
\muLN=\frac{\muL}{\mu_{\scriptscriptstyle W}}
\end{equation}
where $\mu_{\scriptscriptstyle W}=1\unit{mPa.s}$ is standard dynamic viscosity of water; therefore, $\muLN=\muL$ and units can be dropped. 

Not all defined centring metrics can be obtained in every flow-case included here. Regarding the N23 ($\mathrm{i.j}$) dataset, $\yB$ cannot be determined since bubble body photos are not supplied; however, all other introduced centring metrics are measured. Tail-separation\textemdash represented by $\yT$\textemdash is omitted for cases $\{2.\mathrm{j}\mid\forall \mathrm{j}\}$ owing to selective exclusion of tail photos for $\Fr=1.5$ in the original source. For cases $\{\mathrm{A.k}\cup \mathrm{B.k}\mid\forall \mathrm{k}\}$ from S24, all five metrics are quantified using available nose, body and tail photos; for case $\mathrm{C}.1$, only $\yOD$, $\yN$ and $\yT$ can be found due to an atypically short elongated bubble. For K20 cases $\{\Omega.\mathrm{q}\mid\forall \mathrm{q}\}$, $\yOD$, $\yN$ and $\yT$ are the only metrics available for extraction. Pairs of $(\Fr,\muL)$ are designated in all flow-cases; however, not all cases have obtainable values of $\gamma$. Only for cases $\{1.\mathrm{j}\mid\forall \mathrm{j}\}$ in N23 can $\gamma$ be deduced while for S24 and K20, superficial flow ratio is calculable in all cases from $\mathrm{m.k}$ and $\Omega.\mathrm{q}$. 

\begin{figure}[!t]
\centering
\includegraphics[width=\textwidth]{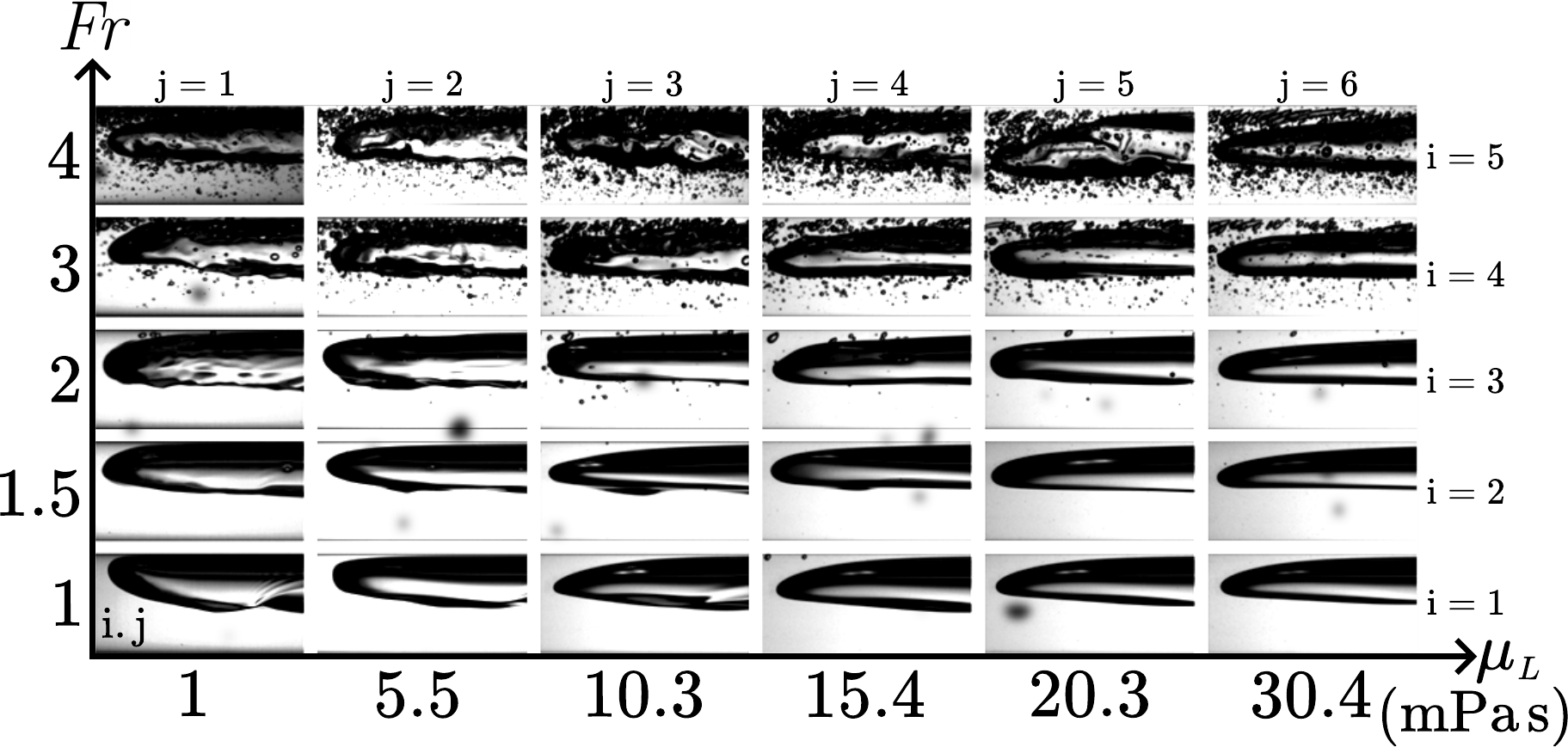}
\caption{Elongated bubble nose region images from \citet{NaidekEA23} (N23 dataset) sorted by mixture Froude number $\Fr$ and liquid viscosity $\muL$, overlaid with adopted case numbers/set notation $\mathrm{i.j}$. Adapted from figure 5 of original source with permission from Elsevier Publishing. Photos were brightened/sharpened using Inkscape's built-in ``Age'' filter (causing grey ``blob'' artifacts); otherwise unaltered. }
\label{fig:04}
\end{figure}

Methodological specifics employed in ascertaining centring data from N23 cases are outlined here; then, notable differences in approach required for S24 and K20 are discussed to follow. The N23 dataset features six $\muL$-cases at each of $\Fr\in\{1,1.5,2,3,4\}$ and a liquid viscosity spread of $\muLN\in\{1,5.5,10.3,15.4,20.3,30.4\}$. In their experiments, water served as the non-viscous baseline and differing mixtures of water and glycerin were utilized to concoct higher-$\muL$ fluid types. Nose photos from N23\textemdash brightened using the ``Age'' filter in Inkscape\textemdash are showcased in figure \ref{fig:04} with corresponding case numbers and $(\Fr,\muL)$ pairs. With respect to tail photos, readers are referred to figure 6 of the original study \citet{NaidekEA23}. Because superficial velocity ratio $\gamma$ can only be determined for cases $\{\mathrm{i.j}\mid\forall\mathrm{j}\}$, the majority of N23 flow-cases are not fully characterized.\footnote{Namely, in N23, superficial flow rate pairs $(\uGS,\uLS)$ are supplied in a table that is separate from long bubble photos (i.e., without cross-identifying labels). For cases $\{1.\mathrm{j}\mid\forall \mathrm{j}\}$, a singular $(\uGS,\uLS)$ pair corresponds to each photographed long bubble; thus $\gamma$ is known. For $\{\mathrm{i.j}\mid \mathrm{i}\ge 2,\forall \mathrm{j}\}$, however, multiple possible $(\uGS,\uLS)$ pairs can be constructed for each long bubble flow-case; hence, $\gamma$ is unknown. An attempt was made to calculate missing $\gamma$-values using supplied values of Weber number ($\We=\rhoL(\uLS)^2 D/\sigma$) in combination with $\Fr$ and thus $\um$; however, a range of $\We$ is given in N23 for each $\Fr$-level (i.e., exact values of $\We$ are unknown\textemdash see figure 6 of their study). Using $\We_\text{avg}$ to determine $\uGS$ leads to significant errors and thus $\gamma$-values could not be reasonably found using this approach.}

To ensure accurate determination of centring metrics, each photographic data-point was calibrated; that is, the pipe's fluid-filled interior was distinguished from the upper and lower edges of solid pipe wall. It is assumed that each collection of photos, as originally published, has identically shared dimensionality, meaning that no geometric alterations were performed that were not implemented for the entire batch. Upon transference of images from source to Inkscape, aspect ratio was retained perfectly; thus, the above assumption permits definition of a constant inner diameter applicable to the entire set, used to normalize centring distances. Regardless, diameter was measured for each photo and an average was taken to negate minor deviations. To do so, image boundaries were inspected at the maximum possible level of magnification (25,600x) and a thin horizontal bar with obviously different pixel density\textemdash relative to the pipe interior\textemdash was discovered. This artifact is presumed to represent solid pipe; therefore, its inward edge was chosen to mark the boundary of inner pipe diameter, applied to both upper and lower limits. Absolute units of length alluded to here are irrelevant outside of the Inkscape session from which they were born, hence the usage of relative metrics which allow scaling to experimental design. Vertical centring distances were collected procedurally using Inkscape's ruler-tool, performed with meticulous care yet still subject to a small degree of human error. When uncertainty was encountered\textemdash for example, in instances where dispersed bubbles obscure visualization of the gas-liquid interface\textemdash a partial-contour was approximately drawn to dictate measurement limits (not shown here).

Also recorded are qualitative measures of interfacial smoothness and small bubble entrainment.\footnote{\ ``Interfacial smoothness'' here refers to the long bubble's lower boundary.} Both classifications have three categories: smooth (S), wavy (W) and transitional (T) for the former; non-negligible (Y), near-negligible (N*) and negligible (N) for the latter. Such distinctions are important; tortuosity along the phasic boundary logically indicates turbulence in the adjacent liquid while significant dispersed bubble content results from large shear forces ``ripping'' gas out of its bulk container. Further, interfacial non-smoothness may affect evaluation of centring metrics; for example, a random peak situated at a designated measurement location would lead to a reduced value of reported separation in comparison to immediately neighbouring regions\textemdash an outlier. Ideally, an averaged bubble contour would be obtained based on a large number of photos taken at identical operational conditions to reduce the impact of stochastic structuring; however, limited data availability necessitates that singular snapshots are assumed to provide sufficient representations of mean interfacial placement. 

\begin{figure}[!t]
\centering
\includegraphics[width=\textwidth]{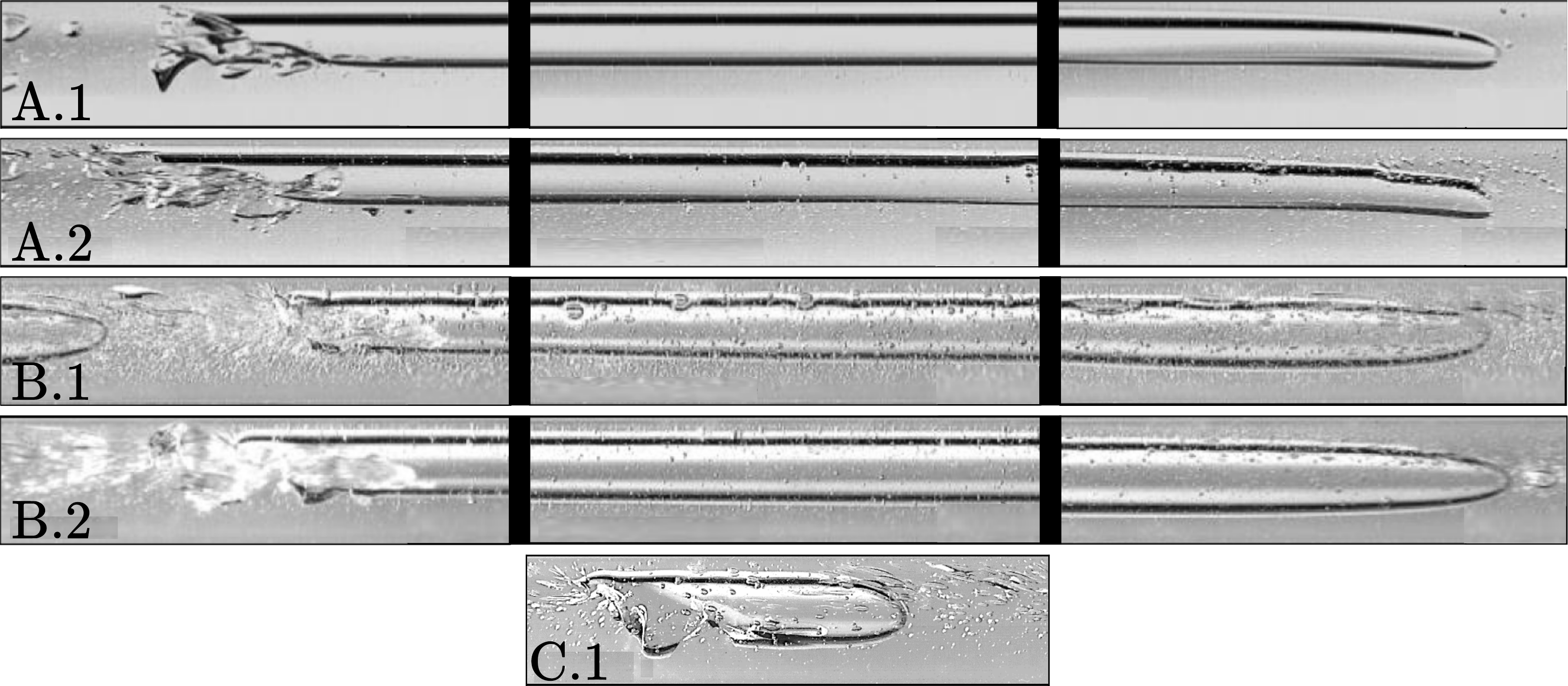}
\caption{Raw, unaltered images of elongated bubbles from \citet{ShinEA24} (S24 dataset) overlaid with case numbers/set notation $\mathrm{m.k}$: bubble nose (right), body (middle) and tail (left) (consolidated for $\mathrm{C}.1$). Case $\mathrm{A}.1$: $\muL=37.7\unit{mPa.s}$; $\Fr=1.92$; $\gamma=3.25$. Case $\mathrm{A}.2$: $\muL=37.7\unit{mPa.s}$; $\Fr=3.23$; $\gamma=0.78$. Case $\mathrm{B}.1$: $\muL=352\unit{mPa.s}$; $\Fr=1.81$; $\gamma=3.00$. Case $\mathrm{B}.2$: $\muL=352\unit{mPa.s}$; $\Fr=2.94$; $\gamma=0.59$. Case $\mathrm{C}.1$: $\muL=352\unit{mPa.s}$; $\Fr=0.68$; $\gamma=0.50$. Black bars separating nose/body/tail photos do not represent scaled length. Adapted from figures 10 and 11b of original source with permission from Elsevier Publishing.}
\label{fig:05}
\end{figure}

The S24 ($\mathrm{m.k}$) dataset features HVL slug flow with two synthetic oil profiles $\muLN\in\{37.7,352\}$, both of which are more viscous than any liquid used in N23.\footnote{The original S24 paper also includes photographic plug flow data, differentiated from slug flow data (see figure 8 of source); however, only designated cases of slug flow were utilized for analysis here, primarily because the majority of plug flow cases stem from ultra-low $\Fr$-values or extremes in $\gamma$ and an entire study could be devoted to characterizing bubble detachment. Further, bubble shapes are, in many cases, atypical, meaning that metrics such as $\yT$ would require redefinition. One plug flow case (figure 8d in S24) is an operational counterpart to case $\mathrm{C.1}$ used here; as such, its centring measurements were taken and are mentioned in brevity in \S\ref{sec:3.1} despite it not being a formal case.} Included are two $\Fr$-values for the lighter oil $(\Fr)_{\muLN=37.7}\in\{1.92,3.23\}$ and three for the heavier oil $(\Fr)_{\muLN=352}\in\{0.68,1.81,2.94\}$; also, $\gamma$-values are available for each S24 flow-case. Rearranged in contrast to their original publication, raw images are displayed in figure \ref{fig:05}. Analyses similar to those described for N23 were performed; however, unlike in N23, long bubble body photos are available and thus values of $\yB$ were measured using a spatial average as shown in figure \ref{fig:03}a. Case $\mathrm{C.1}$ is an anomaly in that a singular photo captures all of nose, body and tail regions. Its unusually short length renders bubble body and tail inseparable; thus, special treatment was implemented. Namely, values of $\yN$, $\yOD$ and $\yT$ were measured normally but\textemdash since the tail location is equivalently $\lambda_{\scriptscriptstyle 1.34D}$\textemdash $\yTD$ does not exist and $\yB$ cannot be properly measured. As such, the $\yT$-value for $\mathrm{C.1}$ is not straightforwardly comparable to those of other cases; regardless, it is kept for completeness. For all other S24 cases $\{\mathrm{m.k}\backslash \mathrm{C.1}\}$, the five defined centring metrics were calculated in the above standardized manner. 

\begin{figure}[!t]
\centering
\includegraphics[width=0.5\textwidth]{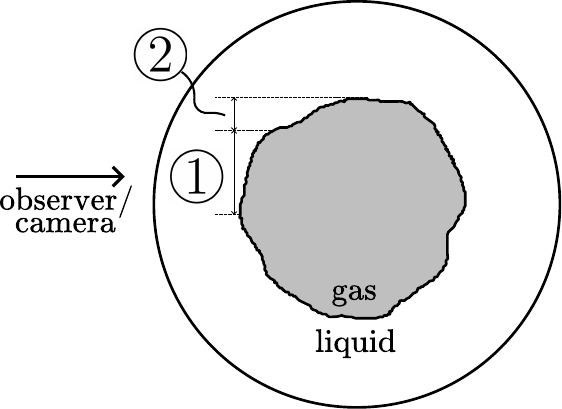}
\caption{Hypothetical depiction of optical phenomenon observed in \citet{ShinEA24} (S24) elongated bubble flow-cases. Region \Circled{1} is a thick black line that, on first glance, appears to be the top of the gas-liquid interface; however, region \Circled{2} represents the actual top. Readers are referred to figure \ref{fig:05} for real examples.}
\label{fig:06}
\end{figure}

An additional layer of complexity presented itself in S24 data analyses; namely, the laboratory-specific illumination system evidently creates\textemdash in resultant photos\textemdash two distinct, differently coloured regions at the long bubble's upper interface: a thick black line underneath a layer of white. Careful discernment is thus required to accurately determine the top-of-interface, necessitating knowledge of a long bubble's 3-dimensional nature. In N23 data, there is less ambiguity since bubble contours are black and plainly contrasted by the grey surrounding liquid. In S24 flow-cases, however, the bubbles' global maxima must be selected according to the top of the aforementioned white region since the black line corresponds to a false top localized by the observer/camera. This issue is least apparent in case $\mathrm{A.1}$, plausibly due to a cross-sectionally flat bubble and thus merging of the two regions. Figure \ref{fig:06} offers a simplistic depiction of the optical phenomenon. 

\begin{figure}[!t]
\centering
\includegraphics[width=0.9\textwidth]{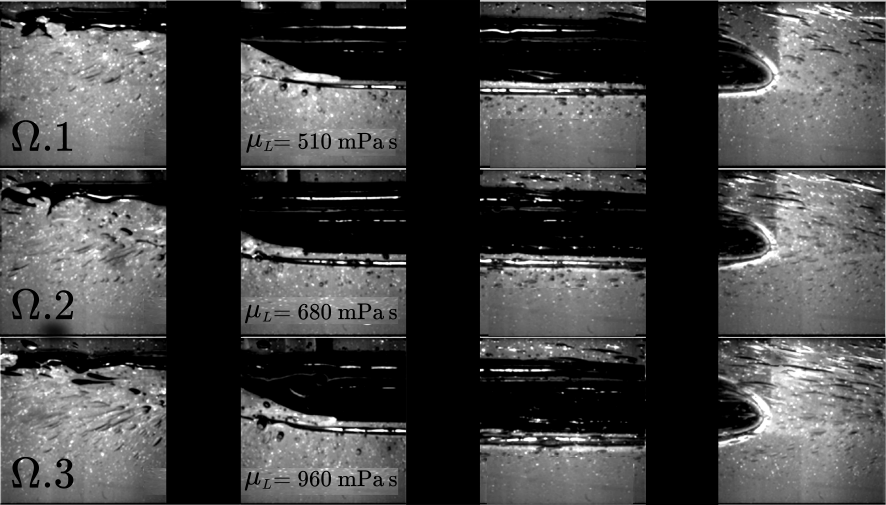}
\caption{Elongated bubble photos from \citet{KimEA20} (K20 dataset) spliced together to form coherent illustrations. Middle dividers and thus total bubble length are not to-scale; left and right dividers are approximately scaled. For all $\Omega.\mathrm{q}$, $\Fr =0.57$ and $\gamma=1.00$; $\muL$-values shown within figure. Adapted from figure 26 of original source with permission from Elsevier Publishing. Photos were brightened/sharpened using Inkscape's built-in ``Age'' filter; otherwise unaltered.}
\label{fig:07}
\end{figure}

Visual data from K20 ($\Omega.\mathrm{q}$)\textemdash simultaneously representing the highest $\muL$-range $\muLN\in\{510, 680, 960\}$ and lowest inertial supply $\Fr=0.57$ studied here (the latter being a constant in $\Omega.\mathrm{q}$ cases with $\gamma=1$)\textemdash were extracted and analyzed using methodology presented above with a few minor logistical deviations. Images were rearranged from their original presentation and are provided here in figure \ref{fig:07} wherein each case is composed of four photographic segments. Black separating bars are designed approximately to-scale at left and right divisions through extrapolation of interfacial evolution; however, since total bubble length is not supplied, middle dividers are non-representative of missing distance. Since required visuality is absent for $\yTD$ and $\yB$ locations\textemdash instead residing somewhere within the middle dividers\textemdash only $\yOD$, $\yN$ and $\yT$ metrics were measured for K20 cases.\footnote{Alternative slug flow cases are shown in K20; however, only the three included here are published in raw format. The others utilize a lighting system at the upper pipe wall which renders centring distances undetectable; thus, they are not used here.}

Lastly, it is worth discussing the role of refraction in two-phase pipe flow visualization. Owing to non-unity refractive indices of the liquid phase and solid pipe (often made from acrylic as in the three datasets investigated here), photographic data generated from flowing pipe-systems may be subject to distortions (e.g., differences between true and observed liquid-levels). Many relevant image processing studies, such as those from \citet{MoralesEA11} and \citet{WidyatamaEA18}, do not account for refraction in their algorithms; however, a select few do. For example, a visual study of the stratified-annular transition from \citet{CherdantsevEA22} utilizes a simplistic refraction model to correct the measured height of the liquid interface to better reflect reality. They consider refraction of light only through the transparent pipe material as it impacts the top of the liquid interface along the wall (e.g., for an annular flow with a fully-wetted pipe-interior, the determined liquid height would be $D/2$ relative to pipe-centreline). To calculate the magnification coefficient ($M=h_\text{obs}/h_\text{true}$) requires inner and outer pipe diameter; further, they 1) assume that rays of light reach the pipe travelling horizontally and 2) use a diffusive screen to achieve uniform illumination. Their model outputs $M_\text{max}=1.34$ (at the upper pipe wall) and $M_\text{avg}=1.17$ (calculated by digitizing figure 3b from their study) while the error in wall film height measurement is reduced\textemdash by incorporating refraction\textemdash only from $\pm0.7\unit{mm}$ to $\pm0.6\unit{mm}$. 

Ideally, the aforementioned refraction model would be applied to bubble centring data extracted from sources considered here; however, there are some substantial problems in this regard:
\begin{enumerate}
\item in all of N23, S24 and K20, the pipe outer diameter\textemdash an important variable in calculating $M$\textemdash is not supplied;
\item the laboratory-specific lighting-system is important in determining the impact of refraction\textemdash none of N23, S24 or K20 have utilized a diffusive screen; nor have they disclosed the details of their lighting-systems (e.g., the angle at which they align with the pipe); and, 
\item the \citet{CherdantsevEA22} model corrects the height of a singular liquid interface (i.e., the height of liquid on the pipe-wall)\textemdash since liquid coats the entire pipe-interior around centred long bubbles, the model would not be useful in analyses performed here and a more complex formulation would be necessitated, requiring an intricate understanding of refractive effects (e.g., without a diffusive screen, the angle of incidence would likely be different at upper and lower interfaces). 
\end{enumerate}

For these reasons, correcting for refraction in the present empirical study is deemed to be not feasible; and, this plausible discrepancy is noted as a potential source of error. In \citet{CherdantsevEA22}, $M>1$ for all pipe regions; therefore, for their particular flow-system, the true liquid interface is situated below what is observed. Were this conclusion applied plainly to centring distances, it would cause a slight additional increase in $\yk$-values; however, given the myriad difficulties listed above, this cannot be stated with any real certainty and subsequent inquiry is required to understand the impact of refraction on centring extent. Even if refractive distortions are significant in N23, S24 and K20 data, it is important to realize that\textemdash since none of the photos are corrected\textemdash relationality within each set would remain unchanged with marginally shifted reference valuations. Further, if we assume each source to have used relatively similar pipe thickness-values and lighting alignment, this preservation of relationality would apply on a broader, generalized scale. 

\subsection{Model derivation}\label{sec:2.2}

Capitalizing on a relatively large sample size, correlative models are construed specifically for N23 data. Derived below, the generalized model outputs centring distances as a function of $\Fr$ and $\muL$, applicable, at minimum, within the range of pipe, fluid and operational conditions used in the original source. A characteristic dimensionless number is utilized, denoted $x$ and defined as 
\begin{equation}
x = \Fr \frac{\muL}{\muG}
\label{eqn:05}
\end{equation}
where $\muG$ is gas viscosity\textemdash assumed to be $\muG=0.0181\unit{mPa.s}$ for air at standard pressure and $T=\SI{19}{\celsius}$.\footnote{Determined using \textit{The Engineering Toolbox} online air viscosity calculator: \url{https://www.engineeringtoolbox.com/air-absolute-kinematic-viscosity-d_601.html}; accessed 2024-12-07.} After some manual exploration, the data were found best fit to a function of the form
\begin{equation}
\oy = \Lambda + \Upsilon x + \Theta\ln{x}
\label{eqn:06}
\end{equation}
where $\oy$ can be $\oyOD$, $\oyTD$ or $\oyN$ and constants $\Lambda$, $\Upsilon$ and $\Theta$ are to be calibrated separately for each form of $\oy$-metric. 

Using methodology described in \citet{Chapra12}, least squares regression was employed to optimize equation \ref{eqn:06} in correspondence with inferred bubble centring data. The summation of squared residuals is given by 
\begin{equation}
S_r = \sum_{n=1}^N \left(\oy - \Lambda - \Upsilon x - \Theta\ln{x}\right)^2
\label{eqn:07}
\end{equation}
where $N$ is the number of data points used for a given regression. This problem is one of minimization; namely, 
\begin{equation}
\min_{\Lambda,\Upsilon,\Theta} S_r
\label{eqn:08}
\end{equation}
must be pursued\textemdash done by determining $\partial S_r/\partial \Lambda$, $\partial S_r/\partial \Upsilon$ and $\partial S_r/\partial \Theta$ and equating each to zero, yielding a system of equations 
\begin{equation}
\begin{bmatrix*}[l]
N  & \sum x & \sum \ln{x} \\
\sum x & \sum x^2 & \sum x \ln{x} \\
\sum \ln{x} & \sum x\ln{x} & \sum \ln^2 x
\end{bmatrix*}
\begin{bmatrix*}[l]
\Lambda \\
\Upsilon \\
\Theta
\end{bmatrix*}
=
\begin{bmatrix*}[l]
\sum \oy \\
\sum x\oy \\
\sum \oy \ln{x}
\end{bmatrix*}
\label{eqn:09}
\end{equation}
that is solved straightforwardly to yield values of $\Lambda$, $\Upsilon$ and $\Theta$ for each form of $\oy$. The outlined model was carried out algorithmically using MATLAB v.R2024a. Owing to limited data availability, correlations were not derived for S24 and K20; thus, the model is readily applicable only for $\muLN\in[1,30.4]$. Extrapolation is plausible but further work is required to verify the model's usefulness and encapsulate higher-$\muL$ liquids (see \S\ref{sec:3.2}). For a mechanistic approach to modelling the bubble centring phenomenon, readers are again referred to the PL20 study.

\section{Results}
\label{sec:3}
\subsection{Empirical observation}\label{sec:3.1}

\begin{figure}[!t] 
\centering
\includegraphics[width=3.3in]{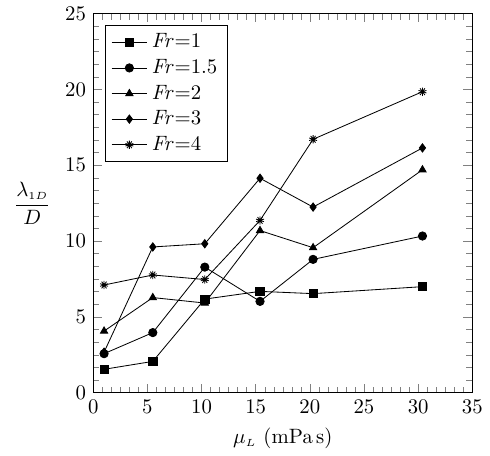}
\caption{Bubble centring data extracted from \citet{NaidekEA23} (N23 dataset): Normalized $\yOD$ (\%) as a function of $\muL$ for fixed values of $\Fr$ and $\um$ (constant $D$).}
\label{fig:08}
\end{figure}

\begin{figure}[!t]
\centering
\includegraphics[width=3.3in]{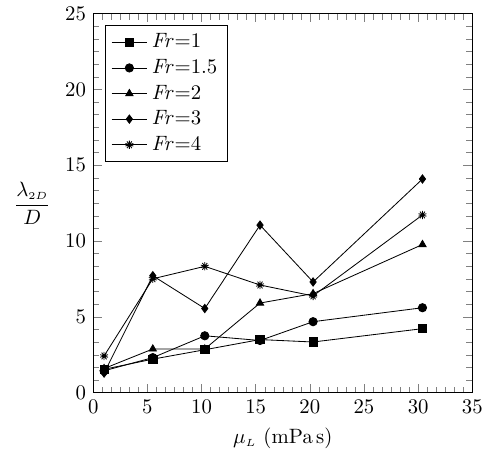}
\caption{Bubble centring data extracted from \citet{NaidekEA23} (N23 dataset): Normalized $\yTD$ (\%) as a function of $\muL$ for fixed values of $\Fr$ and $\um$ (constant $D$).}
\label{fig:09}
\end{figure}

\begin{figure}[!t]
\centering
\includegraphics[width=3.3in]{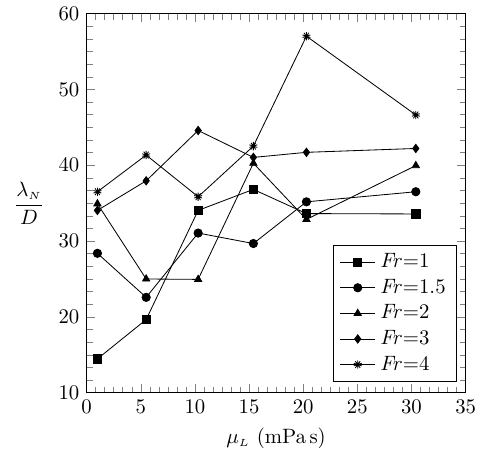}
\caption{Bubble centring data extracted from \citet{NaidekEA23} (N23 dataset): Normalized $\yN$ (\%) as a function of $\muL$ for fixed values of $\Fr$ and $\um$ (constant $D$).}
\label{fig:10}
\end{figure}

Overall, measurements taken from N23, S24 and K20 datasets showcase an intriguing interconnection between bubble centring and liquid viscosity in horizontal slug flow.\footnote{Since the three datasets' $\muL$-ranges do not overlap, the following designations regarding ``levels'' of HVL may prove useful: N23=low; S24=medium; K20=high.} Data from N23 are plotted in two batches of figures: \ref{fig:08}, \ref{fig:09} and \ref{fig:10} in this section; \ref{fig:26}, \ref{fig:27} and \ref{fig:28} in \ref{sec:A2}. The former plots relate $\oyOD$, $\oyTD$ and $\oyN$ to $\muL$ for fixed values of $\Fr$. Using set notation $\mathrm{i.j}$ defined in \S\ref{sec:2.1} wherein fixed $\mathrm{i}$- and $\mathrm{j}$-values correspond to fixed $\Fr$- and $\muL$-values, respectively, these plots display the evolution of $\oy$-metrics for lines of changing $\mathrm{j}$ and constant $\mathrm{i}$. Appended plots show the opposite: $\oy$-metrics versus $\Fr$ for fixed levels of $\muL$; or, lines of constant $\mathrm{j}$ and variable $\mathrm{i}$. Since diameter $D$ is constant within datasets, changing or fixed $\Fr$ in the plots described above represents changing or fixed mixture velocity $\um$ (owing to equation \ref{eqn:01}). Also, because superficial velocity ratio $\gamma$ is unknown for the majority of N23 cases (see \S\ref{sec:2.1}), the impact of selectively changing $\uGS$ or $\uLS$ on centring extent cannot be determined; as such, the plots shown here are the best possible depictions of centring-relationality. Tail-centring ($\oyT$) values from N23 are uniformly either small or null; therefore, they are provided in table \ref{tab:02} using binary designations (i.e., any tail centring or none). For reference, all of the data gathered in this study, along with case-level parameterization, are given in appended table \ref{tab:6}. 

Varying levels of non-linear fluctuation are observed in figures \ref{fig:08}, \ref{fig:09} and \ref{fig:10}; however, in general, there is an obvious tendency for nose-region centring ($z\in[0,2D]$) to enlarge\textemdash for constant mixture inertia and pipe size\textemdash with increasing $\muL$. As originally predicted in PL20, the extent of long bubble centring is a function not only of $\Fr$ and $\gamma$ but of $\muL$. It is further deduced that bubble detachment\textemdash a flow event defined in PL20 synonymous with the initiation of bubble centring\textemdash occurs at lower $\Fr$-values in HVL-containing slug flows than in comparable air-water systems. At $\Fr=1$, for example, $\oyOD$ is near-negligible for $\muLN=1$ yet $>7\%$ for $\muLN=30.4$. 

Analyzing the $\muL$-$\yk$ trends garnered from N23, $\yOD$ and $\yTD$ tell similar stories of different magnitudes\textemdash as is intuitively expected; $\yN$ follows the same generalized proclivity with a superimposed essence of enhanced variability. To understand this important feature, the undulating nature of the bubble nose must be considered. As an elongated bubble translates in an intermittent configuration, its nose-tip will naturally oscillate in the vertical direction to some extent. This is plausibly due to minor fluctuations in velocity and pressure immediately downstream of the nose-tip effecting radial deviations in the non-rigid long bubble structure. A pictorial example of this phenomenon is found in \citet{Diaz16} (paper IV; figure 15; single bubble flow experiment with $\muLN=10$) which elucidates the dynamic behaviour of long bubble shape and positioning in horizontal slug flow. For partially- and fully-centred long bubbles, such fluctuations are potentially more pronounced as the nose-tip has additional space to shift. In this context, ``fluctuation'' could refer to a singular long bubble morphing as it flows or to substantive differences in subsequent long bubble topology and radial placement. % find a better visual example of nose fluctuation with actual slug flow?

Based on above considerations, the amplified variability observed in N23's $\yN$-data (relative to $\yOD$ and $\yTD$) is anticipated since measured nose-tip centring evolution stems from arbitrary snapshots rather than averaged contours. In fact, this could account for non-linearity in all three metrics; and, owing to this concept, $\muL$-$\yk$ curves are expected to flatten somewhat\textemdash while maintaining positive proportionality\textemdash as centring metric location gets farther from long bubble nose-tip. Referring again to figures \ref{fig:08}, \ref{fig:09} and \ref{fig:10}, from $\yN\rightarrow\yOD$ linearity is improved at all $\Fr$-levels; from $\yOD\rightarrow\yTD$, however, curves are flattened for $\Fr\in\{1,1.5,2\}$ while non-linearity is roughly preserved for $\Fr\in\{3,4\}$.\footnote{Another factor in the observed variability of $\muL$-$\yk$ curves is potential changes in $\gamma$ within N23 data-subsets of constant $\Fr$ and $\um$. Such hypothetical changes would not, however, explain the flattening effect noted from $\yN\rightarrow\yTD$.} Generally, $\yOD$ and $\yTD$ are deemed reliable indicators of the incurred degree of long bubble centring while $\yN$ is significantly less relevant due to exaggerated relational volatility. This suggests a paradigmatic change: just as $\yOD$ and $\yTD$ are the empiric focal points in this study, owing to improved spatial-temporal constancy relative to $\yN$, future experimental works must shift away from an evident exclusivity that places nose-tip position as the ultimate measure of bubble centring extent. 

Long bubble tail photos from N23, where available, are assumed to originate from the operational runs of their nose photo counterparts, necessary since their source does not provide $(\uGS,\uLS)$ pairs and instead characterizes images using $(\Fr,\muL)$ labels.\footnote{As mentioned in \S\ref{sec:2.1}, tail photos are not available for N23 cases $\{2.\mathrm{j}\mid\forall \mathrm{j}\}$.} As alluded to earlier, $\yT$-values measured from N23 photos are relatively small, exclusively falling below $\yT<0.06D$; therefore, tail centring results are visualized in binarized form through table \ref{tab:02} (Y $\rightarrow$ any amount; N $\rightarrow$ none). One third of available tail-cases have non-zero $\yT$-values (8 of 24); $88\%$ of those are found in either the highest $\Fr$ ($5.\mathrm{j}$) or $\muL$ ($\mathrm{i}.6$) subsets (7 of 8), shown boldfaced in table \ref{tab:02}. In fact, the majority of N23's highest $\Fr$- or $\muL$-cases $\{5.\mathrm{j}\cup \mathrm{i}.6\}$ exhibit tail centring ($78\%$; 7 of 9). Centring is defined here as the existence of non-negligible liquid separation between bubble and upper pipe wall; therefore, a cut-off is required to classify negligible separation, defined here by $\oyk\leq 2.5\%$. Using this criterion, 7 of 8 cases with non-zero $\yT$ show full-centring; and, of 26 total cases, 15 or $59\%$ are partially-centred by definition (3 water-based; 12 HVL\textemdash at least one for each $\muL$-level). Centring types\textemdash as given in \S\ref{sec:2.1} (FC: full-centring; PC: partial-centring; NC: no-centring)\textemdash are also given in table \ref{tab:02} (bracketed) for all N23 cases, determined using the $2.5\%$ distinction.

\begin{table}[!t]
\begin{center}
\begin{tabular}{lcccccc} 
  & \textbf{1} & \textbf{2} & \textbf{3} & \textbf{4} & \textbf{5} & \textbf{6} \\  \toprule 
\textbf{5}  & N (PC) & \textbf{Y} (FC) & \textbf{Y} (FC) & \textbf{Y} (FC) & \textbf{Y} (FC) & \textbf{Y} (FC) \\
\textbf{4}  & N (PC) & N (PC) & N (PC) & N (PC) & N (PC) & N (PC) \\
\textbf{3}  & N (PC) & N (PC) & Y (FC) & N (PC) & N (PC) & \textbf{Y} (FC) \\
\textbf{2}  & - & - & - & - & - & - \\
\textbf{1}  & N (NC) & N (NC) & N (PC) & N (PC) & N (PC) & \textbf{Y}* (PC) \\ \bottomrule
\end{tabular}
\caption{Extracted tail-centring data for \citet{NaidekEA23} (N23) $\mathrm{i.j}$ cases. Left-most column: $\mathrm{i}$-values; upper-most row: $\mathrm{j}$-values. Y\textemdash tail-centring occurred; Y*\textemdash tail-centring $<2.5\%$; N\textemdash no tail-centring. Bold Y-values correspond to either highest $\Fr$ ($5.\mathrm{j}$) or highest $\muL$ ($\mathrm{i}.6$) subsets. All values observed to be $\yT<0.06D$; numeric results found in \ref{sec:A1}. Shown in brackets are centring types as defined in \S\ref{sec:2.1}, determined using a $2.5\%$ cut-off. Since body photos are unavailable, FC cases are assumed to have $\oyB>2.5\%$.}
\label{tab:02}
\end{center}
\end{table}

\begin{figure}[!t]
\centering
\includegraphics[width=\textwidth]{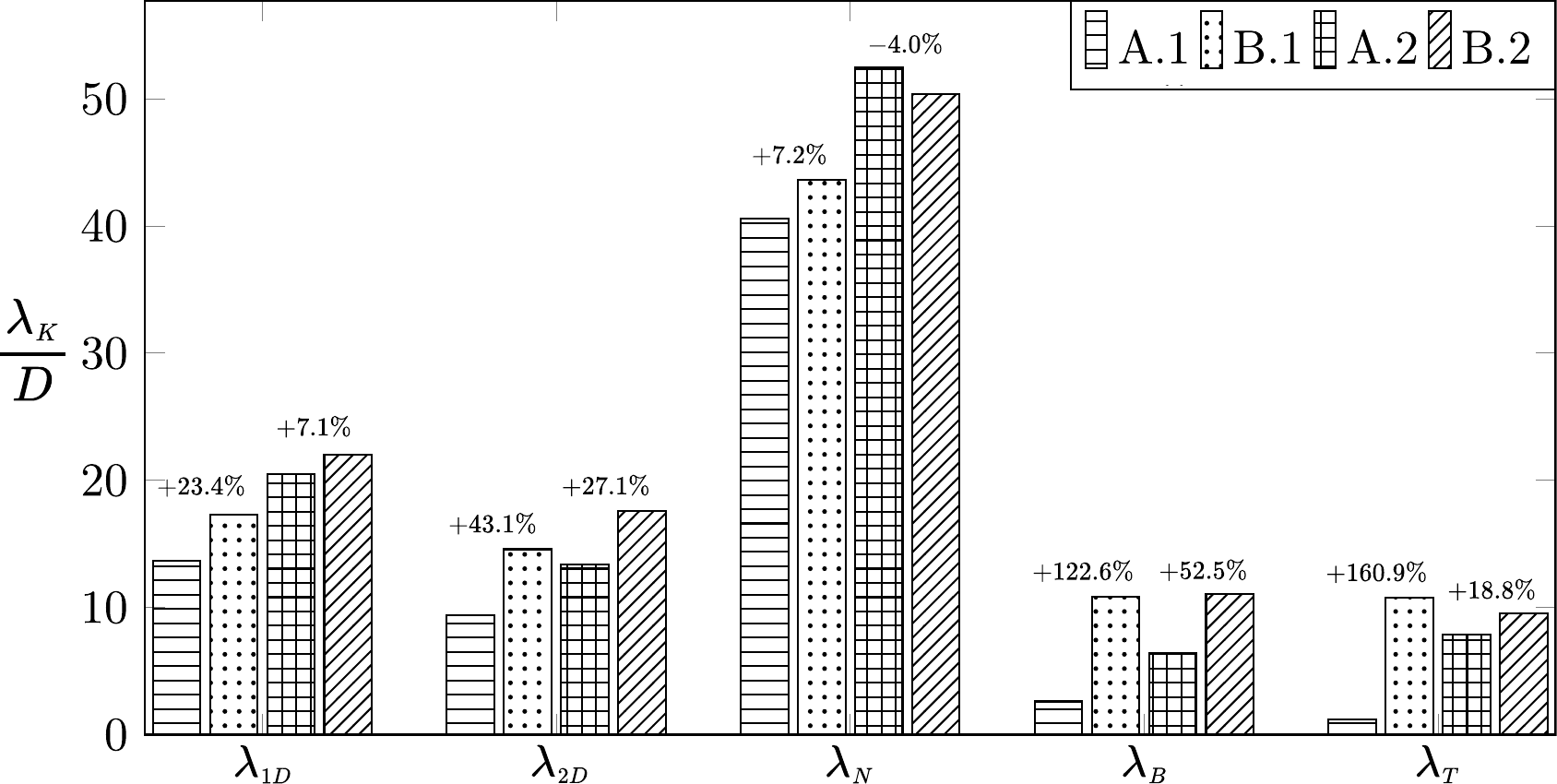}
\caption{Long bubble centring data extracted from \citet{ShinEA24} (S24) $\mathrm{m.k}$ cases. All $\lambda$-values are normalized by $D$ and given as a percentage. Overlaid are $\%$-differences between comparable case-pairs $\{\mathrm{A}.1,\mathrm{B}.1\}$ and $\{\mathrm{A}.2,\mathrm{B}.2\}$ which represent changes in bubble centring extent primarily due to $(\Delta\muL)_{\scriptscriptstyle \mathrm{AB}}=314.3\unit{mPa.s}$. }
\label{fig:11}
\end{figure}

Shifting focus to the medium-HVL designation, long bubble centring data extracted from the S24 ($\mathrm{m.k}$) dataset expands the picture painted by N23 data. Shown in figure \ref{fig:11}, all five presently defined centring metrics\textemdash $\yOD$, $\yTD$, $\yN$, $\yB$ and $\yT$\textemdash are visualized in bar-chart format for cases $\{\mathrm{m.k}\backslash \mathrm{C}.1\}$. Due to anomalous features, case $\mathrm{C}.1$ is treated in isolation. Internally comparable case-pairs are construed here to demonstrate the isolated impact of $\muL$ on centring extent; namely, $\{\mathrm{A}.1,\mathrm{B}.1\}$ and $\{\mathrm{A}.2,\mathrm{B}.2\}$ represent $\muL$-contrasted groupings with similar albeit non-identical underlying operational conditions:

\begin{enumerate}
\item[]\textbf{A.1}: $(\Fr,\muLN,\gamma)=(1.92,37.7,3.25)$
\item[]\textbf{B.1}: $(\Fr,\muLN,\gamma)=(1.81,352,3.00)$
\item[]\textbf{A.2}: $(\Fr,\muLN,\gamma)=(3.23,37.7,0.78)$
\item[]\textbf{B.2}: $(\Fr,\muLN,\gamma)=(2.94,352,0.59)$
\end{enumerate}

Under this logic, figure \ref{fig:11} is designed to elucidate changes in centring distances due primarily to $(\Delta\muLN)_{\scriptscriptstyle \mathrm{AB}}=314.3$ in that $\{\mathrm{A}.1,\mathrm{B}.1\}$ and $\{\mathrm{A}.2,\mathrm{B}.2\}$ pairs are arranged in juxtaposition and overlaid with \%-difference values ($2\Delta\yk/\Sigma\yk$) calculated from $\mathrm{A.k}\rightarrow \mathrm{B.k}$ for each measured centring metric. As was generally observed in N23 analyses, S24 data show that\textemdash for otherwise comparable slug flow systems\textemdash increasing $\muL$ results in significantly amplified long bubble centring tendencies. A near-perfect positive proportionality is found\textemdash all of $\yOD$, $\yTD$, $\yB$ and $\yT$ increase non-negligibly within case-pairs $\{\mathrm{A}.1,\mathrm{B}.1\}$ and $\{\mathrm{A}.2,\mathrm{B}.2\}$, especially so in the former at body and tail locations wherein $+122.6\%$ and $+160.0\%$ differences are seen, respectively. The only exception lies in $\yN$ measurements. From $\mathrm{A}.1\rightarrow \mathrm{B}.1$, $\yN$ incurs a $+7.2\%$ difference; however, from $\mathrm{A}.2\rightarrow \mathrm{B}.2$, a $-4.0\%$ difference is found. This is clearly a stochastic artifact stemming from the long bubble nose-tip's susceptibility to flow-induced oscillation\textemdash established earlier in this section. Observing that $\mathrm{A}.2$ and $\mathrm{B}.2$ have $\yN$-values near $0.5D$\textemdash the theoretical limit for bubble centring\textemdash the discrepancy becomes explicable due to undulations of the nose-tip about the pipe centreline; thus, such divergence is noncontradictory to the overall derived relationality between $\muL$ and centring extent. 

Centring measurements for S24's case $\mathrm{C}.1$ are given by $(\oyOD,\oyN,\oyT)=(11.9,43.9,11.8)\%$. For reasons described in \S\ref{sec:2.1}, these values are not straightforwardly compared to those obtained for $\mathrm{A.k}$ and $\mathrm{B.k}$ cases; for example, $12\%$ tail-centring does not indicate that all system-specific slug flows at $\Fr=0.68$ will experience the same because $\yT=\lambda_{\scriptscriptstyle 1.34D}$ for $\mathrm{C}.1$. Since $\gamma=0.5$ or, equivalently, $\uGS=0.5 \uLS$, increasing gas supply while holding $\Fr$ constant\textemdash thus reducing liquid supply\textemdash may result in a far longer bubble and, by extension, an altered tail-centring profile.\footnote{Confirmation of $\mathrm{C}.2$'s outlier status is found by comparing its $\yOD$- and $\yT$-values to those from $\mathrm{B.k}$ cases (which share $\muLN=352$); namely $(\yOD)_{\mathrm{C}.1}<(\yOD)_{\mathrm{B.k}}$ and $(\yT)_{\mathrm{C}.1}>(\yT)_{\mathrm{B.k}}$ for all $\mathrm{k}$. Since $(\Fr)_{\mathrm{C}.1}<(\Fr)_{\mathrm{B.k}}$, the former is expected and the latter is not, thus confirming that $\yT$-values cannot be compared and $\mathrm{C}.1$ is indeed exceptional.} Regardless, the case serves as an intriguing theoretical quandary, demonstrating that full-centring can occur in HVL slug flows with extremely low inertial input. For reference, full-centring in air-water flows is associated with $\Fr\geq 3.5$ \citep{Bendiksen84} while initial detachment is found approximately within $2.2\leq\Fr\leq 2.4$ as interpreted using experiment \citep{deOliveiraEA15} and modelling efforts \citep{PerkinsLi20}. That full-centring realization is possible for diminutive operational flow rates is a remarkable feature of high-$\muL$ horizontal slug flow, one that solidifies a novel empirical precedent in the widening mechanistic differential between HVL- and water-based systems.\footnote{The lighter-oil plug flow analog to case $\mathrm{C}.1$\textemdash mentioned in footnote 16 (\S\ref{sec:2.1}) and found in figure 8d of S24's original paper\textemdash is not formally included here; regardless, its centring metrics are given as follows: $\oyOD=6.60\%$, $\oyTD=5.10\%$, $\oyN=35.29\%$ and $\oyT=4.68\%$ where $\yT=\lambda_{\scriptscriptstyle 3.72D}$. Compared with $\mathrm{C}.1$, a significant generalized increase in centring is incurred due to $(\Delta\muL)_{\scriptscriptstyle \mathrm{AB}}$. Using the $2.5\%$ negligibility criterion, the plug flow case demonstrates full-centring, suggesting that low-$\Fr$ tendencies displayed by $\mathrm{C}.1$ hold true also for the lighter-oil (however, case $\mathrm{A}.1$ shows partial-centring). For reference, the case in question is derived from $\Fr=0.72$, $\gamma=0.6$ and $\muLN=37.7$.} 

Regarding centring-type designation for other S24 data, 3 of 4 flow-cases $\{\mathrm{A.k}\cup \mathrm{B.k}\backslash \mathrm{A}.1\}$ exhibit full-centring in compliance with the $2.5\%$ negligibility threshold. Only for $\mathrm{A}.1$\textemdash the lowest $\Fr$-valued run at $\muLN=37.7$\textemdash was the cut-off invoked since $\oyT=2.64\%$ and $\oyB=1.17\%$. A useful metric which gauges the uniformity of incurred centring is $\oyT/\oyOD$ in that $\oyT/\oyOD\rightarrow 1$ suggests equidistant centring throughout the entire long bubble while $\oyT/\oyOD\rightarrow 0$ represents the edge-case of partial-centring. For S24 cases, $\oyT/\oyOD$ values of 0.09, 0.62, 0.38 and 0.43 are found for cases $\mathrm{A}.1$, $\mathrm{B}.1$, $\mathrm{A}.2$ and $\mathrm{B}.2$, respectively, numerics which yet again increase within comparable case-pairs and thus as a positive function of $\muL$. For pairs of constant $\muL$ with increasing $\Fr$ and decreasing $\gamma$, however, $\mathrm{A}.1\rightarrow \mathrm{A}.2$ shows an increase in $\oyT/\oyOD$ while $\mathrm{B}.1\rightarrow \mathrm{B}.2$ shows a marked decrease.\footnote{While pairs $\{\mathrm{A.1},\mathrm{A.2}\}$ and $\{\mathrm{B.1},\mathrm{B.2}\}$ can, in general, highlight the impact of changing both $\Fr$ and $\gamma$ on centring metrics (with $\muL$ fixed), they cannot convey the impact of changing either $\Fr$ or $\gamma$ individually\textemdash a limitation rooted in data availability within S24.} Interestingly, while $(\oyOD)_{\scriptscriptstyle \mathrm{B}.2}>(\oyOD)_{\scriptscriptstyle \mathrm{B}.1}$, it also holds true that $(\oyT)_{\scriptscriptstyle \mathrm{B}.2}<(\oyT)_{\scriptscriptstyle \mathrm{B}.1}$, meaning that the nose-extent of centring in $\mathrm{B}.2$\textemdash the higher-$\Fr$ case\textemdash is larger but the evenness of centring is enhanced in $\mathrm{B}.1$. Both $\mathrm{B.k}$ cases tend toward perfect-centring realization (see \S\ref{sec:2.1}), more so, in fact, than any others studied here. For example, case $\mathrm{B}.2$ shows $22.0\%$ detachment at $z=D$ while, at the same location, the bubble occupies $54.8\%$ of inner pipe height; this leaves $23.2\%$ or $0.232D$ underneath the bubble, meaning that $\oyOD$ is merely $0.6\%$ less than conditions resembling near-nose perfect symmetry. At the bubble tail, there is $11.0\%$ detachment and a bubble height of $54.2\%$, yielding a space of $34.8\%$ or $0.348D$ beneath; thus, if $\oyT$ were increased $11.9\%$ in this case, perfect symmetry would be approximately realized for the entirety of the film region.\footnote{However, a peculiar observation arises in comparing body- to tail-centring for S24 cases: $\oyB<\oyT$ for $\{\mathrm{A}.1,\mathrm{B}.2\}$; $\oyB\approx\oyT$ for $\mathrm{B}.1$; and, $\oyB>\oyT$ for $\mathrm{A}.2$. This suggests that $\oyT/\oyOD$ is only a rough measure of centring evenness in that separation distances do not always decrease linearly from nose-tip to tail.}

Another notable observation\textemdash not seen as clearly in N23 data\textemdash is garnered from the S24 dataset; namely, increasing $\muL$ drastically impacts elongated bubble shape, particularly at the nose region. In the lighter-oil cases $\{\mathrm{A.k}\mid\forall \mathrm{k}\}$, the bubble nose is thin and markedly curved toward the downward direction, resulting in plainly visible asymmetry; in the heavier-oil cases $\{\mathrm{B.k}\mid\forall \mathrm{k}\}$, however, the bubble nose is thick and symmetrical, especially in the higher-$\Fr$ version $\mathrm{B}.2$ wherein perfect-centring is nearly realized for at least $4D$ from nose-tip, as discussed above. Within comparable case-pairs $\{\mathrm{A}.1,\mathrm{B}.1\}$ and $\{\mathrm{A}.2,\mathrm{B}.2\}$, $\Fr$ and $\gamma$ are numerically similar; since roughly equal volumes of gas and liquid are supplied amidst pairings, it is evident that $(\Delta\muL)_{\scriptscriptstyle \mathrm{AB}}$ alters the 3-dimensional distribution of phases. S24's authors mention that higher-$\muL$ results in curtailing of bubble length. For the lighter-oil cases, gas is stretched over a longer distance and thus thinner\textemdash viewed from a vantage perpendicular to the direction of flow\textemdash than in the heavier-oil cases. Improvements in symmetry signify greater axial coherence of centring mechanisms or downward suction applied to the bubble. That such an enhancement is correlated with increasing $\muL$ is conducive to the overall narrative derived here. To further comprehend the impact of $\muL$ on long bubble shape and centring manifestation, cross-sectional photos\textemdash obtained using specialized experimental design such as that utilized in \citet{JamariEA08}\textemdash are necessary.

\begin{figure}[!ht]
\centering
\includegraphics[width=0.75\textwidth]{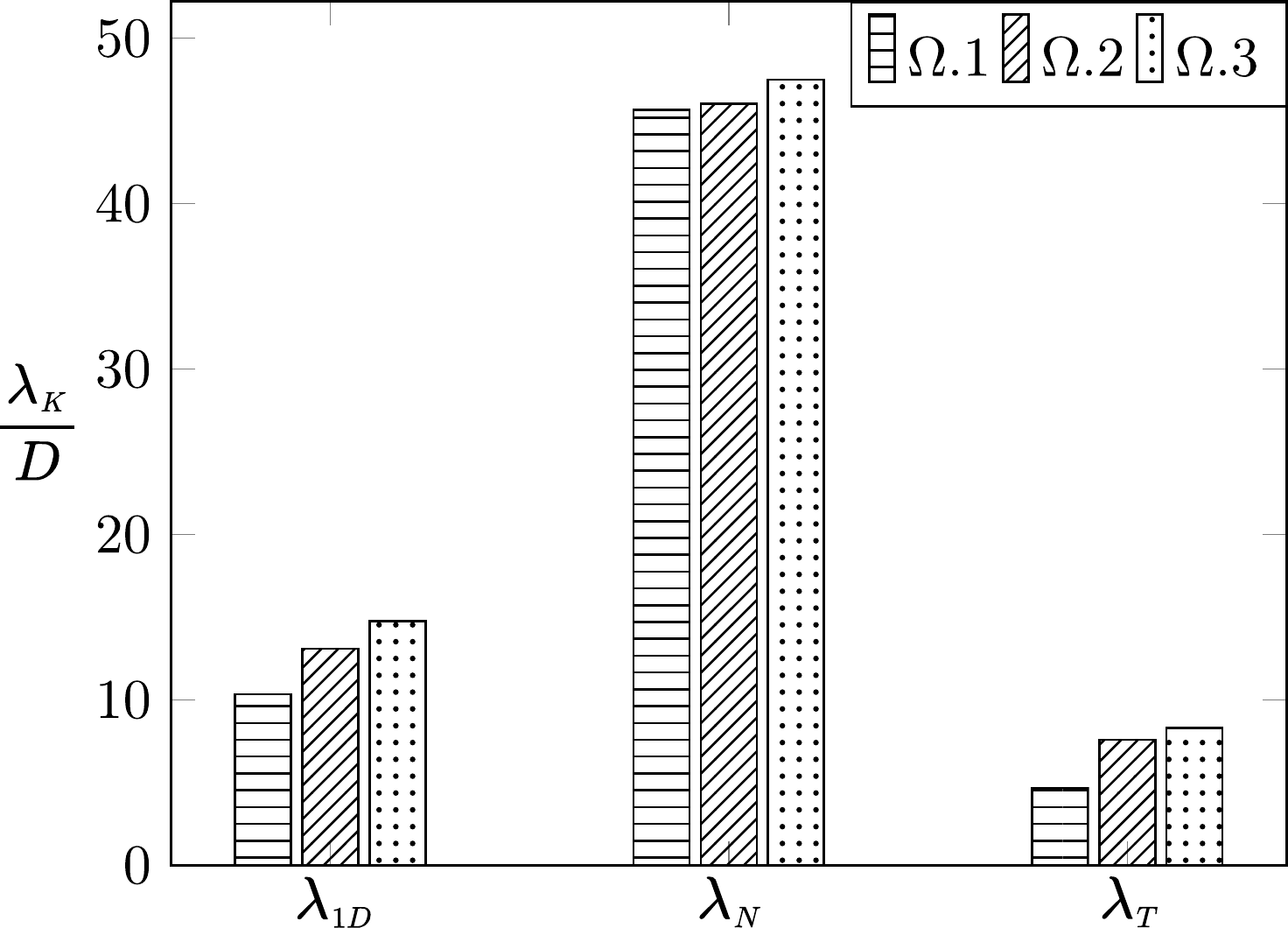}
\caption{Long bubble centring data extracted from \citet{KimEA20} (K20) $\Omega.\mathrm{q}$ cases. All $\lambda$-values are normalized by $D$ and given as a percentage. Here, $\muL\uparrow$ as  $\mathrm{q}\uparrow$.}
\label{fig:12}
\end{figure}

Bubble centring measurements extracted from K20\textemdash representing the highest range of $\muL$-values studied here with $\muLN\in\{510,680,960\}$\textemdash are shown in figure \ref{fig:12}. All three $\Omega.\mathrm{q}$ cases feature equivalent operational conditions: $(\Fr,\gamma,D)=(0.57,1,50.8\unit{mm})$. Since non-negligible $\yT$-values are measured in each, these flow-cases remarkably exemplify the tendency of HVL slug flow systems to exhibit full-centring under relatively low inertial input. As in S24's $\mathrm{C}.1$, $\Omega.\mathrm{q}$ cases display complete detachment from the upper pipe wall for $\Fr<0.7$. Unlike $\mathrm{C}.1$, however, K20 cases are observed with relatively standard bubble lengths, rendering them non-anomalous and thus stronger examples of the aforementioned low-$\Fr$ tendency in HVL flows. As outlined in \S\ref{sec:2.1}, only $\yOD$, $\yN$ and $\yT$ centring metrics are available for K20 flow-cases; for all three, a positive change is incurred across each of the incremental $\muL$-increases: $(\Delta\muLN)_{\scriptstyle 12}=170$ from $\Omega.1\rightarrow\Omega.2$ ($+28.6\%$) and $(\Delta\muLN)_{\scriptstyle 23}=280$ from $\Omega.2\rightarrow\Omega.3$ ($+34.1\%$). Table \ref{tab:03} provides \%-difference values for both the two $\muL$-shifts and for total change: $(\Delta\muLN)_{\scriptstyle 13}=450$ from $\Omega.1\rightarrow\Omega.3$ ($+61.2\%$). The largest step-wise differences are seen in $\yT$ and $\yOD$ for $\Omega.1\rightarrow\Omega.2$ with $+47.4\%$ and $+23.7\%$, respectively, while the smallest differences are observed in $\yN$ with $\leq+3.1\%$ for both $\muL$-shifts. 

\begin{table}[!t]
\begin{center}
\begin{tabular}{lcccc} 
  & $\muL$ & $\yOD$ & $\yN$ & $\yT$  \\  \toprule 
$\Omega.1\rightarrow\Omega.2$  & $+28.6$ & $+23.7$ & $+0.8$ & $+47.4$ \\
$\Omega.2\rightarrow\Omega.3$  & $+34.1$ & $+11.9$ & $+3.1$ & $+9.0$ \\
$\Omega.1\rightarrow\Omega.3$  & $+61.2$ & $+35.4$ & $+3.9$ & $+55.8$ \\ \bottomrule
\end{tabular}
\caption{\%-difference values for \citet{KimEA20} (K20) bubble centring measurements, representative of changes induced by corresponding $\Delta\muL$. $\Omega.1\rightarrow\Omega.2$: $(\Delta\muL)_{\scriptstyle 12}=170\unit{mPa.s}$; $\Omega.2\rightarrow\Omega.3$: $(\Delta\muL)_{\scriptstyle 23}=280\unit{mPa.s}$; $\Omega.1\rightarrow\Omega.3$: $(\Delta\muL)_{\scriptstyle 13}=450\unit{mPa.s}$. Values given in \%, calculated using standard $2\Delta\yk/\Sigma\yk$ or $2\Delta\muL/\Sigma\muL$.}
\label{tab:03}
\end{center}
\end{table}

As mentioned, all three K20 cases exhibit full-centring under the $\oyk>2.5\%$ negligibility cut-off. The generalized extent of centring is significantly less in $\{\Omega.\mathrm{q}\mid\forall \mathrm{q}\}$ than in S24's $\{\mathrm{B.k}\mid\forall \mathrm{k}\}$ cases as evidenced by $\oyOD$ and $\oyT$ measurements; however, this is expected owing to the rather low $\Fr$-value used in available K20 flow-cases. Values of approximate centring evenness $\oyT/\oyOD$ are calculated to be 0.45, 0.58 and 0.56 for cases $\Omega.1$, $\Omega.2$ and $\Omega.3$, respectively, suggesting relatively coherent uniformity of centring in all instances. Across $(\Delta\muLN)_{\scriptstyle 12}$, an increase in $\oyT/\oyOD$ is incurred ($+24.4\%$); across $(\Delta\muLN)_{\scriptstyle 23}$, however, a near-negligible decrease is observed ($-2.9\%$), suggesting that centring evenness stabilizes for $\muLN\geq 680$ in the system/conditions of interest.\footnote{Also worth noting is that all three K20 flow-cases feature a smooth gas-liquid interface, suggestive of laminarity in the liquid film. The same is observed in S24 cases; however, in $\mathrm{A}.2$ and $\mathrm{B}.1$ minor non-linearities are found on the upper (near-nose) bubble boundary. A variety of interfacial classifications are found in N23 cases. As is made clear in \S\ref{sec:4} and PL20, the existence of film-laminarity is critical in the mechanics of bubble centring.}

Next, the continuity of utilized datasets is investigated. In this context, ``continuity'' is the potentiality for seamless extrapolation of data-based trends between differing sources. Were it not established, a question would arise as to whether empirically derived conclusions stem from an irreproducible environment\textemdash due, for instance, to group-dependent methods or parameterization\textemdash thus making inter-laboratory integration unfeasible. To begin, certain differences between N23, S24 and K20 datasets must be clarified; most notably, all were born using different pipe sizes. N23 and S24 are most similar in this regard with a $26.1\%$ difference; with respect to K20, there are $87.0\%$ and $64.6\%$ differences in contrast with N23 and S24 pipe diameters, respectively. Another discrepancy lies in employed liquid-types: N23 used water-glycerin mixtures while both S24 and K20 used synthetic oils with differing values of liquid density $\rhoL$ and surface tension $\sigma$, as reported in \S\ref{sec:2.1}. The role of non-$\muL$ fluid property diversity in bubble centring realization is yet to be studied in-depth; however, increasing $\rhoL$ is intuitively counterproductive to centring as a denser liquid is more inclined to remain underneath the long bubble owing to gravity.\footnote{The exact function of $\rhoL$ is less simplistic and can be understood via the PL20 centring model (e.g., it shows up in the underlying TB90 algorithm and thus affects film-height and -velocity profiles). For reference, PL20 also offers a detailed discussion of the phenomenological role of gravity and buoyancy in centring mechanics.} The impact of differing $\sigma$ on centring extent is less clear albeit plausibly more important for smaller pipes. Despite significant differences in non-$\muL$ liquid properties, it is assumed that N23 and S24 form\textemdash given their similar values of $D$\textemdash an optimal pair for the purpose of probing empiric continuity. K20 is ignored in this endeavour due to its non-conforming pipe size and uniformly low values of $\Fr$, in that the majority of N23/S24 cases do not have similar $\Fr$-cases in K20 to be compared with. 

Judging by case-level contrasting, the combined results of N23 and S24 bubble centring inquiry craft a tapestry of robust connectivity. For example, $\mathrm{A}.1$ of S24 is best compared to N23's cases $2.6$ and $3.6$: $\mathrm{A}.1$ features $(\Fr,\muLN)=(1.92,37.7)$ while $2.6$ and $3.6$ are characterized by $(\Fr,\muLN)=(1.5,30.4)$ and $(\Fr,\muLN)=(2.0,30.4)$, respectively, meaning that $(\Fr)_{\scriptscriptstyle 2.6}<(\Fr)_{\scriptscriptstyle \mathrm{A}.1}<(\Fr)_{\scriptscriptstyle 3.6}$ and $(\muLN)_{\scriptscriptstyle \mathrm{A}.1}>(\muLN)_{\scriptscriptstyle 2.6/3.6}$ (representing a difference of $\Delta\muLN=7.3$ or $21.4\%$). Centring analyses for $\mathrm{A}.1$ show that $(\oyOD,\oyTD)=(13.69,9.41)\%$ whereas $(\oyOD,\oyTD)=(10.33,5.60)\%$ and $(\oyOD,\oyTD)=(14.70,9.77)\%$ are obtained for $2.6$ and $3.6$. Thus it stands that 
\begin{equation} \label{eqn:10}
\left(\oyOD\right)_{\scriptscriptstyle 2.6} < \left(\oyOD\right)_{\scriptscriptstyle \mathrm{A}.1} <  \left(\oyOD\right)_{\scriptscriptstyle 3.6}
\end{equation}
and
\begin{equation} \label{eqn:11}
\left(\oyTD\right)_{\scriptscriptstyle 2.6} < \left(\oyTD\right)_{\scriptscriptstyle \mathrm{A}.1} <  \left(\oyTD\right)_{\scriptscriptstyle 3.6}
\end{equation}
wherein the $\mathrm{A}.1$ centring profile is far more similar to that of $3.6$ than of $2.6$. Considering the three cases' $\Fr$-inequality and comparable $\muL$-valuation, the above outcomes are reasonably aligned with continuity. S24's higher-$\muL$ cases $\mathrm{B}.1$ ($\Fr=1.81$) and $\mathrm{B}.2$ ($\Fr=2.94$), both featuring $\muLN=352$, are logically compared to N23's $3.6$ ($\Fr=2$) and $4.6$ ($\Fr=3$) which utilize $\muLN=30.4$ ($\Delta\muLN=321.6$ or $168.2\%$ difference). Centring measurements reveal that 
\begin{equation} \label{eqn:12}
\begin{split}
\left(\oyOD\right)_{\scriptscriptstyle \mathrm{B}.1} &> \left(\oyOD\right)_{\scriptscriptstyle 3.6} \\
\left(\oyTD\right)_{\scriptscriptstyle \mathrm{B}.1} &> \left(\oyTD\right)_{\scriptscriptstyle 3.6}
\end{split}
\end{equation}
and
\begin{equation} \label{eqn:13}
\begin{split}
\left(\oyOD\right)_{\scriptscriptstyle \mathrm{B}.2} &> \left(\oyOD\right)_{\scriptscriptstyle 4.6} \\
\left(\oyTD\right)_{\scriptscriptstyle \mathrm{B}.2} &> \left(\oyTD\right)_{\scriptscriptstyle 4.6}
\end{split}
\end{equation}
which is expected\textemdash in light of the correlative framework presented thus far\textemdash because $(\muL)_{\scriptscriptstyle \mathrm{B.k}}\gg(\muL)_{\scriptscriptstyle \mathrm{i}.6}$. In fact, exhaustive data analyses for sets $\mathrm{i.j}$ and $\mathrm{m.k}$ demonstrate that
\begin{equation} \label{eqn:14}
\max_{\mathrm{i.j},\mathrm{m.k}} \left(\oyOD\right) = \left(\oyOD\right)_{\scriptscriptstyle \mathrm{B}.2}
\end{equation}
and
\begin{equation} \label{eqn:15}
\max_{\mathrm{i.j},\mathrm{m.k}} \left(\oyTD\right) = \left(\oyTD\right)_{\scriptscriptstyle \mathrm{B}.2}\text{;}
\end{equation}
that is, maximal values of $\oyOD$ and $\oyTD$ are encountered in the case of globally maximized $\muL$ (keeping in mind that K20 is not included at present). Furthermore, excluding anomalous case $\mathrm{C}.1$, it holds true that
\begin{equation} \label{eqn:16}
\max_{\mathrm{i.j},\mathrm{m.k}\backslash \mathrm{C}.1} \left(\oyT\right) = \left(\oyT\right)_{\scriptscriptstyle \mathrm{B}.2}\text{.}
\end{equation}

Regardless of systemic differences, relationality derived above suggests that an acceptable level of continuity exists between N23 and S24 datasets. Therefore, the generalized conclusions discovered here are readily extrapolated to other HVL-containing systems with similar parameterization (see table \ref{tab:01}); namely, increasing $\muL$ enlarges the probabilistic likelihood and extent of long bubble centring in horizontal slug flow. This investigation is, however, preliminary. Additional work is required to understand the functional role of $\gamma$, for example, a crucial ratio which is missing for the majority of N23 cases. 

An illuminating consequence arises from the continuity established between N23 and S24 datasets: cumulated tail-centring results yield insight into a critical $\muL$-value, above which full-centring is anticipated for a broad range of $\Fr$-values. For N23 data, a maximum of $\oyT=6\%$ is obtained; regarding S24's $\{\mathrm{m.k}\backslash \mathrm{C}.1\}$, however, $75\%$ of cases have $\oyT>6\%$ and $50\%$ have $\oyT>10\%$, the latter of which stem exclusively from the larger $\muL$ subset. Therefore, a conservative cut-off for statistically likely full-centring occurrence is $\muLN\geq 352$ while the actual boundary lies within $37.7<\muLN<352$ as gleamed from the evident blossoming of tail-centring seen at $\muLN=37.7$. Considerable further inquiry is necessary to truly affirm this delineation; however, given the contextual structure outlined here, the aforementioned range serves as an adequate heuristic. 

Overall, data extraction and analyses delved into here act as a testament of correlative potential, demonstrating straightforwardly that increasing liquid viscosity has an expanding impact on elongated bubble centring in horizontal gas-liquid slug flow. For reference, all discussed results and parameters are found in \ref{sec:A1}.

\subsection{Correlational analysis}\label{sec:3.2}

Presented in \S\ref{sec:2.2}, an empirical model\textemdash described by equation \ref{eqn:06}\textemdash was calibrated using least squares regression to fit N23 bubble centring data, specifically for $\oyOD$, $\oyTD$ and $\oyN$. Derived coefficients and $R^2$-values are included in table \ref{tab:04}. The most successful global match is obtained for $\oyOD$ with $R^2=0.879$. Acceptable conformance is found also for $\oyTD$, for which $R^2=0.739$ is calculated, while the least adequate fit is observed for the nose-tip metric $\oyN$ with $R^2=0.519$. The latter finding aligns with discussion provided in \S\ref{sec:3.1} regarding dynamical variability of the long bubble nose-tip; namely, $\yN$ is the least reliable metric used here for gauging the extent of centring incurred during slug flow. 

\begin{table}[!ht]
\begin{center}
\begin{tabular}{lcccc} 
  & $\Lambda$ & $\Upsilon$ & $\Theta$ & $R^2$ \\  \toprule 
$\oyOD$  & -1.844 & 0.00194 & 1.005 & 0.879 \\
$\oyTD$  & -2.497 &  0.00120 & 0.820 & 0.739 \\
$\oyN$  & 15.106 & 0.00217 & 2.343 & 0.519 \\ \bottomrule
\end{tabular}
\caption{Results of least squares regression applied to bubble centring data from \citet{NaidekEA23} (N23) dataset, as per equations \ref{eqn:06}, \ref{eqn:07}, \ref{eqn:08} and \ref{eqn:09}.}
\label{tab:04}
\end{center}
\end{table}

\begin{figure}[!t]
\centering
\includegraphics[width=\textwidth]{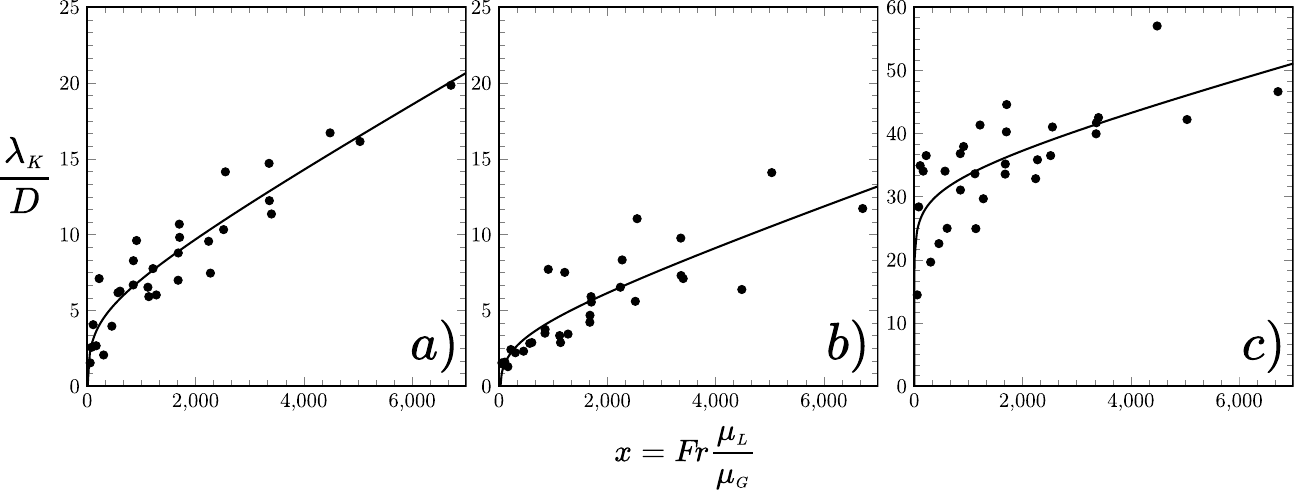}
\caption{Correlative modelling results for \citet{NaidekEA23} (N23) bubble centring data (values given in \%): $\oyk=\yk/D$ vs $x=\Fr \muL/\muG$ for a) $\oyOD$, b) $\oyTD$ and c) $\oyN$. Solid lines: modelling predictions; discrete points: extracted measurements.}
\label{fig:13}
\end{figure}

Figure \ref{fig:13} shows the three obtained correlative models overlaid with measured $\oyOD$, $\oyTD$ and $\oyN$ data from N23. Defined in \S\ref{sec:2.2}, the independent variable used is $x=\Fr \muL/\muG$ which represents a simple coalescence of dynamical influences: mixture inertial input and liquid viscousness, both of which have been shown to affect the degree of realized bubble centring in horizontal slug flows. A factor of $\muG$ is included for the purpose of normalization; $\mu_{\scriptscriptstyle W}$ could have been used identically yielding different coefficient values. The nature and efficacy of equation \ref{eqn:06}\textemdash namely, that centring data can be modelled using a superposition of linear and logarithmic contributions of $x$\textemdash sheds light on future engineering approaches to incorporating the phenomenon into design protocols. These particular calibrations are limited in scope to $\muLN\in[1,30.4]$ and $D=26\unit{mm}$; however, rederivation is straightforward assuming that ample system-specific photographic data are available. Also, going forward, $\gamma$ ought to be merged into the dimensionless group and thus the correlative process. 

To understand potential applications of the presented form of bubble centring correlations, a crucial distinction from the aforesaid PL20 study must be invoked; that is, the difference between long bubble detachment (i.e., centring) and the existence of a thin liquid film above the non-centred gas phase. In PL20, the latter is deemed to be a prerequisite for the initiation of the former. If gas is directly in contact with the upper pipe wall, the downward force\textemdash which is otherwise causal in detachment\textemdash cannot be transmitted onto the bubble. This is because a change in vertical pressure differential is required to initiate centring, not simply a change in pressure underneath the bubble. Introduced in \S\ref{sec:3.1}, a threshold value of $2.5\%$ or $0.025D$ is used to delineate the two unique flow scenarios.\footnote{The $2.5\%$ threshold is roughly based on what the naked eye can plainly see as liquid separation in photographic centring data. Values below $2.5\%$ cannot be measured without magnification and careful discernment. Readers are referred again to the PL20 study for a comprehensive discussion of thin upper film phenomena related to the centring process.} This is summarized as follows:

\begin{enumerate}
\item[]$\mathbf{\oyk>2.5\%}$: bubble centring
\item[]$\mathbf{\oyk\leq2.5\%}$: thin upper liquid film
\end{enumerate} 

Provided a calibrated correlative model constructed using a sufficient amount of bubble centring data, the above cut-off can be utilized to predict the lowest operational rates or $\Fr$-value (for a fixed $D$-value) at which detachment will occur. Using the analytic structure presented in \S\ref{sec:2.2}, an iterative (or graphical) solution is necessary to calculate the critical value of $\Fr$. This approach is tested using coefficients determined here. The Newton-Raphson method, as given in \citet{Chapra12}, was implemented in MATLAB to solve for detachment Froude number $\FrD$ corresponding to $\oyOD=2.5\%$ and $\oyTD=2.5\%$ at all of $\muLN\in\{1,5,10,20,30\}$. Results are presented in table \ref{tab:05}. 

\begin{table}[!t]
\begin{center}
\begin{tabular}{c|c|ccccc} 
%\toprule
 & $\muLN$ & 1 & 5 & 10 & 20 & 30 \\
\midrule
 $\oyOD=2.5\%$ & $\FrD$ & 1.20 & 0.24 & 0.12 & 0.06 & 0.04 \\
\midrule
$\oyTD=2.5\%$ & $\FrD$ & 5.25 & 1.05 & 0.52 & 0.26 & 0.17 \\
\end{tabular}
\caption{Values of detachment Froude number $\FrD$ predicted by N23 correlative models at arbitrary values of $\muLN$. $\FrD$ represents threshold inertial input required to initiate long bubble centring in horizontal slug flow. Delineating separation thickness is defined as $0.025D$. Shown for both $\oyOD$ and $\oyTD$ models.}
\label{tab:05}
\end{center}
\end{table}

For a location $1D$ upstream of long bubble nose-tip, an upper film with $0.025D$ thickness is predicted at $\FrD=1.20$ for air-water slug flow\textemdash an underprediction relative to existing literature yet reasonable regardless. For $\muLN\in\{5,10,20,30\}$, $\FrD$ calculated based on $\oyOD$ is severely low in that obtained values may not represent physical slug flow solutions. This elicits two possible inferences; namely, the existence of either 1) a critical weakness within the correlative model (e.g., too few input data or a weakly formulated independent variable) or 2) a thin upper liquid film by default for the $\muL$-cases of inquiry, logically maintaining a thickness of $\oyOD>2.5\%$. At $2D$ upstream of nose-tip, $\FrD$ is overpredicted for $\muLN=1$, reasonably predicted for $\muLN\in\{5,10\}$ and yet again underpredicted for $\muLN\in\{20,30\}$. That calculated $\FrD$-values are larger for $\oyTD$ than they are for $\oyOD$ is expected since increased inertial supply is necessary to induce more extensive long bubble detachment (``extent'' referring to distance measured upstream from nose-tip). Despite showing promise with respect to outputted order-of-magnitude and underlying relationality, the correlative approach utilized here is a first-order attempt at empirically modelling the bubble centring phenomenon and thus demands further development to become reliably useful. 

\begin{figure}[!t]
\centering
\includegraphics[width=3.3in]{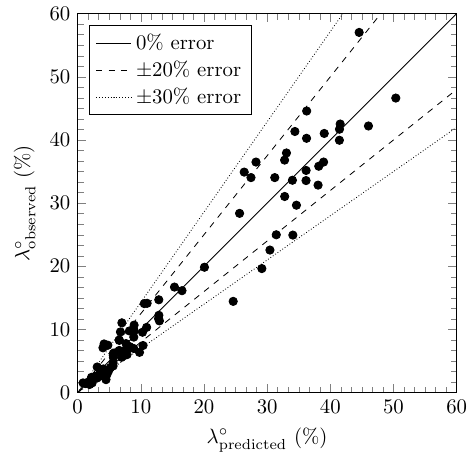}
\caption{Predicted versus observed values of all $\oyOD$, $\oyTD$ and $\oyN$ bubble centring measurements from \citet{NaidekEA23} (N23) dataset overlaid with $0\%$, $\pm 20\%$ and $\pm 30\%$ error lines.}
\label{fig:14}
\end{figure}

Finally, figure \ref{fig:14} shows every $\oyOD$-, $\oyTD$- and $\oyN$-value extracted from N23 flow-cases, both as measured (ordinate) and as predicted by subsequently derived correlative models (abscissa), overlaid with $0\%$, $\pm20\%$ and $\pm30\%$ error lines. Of the 90 centring measurements collected, $58.9\%$ and $76.7\%$ are predicted with $<20\%$ and $<30\%$ error, respectively. Specifically, $\oyOD$ is predicted at $56.7\%$ and $80\%$ accuracy for $<20\%$ and $<30\%$ error; $\oyTD$ at $46.7\%$ and $63.3\%$; and $\oyN$ at $73.3\%$ and $86.7\%$. Considering that long bubble centring has a stochastic element, the generalized predictive capacity of empirical models crafted here\textemdash when applied to data they were born out of\textemdash is deemed satisfactory.  

\section{Theory}
\label{sec:4}
\subsection{Centring mechanics}\label{sec:4.1}

Having established an unambiguous depiction of correlativity, the governing structure of causality must be investigated. Regarding the flow of gas and liquid in a horizontally-oriented pipe, the presence of an HVL introduces a layer of dynamical complexity not encountered in air-water systems. Postulated here is an understated criticality corresponding to the bubble centring phenomenon, in that it must be afforded careful attention in formulating a theoretical framework which encapsulates the mechanistic workings of HVL slug flow. This notion is empirically founded in \S\ref{sec:3.1} which demonstrates, in general, that centring distances expand proportionally with increasing liquid viscosity. The PL20 study pioneered a phenomenological explanation for the prevalence of centring in HVL slug flow; namely, film region liquid laminarity\textemdash put forth as a condition necessary for long bubble detachment\textemdash is probabilistically likely in high-$\muL$ flow-systems. Specifically, coherent streamlines\textemdash flowing in the upstream ($+z$) direction relative to unit-cell translation\textemdash must exist immediately beneath the long bubble for a downward force to be generated. Turbulence can, however, be found simultaneously in other sections of the slug flow architecture; for example, in the mixing region upstream of long bubble tail or adjacent to the film region's lower pipe wall. The laminarity requirement refers solely to liquid directly underneath the long bubble. 

To fully understand the aforementioned streamline concept, the partial-centring configuration must be explored. Air-water and low-HVL slug flows often display partial-centring in that some length of long bubble $z\in[0,\lD]$ flows detached from upper pipe wall while the remainder $z\in(\lD,\lB]$ presents in traditional (non-detached) conformation.\footnote{In conventional literature, $\lf$ is used to describe film region length; however, following PL20, $\lB$ is utilized here to denote elongated bubble length. Generally, the two are equivalent: $\lf=\lB$.} PL20 describes the existence of a thin upper liquid film (defined numerically in \S\ref{sec:3.2}) as an additional phenomenological prerequisite in the initiation of long bubble centring; axial extent of centring, then, is limited by the sustained extent of thin upper film\textemdash represented by $z=\luf$. Film region laminarity is assumed in PL20 analyses and two distinct cases of thin upper film realization are delineated: 1) $\luf=\lB$ with rigid body mechanics applied to the long bubble, such that total pressure differential generates a singular downward force on a non-deformable entity (full-centring or none) and 2) $\luf<\lB$ under non-rigid bubble dynamics, such that downward forces apply only on $z\in[0,\luf]$ (partial-centring or none). Here, a modified hypothetical approach to characterizing partial-centring is utilized; namely, granted that a thin upper liquid film stretches over the entire long bubble $\luf=\lB$, a regime change partway through the film region acts to modulate partial-centring manifestation. 

At this point, it is vital to recognize that liquid velocity distribution in the film region is a predominant driving factor in the production of pressure alterations needed to initiate long bubble detachment in horizontal slug flow. As such, a brief overview of slug flow mechanics is provided here based on the unit-cell assumption\textemdash wherein slug and film units flow in alternating sequence with equivalent geometric and dynamic configuration\textemdash as per the \citet{TaitelBarnea90} (TB90) model.\footnote{TB90: \citet{TaitelBarnea90} unit-cell slug flow model. Note that TB90 did not cover or incorporate the bubble centring phenomenon.} Adhering operational vantage to that of the moving unit-cell\textemdash or, equivalently, to elongated bubble translational velocity $\ut$\textemdash liquid in the film appears, illusorily, to move backward or upstream. A liquid parcel located within the film at $t_0$ is picked-up by the preceding or upstream slug region at some $t_1>t_0$; then, the parcel is shed from the slug to be absorbed by the following film region at $t_2>t_1$. This cyclic process unfolds because the unit-cell's interfacial structure translates faster than all of the present liquid.\footnote{A similar cycle occurs for the gas phase in that gas exits the long bubble's tail (pick-up) before entering the subsequent bubble at its nose (shedding).} 

\begin{figure}[!t] 
\centering
 \includegraphics[width=\textwidth]{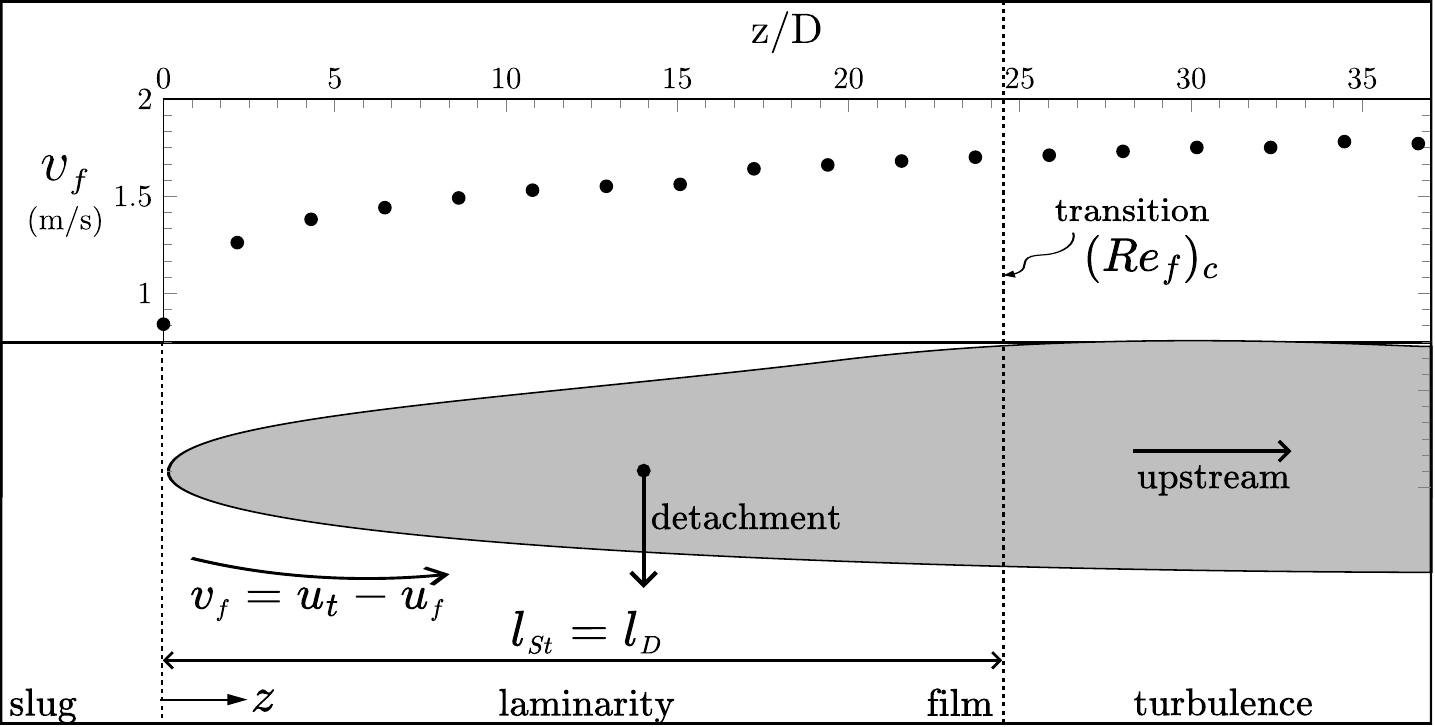}
\caption{Illustration of a novel hypothesis for partial-centring in horizontal slug flow: film liquid velocity\textemdash relative to unit-cell translation\textemdash creates two regimes within the film: 1) a downstream laminar region wherein a vertical force can be applied to the corresponding section of bubble, potentially causing detachment and 2) an upstream turbulent region wherein the mode of vertical force transmission is broken. Overlaid are air-water slug flow data from \citet{BeltLeinan15} which demonstrate a $\vf$-profile in the film; however, centring-configuration and critical film-Reynolds number $\Reyfc$ shown are not empirically based (included for interpretive purposes only).}
\label{fig:15}
\end{figure}

As predicted by the TB90 analytical model, absolute velocity of liquid in the film $\uf$ decreases non-linearly with increasing distance from bubble nose-tip $z$.\footnote{Indicated also by experiment for air-water slug flow; see figure 8 in \citet{BeltLeinan15}.} Therefore, relative film velocity $\vf=\ut-\uf$ increases with $z$, leading to the following conclusion in PL20: changes in $\vf$ along the long bubble's lower interface create\textemdash owing to Bernoulli's principle\textemdash a pressure differential between the thin upper film and liquid beneath the gas, mediated by film region laminarity. Since, under the unit-cell simplification, the film region is an inertial or non-accelerating system, a logical argument is made for the potential existence of a dual-regime film configuration: laminarity near-nose and turbulence near-tail, segregated by a transitive spatial marker at which flow forces overwhelm viscous regularity due to increasing relative motion $\Delta\vf$. This is analogous to transitional distance beyond the inlet of single-phase pipe flow. Considering the context outlined above, this concept\textemdash demonstrated visually in figure \ref{fig:15}\textemdash is hypothesized here to hold a causal role in the realization of partial-centring. 

Summarizing figure \ref{fig:15}, the defined streamline assumption remains valid for a distance $\lC$ from bubble nose-tip; beyond that, due to a critical relative velocity being reached, it disintegrates as smooth flow trajectories morph into diffusive chaos. As such, only for $z\in[0,\lC]$ can centring mechanisms be applied; and, plausibly, the transitional location marks also the length of potential bubble detachment $\lD$.\footnote{In this theory, there is a possible scenario in which $\lC\neq\lD$. Envisioning an incremental\textemdash rather than summative\textemdash vertical force balance applied to the bubble, partial-centring could, hypothetically, occur with $\lD<\lC$ if $\Sigma F_y(z)$ changes sign before the laminar region desists (with respect to $z$). Thus, the two length metrics are defined separately; however, unless otherwise invoked, $\lC=\lD$ can generally be assumed.} In the turbulent film segment $z\in(\lC,\lB]$, postulated centring events are blocked and detachment cannot occur, thus actualizing the partial-centring configuration. Regardless of its axial extent, film region laminarity does not guarantee the occurrence of long bubble centring\textemdash it is simply a necessary condition. As formulated in PL20, centring occurs if the catalyzed downward force is greater than the counteracting buoyant tendencies acting upon the bubble.\footnote{This explains why superficial rate ratio $\gamma$ affects realization of bubble centring: input gas supply\textemdash and, by proxy, long bubble volume and buoyant forces\textemdash could be increased while mixture inertia is maintained. Doing so would decrease liquid input and thus velocities relative to bubble motion; however, since the downward force is a function of $\Delta\vf$, not $\vf$, the exact impact of changing $\gamma$ on centring is not immediately evident, thus elucidating the inherent complexity present in slug flow.} Film laminarity thus serves as a modulator in facilitating the centring phenomenon\textemdash should the right dynamical conditions exist\textemdash as does the existence of a thin upper liquid film. 

The outlined theory of partial-centring is seamlessly extrapolated to the case of full-centring; namely, realization of the latter necessitates\textemdash within the liquid adjacent to the lower gas-liquid interface\textemdash laminarity throughout the entire axial span of the film region $z\in[0,\lB]$. For practical and theoretical purposes, an adapted form of Reynolds number\textemdash characterizing the laminar-turbulent transition in the film region's flowing liquid\textemdash is desirable. Since, in the system of interest, the frame-of-reference is affixed to the steadily translating, structurally constant slug unit-cell, apparent liquid motion traverses the film region from long bubble nose-tip to tail. Therefore, a plausible approach to modelling the dynamic balance between viscous and inertial forces within the film is given by
\begin{equation} \label{eqn:17}
\Reyf(z) = \frac{\rhoL (\ut-\uf) \hf}{\muL}
\end{equation}
where $\hf=\hf(z)$ is the film height profile. To determine a numeric value for critical film-Reynolds number $\Reyfc$, considerable experimental inquiry is needed; regardless, its conceptualization is a valuable tool in differentiating potentialities of partial- and full-centring manifestation. For example, full-centring is realized only if 
\begin{equation} \label{eqn:18}
\forall z\in[0,\lB]\colon\Reyf < \Reyfc 
\end{equation}
while partial-centring occurs according to
\begin{equation} \label{eqn:19}
\begin{split} 
\forall z\in[0,\lD)\colon&\Reyf < \Reyfc \\
\forall z\in(\lD,\lB]\colon&\Reyf > \Reyfc \text{.}
\end{split}
\end{equation}
Using this logic, the demonstrable prevalence of bubble centring in HVL-containing slug flow systems becomes clearly comprehensible. Imagine, as an exemplar, a $\muL$-elevation of $\mathcal{O}(10^3)$ relative to the water baseline. Assuming that order-of-magnitude remains constant for $\rhoL$, $\sigma$ and spatial averages $\overline{\vf}$ and $\overline{\hf}$, average film-Reynolds number $\overline{\Reyf}$ will equivalently experience an $\mathcal{O}(10^{-3})$ diminishment owing to equation \ref{eqn:17}. Consequently, the likelihood of equation \ref{eqn:18} holding true\textemdash necessary for full-centring realization\textemdash will increase monumentally; meaning, in conditions otherwise producing an adequately large downward force, film region laminarity enables the centring process to proliferate across the long bubble's entirety. 

\subsection{Boundary layer theory}\label{sec:4.2}

Complications arise in the systemic treatment of relative liquid motion and shear in the film region of horizontal slug flow. Because shear emerges from absolute motion in a wall-bounded flow-system, careful justification must be invoked in utilizing equation \ref{eqn:17}; namely, gas-liquid and liquid-wall contacts in the film region must be considered.\footnote{More accurately, shear arises according to velocity relative to the wall; that is, we could technically utilize the bubble-frame (i.e., "relative velocity") to analyze shear if a negative (upstream) velocity were imposed on the wall boundary\textemdash in the absence of such a transformation, the lab-frame (i.e., "absolute velocity") must be used.} Triviality is encountered in the case of liquid inviscidity wherein friction and thus shear are nullified such that relative and absolute motions can be treated equivalently. This may be a plausible approximation for water-based flows yet is difficultly reasoned for HVLs. PL20 approached this quandary by defining negligible interfacial shear as a phenomenological simplification while stipulating that liquid-wall shear could remain present in the film. Regarding the gas-liquid boundary as an inviscid surface, a Bernoulli energy balance can be applied in a manner inspired by \citet{Benjamin68}, using relative velocities (outputted by TB90) to calculate the pressure profile beneath the bubble. For systems featuring an HVL, informed discernment is necessary in modelling the phasic interface. Liquid-wall and gas-wall shear clearly increase with viscosity regardless of flow-regime (e.g., under Blasius-type friction models\textemdash see TB90); interfacial shear, however, is strongly dependent on wave presence as investigated by \citet{TzotziAndritsos13}. From their study on stratified flow, there are three distinct interfacial configurations to consider, the realization of which are contingent on input superficial gas rate $\uGS$ compared to critical transition rates $(\uGS)_\alpha$ and $(\uGS)_\beta$ ($f_i/f_{\scriptscriptstyle G}$: ratio of interfacial-to-gas friction factors):\footnote{The film region in slug flow can be treated as a stratified flow with variable liquid height $\hf(z)$. This complicates the numerical approach in that $f_i/f_{\scriptscriptstyle G}$ must be treated iteratively\textemdash changing with $z$\textemdash particularly near bubble nose. Usage of gas velocity in friction factor models is tricky. Plausibly, $u_{\scriptscriptstyle G}$\textemdash average gas velocity above the interface\textemdash makes sense; however, since calculating $u_{\scriptscriptstyle G}$ requires values of $f_i/f_{\scriptscriptstyle G}$, $\uGS$ is the simplest option.}

\begin{enumerate}
\item[]\textbf{Smooth interface}: no wave disturbances present; occurs for low input gas rates $\uGS<(\uGS)_\alpha$; modelled by
	\begin{equation} \label{eqn:20}
	\frac{f_i}{f_{\scriptscriptstyle G}} = 1
	\end{equation}
\item[]\textbf{2D waves}: small-amplitude, short-wavelength, regular waves; occurs for $(\uGS)_\alpha<\uGS<(\uGS)_\beta$ only if $\muLN<20$; modelled by
	\begin{equation} \label{eqn:21}
	\frac{f_i}{f_{\scriptscriptstyle G}} = 1+0.35\left(\frac{\hL}{D}\right)^{0.5} \big(\uGS-(\uGS)_\alpha\big)
	\end{equation}
\item[]\textbf{Large amplitude waves}: irregular, large waves (also called roll-waves or Kelvin-Helmholtz waves); occurs for $\uGS>(\uGS)_\beta$; modelled by
	\begin{equation} \label{eqn:22}
	\frac{f_i}{f_{\scriptscriptstyle G}} = 2\left(\frac{\hL}{D}\right)^{0.1}\big(\muLN\big)^{0.1} + 4 \left(\frac{\hL}{D}\right)^{0.5} \big(\uGS - (\uGS)_\beta\big)
	\end{equation}
\end{enumerate}

Summarizing, if the interface is smooth, it can be reasonably modelled using ideal flow theory as $f_i=f_{\scriptscriptstyle G}$ and $f_{\scriptscriptstyle G}$ is small since $\muG\ll\muL$; if waves are present, however, there are two possibilities with respect to interfacial shear for HVL systems: 1) if $\muLN<20$, $f_i$ is $\muL$-independent for low $\uGS$ yet $\muL$-dependent for high $\uGS$ and 2) if $\muLN\geq20$, $f_i$ is a function of $\muL$ for all $\uGS$. PL20 bubble centring methodology can be updated for HVL-inclusion by integrating the outlined interfacial shear model into the underlying unit-cell slug model (i.e., TB90), thus altering output of $\vf$ and $\hf$ profiles for use in the vertical force balance. In doing so, an inviscid energy balance arguably remains applicable since governing slug flow dynamics are transformed to reflect high-$\muL$ conditions. Another approach, potentially useful in tandem, is to introduce an additional head-loss term to account for frictional energy losses, analogous to what is done in \citet{GokcalEA09} wherein analysis from \citet{Benjamin68} is extended to model long bubble drift velocity for HVL slug flows.\footnote{One necessary difference, however, is that the head loss term\textemdash denoted $\Delta$ in \citet{GokcalEA09}\textemdash would presumably require opposite signage due to the usage of relative velocity in energy conservation.} Based on data examined here, a smooth interface is typically observed for $\muLN\geq 37.7$ (S24 and K20) while smooth or wavy interfaces are observed for $\muLN\leq 30.4$ (N23); however, a wider range of $\Fr$-values should be studied to better comprehend trends in boundary configuration. 

Less straightforward to resolve is the role of liquid-wall shear in bubble centring phenomenology\textemdash especially for systems featuring an HVL. In verity, a boundary layer (BL) of variable thickness $\delta$ will form along the inner pipe wall during slug flow, creating two dynamically unique regimes within the film region: 1) an outer flow immediately underneath the long bubble governed by relative motion and 2) an inner near-wall flow governed by absolute motion and viscous effects.\footnote{BL: boundary layer.} For centring to initiate, then, the two flows must remain distinctly separate; meaning, the film region BL growth rate must be minimal as to not constrict and nullify the outer flow, at least long enough for centring mechanisms to occur. Under this dichotomy, it is reasonable to classify the outer flow as approximately inviscid\textemdash particularly if a smooth interface persists. 

\begin{figure}[!t] 
\centering
 \includegraphics[width=\textwidth]{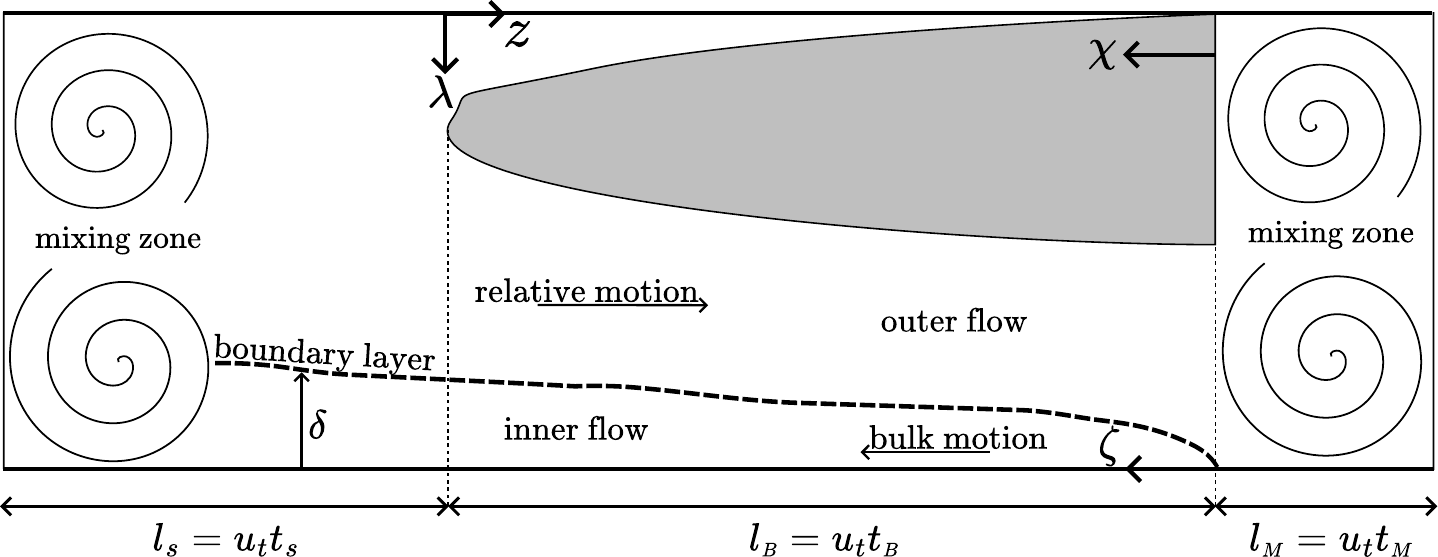}
\caption{Boundary layer (BL) theory for horizontal gas-liquid slug flow. Depicted are two dynamically unique regions within the film: 1) an outer flow governed by relative motion and 2) an inner flow governed by near-wall viscous effects and bulk/absolute motion. Each unit-cell has an identically formed BL which begins at the first instance of mixing region emergence. Snapshot corresponds to $\chi=\zeta$.}
\label{fig:16}
\end{figure}

BL development in horizontal gas-liquid slug flow is complex, nuanced and\textemdash to the author's knowledge\textemdash not yet theoretically grounded within the relevant base of literature. Even in temporally evolved flow-systems, BL expansion is a function of time. At an arbitrary, spatially-fixed location, the velocity profile above the lower pipe wall\textemdash which dictates BL height akin to an overbearing free-stream\textemdash changes depending on which slug flow region encapsulates the point. Two alternative spatial metrics are introduced here: 1) the lateral inverse of $z$, defined as
\begin{equation} \label{eqn:23}
\chi=\lB-z\text{,}
\end{equation}
which represents film region distance measured downstream of long bubble tail and 2) an axial datum $\zeta$ fixed near the pipe control volume (PCV) inlet which marks the most upstream position, or starting point, of an arbitrary slug unit's BL.\footnote{PCV: pipe control volume\textemdash pipe region from phasic inlet to outlet.} BL height is thus expressed as 
\begin{equation} \label{eqn:24}
\delta=\delta(\zeta,t)\text{.}
\end{equation}
The question as to proper placement of $\zeta$ must be explored. The intuitive approach is to align $\zeta$ precisely with the PCV inlet. However, a novel hypothesis is formulated here: emergence of a turbulent mixing region\textemdash formed due to a hydraulic jump at the connection of slug and long bubble tail\textemdash must generate enough turbulent kinetic energy to disrupt and reset BL manifestation. Stated differently, the accelerative pick-up process must convey enough momentum to overwhelm and deconstruct coherent structures initiated at PCV inlet and otherwise sustained within the viscous BL. Therefore, the first\textemdash as in, most upstream\textemdash instance of mixing zone spatiality is defined here as ground-zero for BL development in a slug unit-cell; or, $\{\delta=0\mid\forall t\}$ at $\zeta=0$.\footnote{This is, of course, a simplification; considering the complexity of slug formation\textemdash as, for example, discussed in \citet{TaitelDukler77}\textemdash near-wall flow structures may be disturbed at a location upstream of the first fully-developed mixing zone or, alternately, BL destruction may not be complete and artifacts of near-PCV-inlet BL may survive the hydraulic jump.} Conforming to the notion that every unit-cell is equivalent with respect to geometry and dynamics, mixing regions of conjoining cells are postulated to be segregating limits for BL axial span, as depicted in figure \ref{fig:16} for $\chi=\zeta$\textemdash conditions at which long bubble tail synchronizes with BL origin. Alternatively viewed, a BL break-up mechanism\textemdash such as that put forth above\textemdash is entirely necessary in maintaining ideological sanctity of the unit-cell assumption, in that it would dissolve without.

Established above, separation of inner and outer film flow regions is a fundamental requirement in the manifestation of elongated bubble centring; that is, streamlines of relative motion must not be hindered by near-wall, absolute motion-driven viscous effects. Inner flow laminarity is not necessary\textemdash BL turbulence and bubble detachment could simultaneously prevail so long as dynamical regularity is sustained in the outer flow. Figure \ref{fig:16} illustrates 2-dimensional BL growth in slug flow viewed perpendicular to flow direction; however, pipe geometry requires that 3-dimensional effects must be considered. In the film region, BL growth is cross-sectionally concave in accordance with the lower pipe wall\textemdash analogous to an open-channel flow\textemdash which, in effect, reduces contact-area between a non-centred long bubble and the outer flow. Assuming that velocity distribution remains unaffected, this reduction impacts centring in that the generated downward force applied to the long bubble decreases in proportion to the resultant decrease in interfacial contact area; thus, detachment conditions $\FrD$ will be underpredicted without incorporating the inner/outer film flow distinction into modelling methodology.\footnote{Proof of BL impact on $\FrD$: case 1 $\rightarrow$ no BL assumed; case 2 $\rightarrow$ BL inclusion. For a fixed $\Fr$, assume that $\vf(z)$, $\hf(z)$, $\lB$ and thus $\Delta p$ beneath the bubble are equal in both cases. In simple terms, $F=\Delta p A$ where $F$ is downward force and $A$ is contact-area between streamlines and long bubble; thus, $F_1=\Delta p A_1$ and $F_2=\Delta p A_2$. Since the film BL decreases contact-area, $A_2<A_1$ and $F_2<F_1$; therefore, if $\Fr=\FrD$ is calculated under the pretence of case 1 (no BL assumed), then an additional $\Delta\Fr$ is needed, in actuality (case 2), to cause detachment. As such, $\FrD$ is underpredicted.} 

\begin{figure}[!t]  
\centering
 \includegraphics[width=\textwidth]{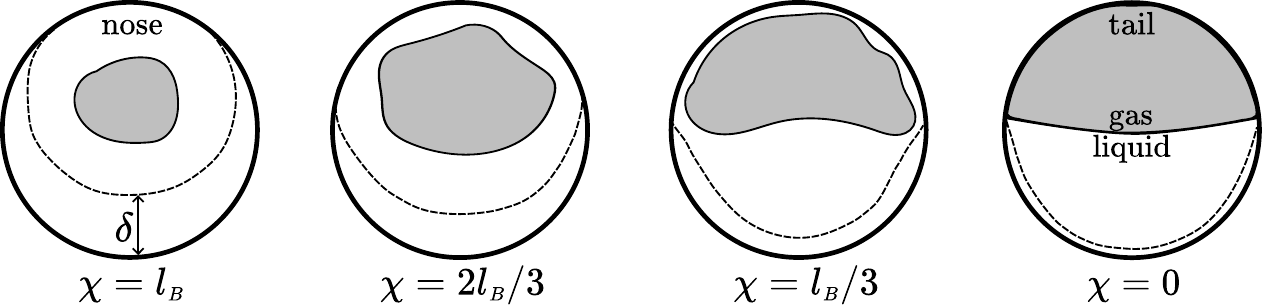}
\caption{Cross-sectional evolution of hypothetical film region boundary layer at four points throughout a partially-centred elongated bubble for $\chi=\zeta+\xi$ where $\xi$ is a small distance.}
\label{fig:17}
\end{figure}

Figure \ref{fig:17} provides a speculative depiction of cross-sectional BL development at four equally spaced locations within the film region of a partially-centred long bubble, from tail ($\chi=0$) to nose ($\chi=\lB$) for $\chi=\zeta+\xi$ where $\xi$ is a small distance. Evidently, there is an axially-inverse relationship between gas-liquid interfacial area and BL extent $\delta$ in that long bubble nose and tail are exposed to the thickest and thinnest segments of BL, respectively. Thus, as illustrated visually, the tendency of side-wall BL growth to reduce contact-area is minimal at the bubble nose; and, since detachment first occurs at $\chi=\lB$, film BL configuration is, seemingly, conducive to centring mechanism transmission. Further, as mentioned earlier, it is crucial to realize that an environment of inner/outer film layer separateness is needed only momentarily, for centring can manifest and then by maintained\textemdash by different mechanisms\textemdash while BL growth extinguishes the outer flow. 

\begin{figure}[!t] 
\centering
 \includegraphics[width=0.9\textwidth]{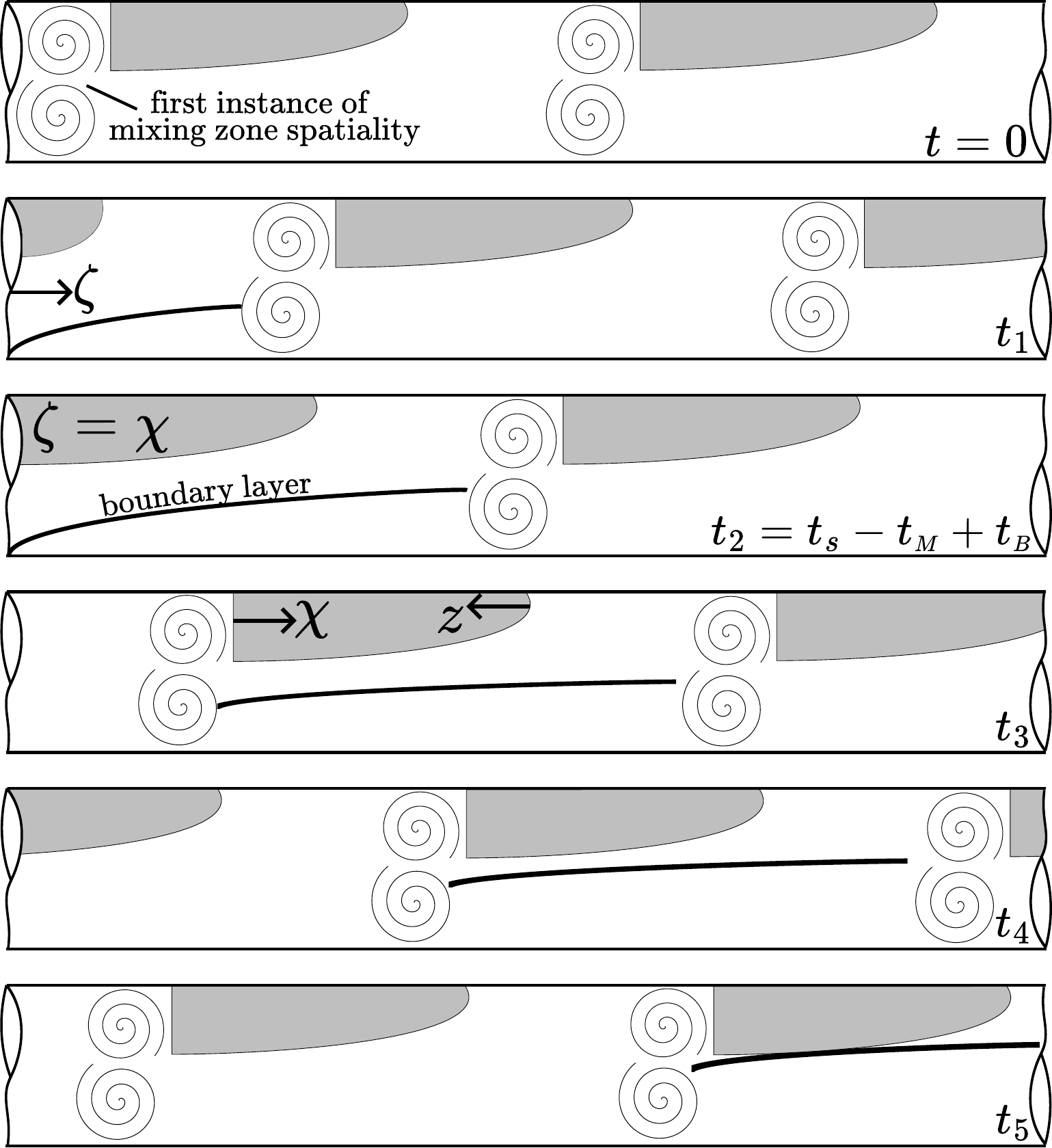}
\caption{Theoretical depiction of slug flow boundary layer development in time and space for an arbitrary unit-cell.}
\label{fig:18}
\end{figure}

Primarily explored thus far is a temporal snapshot at which long bubble tail aligns with BL origin: $\chi=\zeta$. To comprehend the dynamical landscape for all times, the relationship between $\chi$ and $\zeta$ must be solidified. Here, an initial time $t=0$ is defined in reference to the most upstream genesis of a coherent turbulent mixing region for a singular, arbitrary unit-cell, such that a cumulative stretch of a slug region, less its mixing zone, and a film region pass for $\chi$ and $\zeta$ to align; or, 
\begin{equation} \label{eqn:25}
\chi=\zeta\quad\text{if }t = \ts-\tm+\tB
\end{equation}
where $\tB$, $\ts$ and $\tm$ are times-of-passage for film (long bubble), slug and mixing regions. Dynamical configurations at $t=0$ and $t =\ts-\tm+\tB$ are both depicted in figure \ref{fig:18} wherein, for the former, the unit-cell of interest is not yet in the observational frame. From the above definition of $\chi=\zeta$, it stands that 
\begin{equation} \label{eqn:26}
\chi=\zeta+\ut\left(t-\tB-\ts+\tm\right)
\end{equation}
or, since $\ut=\lB/\tB=\ls/\ts=\lm/\tm$,
\begin{equation} \label{eqn:27}
\chi=\zeta+\ut t - \lB - \ls + \lm \text{.}
\end{equation}

Within the paradigm constructed here, the film region BL is formed prior to the long bubble's arrival, temporally speaking, for early times $t\in[0,\ts-\tm+\tB]$ as defined. Depicted in figure \ref{fig:18} at $t=t_1$, this phenomenon is understood through careful differentiation between absolute and relative motions in slug flow. Because the turbulent mixing region acts to destroy near-wall flow-facets\textemdash as foundationally assumed here\textemdash it appears that BL generation occurs in its immediate upstream wake. Since the entire unit-cell structure translates faster than any liquid present, however, the free-stream velocity which first services a fixed BL location $u_{\scriptscriptstyle L}$ (slug region average liquid velocity) is inevitably replaced by a slower, axially-decreasing free-stream motion $\uf(z)$ (film region liquid velocity profile). This suggests that, for a point of spatial constancy, $\delta$ will decrease over time from its originally sustained height\textemdash until, that is, it is destroyed entirely by the forthcoming slug mixing region. Beginning at its origin, the BL is consumed by the approaching turbulent bridge which acts to dissipate inner flow buildup by lifting liquid away from the lower pipe wall\textemdash elucidated in figure \ref{fig:18} for $t>\ts-\tm+\tB$. This occurs progressively\textemdash as the long bubble translates toward PCV outlet, an equivalent distance of BL is recycled as if being fed through a fluid dynamic shredder; simultaneously, another BL is similarly constructed in the preceding unit-cell, preserving the integrity of intercellular sameness. Despite being disrupted at its upstream end, however, BL growth continues until the outer flow is compressed into nonexistence. Such an event represents steady-state developmental status for film BL evolution, analogous to BL convergence beyond a critical distance in a single-phase pipe flow.

In consideration of slug flow BL theory construed here, an alternative form of film-Reynolds number is defined\textemdash in replacement of equation \ref{eqn:17}\textemdash to predict laminarity in the outer flow; namely,
\begin{equation} \label{eqn:28}
\Reyf(z) = \frac{\rhoL (\ut-\uf) (\hf-\delta)}{\muL}\text{.}
\end{equation}
In parallel, representative logical conditions for non-impingement of the outer flow region\textemdash necessary for long bubble detachment\textemdash are straightforwardly formulated; for example, full-centring requires that 
\begin{equation} \label{eqn:29}
\forall\chi\in [0,\lB]\colon\delta<\hf
\end{equation}
while partial-centring corresponds to
\begin{equation} \label{eqn:30}
\begin{split}
\forall\chi\in (\lB-\lD,\lB]\colon&\delta<\hf\\
\forall\chi\in[0,\lB-\lD]\colon&\delta=\hf \text{.}
\end{split}
\end{equation}
Obviously, the above conditions alone, if satisfied, do not guarantee perpetuation of centring mechanisms and must be paired with equations \ref{eqn:18} and \ref{eqn:19}\textemdash analog counterparts which ascertain outer flow laminarity. Further, \ref{eqn:30} does not represent the only probable conditions for partial-centring; for example, \ref{eqn:29} could hold true despite the outer flow being axially-divided into laminar and turbulent regions, as in figure \ref{fig:15}. Such expressions coincide with a 2D approximation of 3D phenomena in that a cross-sectionally curved BL is quantified using a scalar-valued height $\delta$. This is not unreasonable, however, in comparison to historic analytical slug flow models\textemdash such as TB90\textemdash which utilize unidimensional averaging such that momentum and mass conservation can be simply applied. 

To investigate the validity of BL theory put forth here, a case of air-HVL slug flow from \citet{KimKim23} is utilized. Parameterization is as follows: $D=40\unit{mm}$; $\muL=37\unit{mPa.s}$; $\rhoL=878\unit{kg/m^3}$; and $(\uGS,\uLS)=(0.6,1.2)\unit{m/s}$. The full-form (variable $\hf$) TB90 slug flow model\textemdash programmed in MATLAB and validated using air-water data (see figures 11 and 12 of PL20)\textemdash was used to calculate film region profiles $\hf(z)$ and $\uf(z)$ such that elementary BL theory could be employed. The published source includes film and slug length data $\lB$ and $\ls$, respectively; thus, calibration was performed using $\ls$ and HVL-based closure models as inputs for TB90 and $\lB$ to gauge output accuracy, in this case obtaining a $6.96\%$ difference between experimental and modelling values. Utilized closure models include: 

\begin{enumerate}
\item[]$\mathbf{\HLs}$, slug region holdup: \citet{KoraEA11}
\item[]$\mathbf{\ud}$, long bubble drift velocity: \citet{JeyachandraEA12}\footnote{Parameterization ranges used to derive correlations for $\HLs$ and $\ud$ used here ($D=50.8\unit{mm}$/$\muLN\in[181,587]$ and $D\in[50.8,152.4]\unit{mm}$/$\muLN\in[154,594]$, respectively) do not include values used by \citet{KimKim23}; however, extrapolation is assumed to yield reasonable accuracy, confirmed through successful prediction of $\lB$ (TB90 is highly-sensitive to closure model inputs).}
\item[]$\mathbf{C}$, ratio of max-to-average slug region liquid velocity: \citet{DuklerHubbard75}
\item[]$\mathbf{f_i/f_{\scriptscriptstyle G}}$, ratio of interfacial-to-gas friction factors: \citet{TzotziAndritsos13}
\item[]$\mathbf{\ut}$, long bubble translational/unit-cell structure velocity: \citet{NicklinEA62}\textemdash standard model, given by
\begin{equation} \label{eqn:31}
\ut=C\left(\uGS+\uLS\right) + \ud
\end{equation}
\end{enumerate}

Given that BL development in horizontal slug flow is uncharted phenomenological territory, there is an evident lack of appropriate models to predict both $\delta$ and the distinction between laminar and turbulent near-wall manifestation. Therefore, analysis presented here is a first-order approximation designed primarily to provide proof-of-concept. For example, that a fixed location's free-stream velocity evolves with time suggests that a leading-edge Reynolds number\textemdash analogous to that used in flat-plate BL theory\textemdash is difficult to define. The simplest approach is to capture instantaneous BL behaviour; namely, to investigate the probable BL profile at $\chi=\zeta$. This allows for straightforward allotment of sectional free-stream velocities: $\uf$ for $\chi\in[0,\lB]$ and $\uL$ for $\chi\in(\lB,\lB+\ls]$. Under this pretence, a leading-edge Reynolds number for slug flow BL growth is defined as 
\begin{equation} \label{eqn:32}
\Rey_{\chi}=\frac{\rhoL \chi u_{\scriptscriptstyle \chi}}{\muL}
\end{equation}
where
\begin{equation} \label{eqn:33}
u_{\scriptscriptstyle \chi} =
	\begin{cases}
	\uf(\chi), & \chi\in[0,\lB] \\
	\uL, & \chi\in(\lB,\lB+\ls]\text{.}
	\end{cases}
\end{equation}

In regard to a critical value of $\Reychi$ which differentiates laminar and turbulent BL realization and a calculatory procedure for $\delta$, there exists no precedent for horizontal slug flow; therefore, flat-plate BL theory is used here to roughly probe the given conceptual framework's potential validity. For a constant free-stream velocity flowing over a leading-edge, horizontally-oriented flat-plate, a range for transitional $\Reychi$ is given by $\num{3e5}\leq\Reychic\leq\num{3e6}$, depending on outer-flow perturbation, while the most commonly used criterion is
\begin{equation} \label{eqn:34}
\begin{split}
\Reychic<\num{5e5}\colon&\quad\text{laminar BL}\\
\Reychic>\num{5e5}\colon&\quad\text{turbulent BL}
\end{split}
\end{equation}
as per \citet{Schlichting00} (SG00).\footnote{SG00: \citet{Schlichting00} text on boundary layer theory fundamentals.} Applying \ref{eqn:32} and \ref{eqn:33} to the aforementioned HVL slug flow case, maximum and average values of $\Reychi$ in the film region are found to be $\num{13136}$ and $\num{5596}$, respectively. Regardless of the precise numerical value of $\Reychic$, these numbers are indicative of a laminar BL; although, considering that that a long bubble travels a significant distance between PCV inlet and outlet, it is plausible that BL growth will adopt turbulence later in the pipe. This should not affect the centring phenomenon, however, since the detachment process likely occurs in early (upstream) sections of the PCV, some distance beyond slug formation. 

To follow, laminar flat-plate BL theory from SG00 is used\textemdash in tandem with the calibrated TB90 unit-cell slug model\textemdash to approximate BL thickness for $\chi=\zeta$ and, by extension, to determine whether an outer film flow can be maintained so that centring may initiate.\footnote{An attempt was made to upscale laminar flat-plate BL theory from SG00 to an open-channel pipe flow; however, despite being set-up with appropriate geometry, the results were too complicated to feasibly be incorporated into this study. As such, a novel BL model for slug flow is left for future publication.} A concise overview of relevant BL theory from SG00 is given here using appropriate film region parameterization. Inertial force per unit volume of liquid is given generally by 
\begin{equation} \label{eqn:35}
F_{\scriptscriptstyle I}=\rhoL \frac{\mathrm{D}\uf}{\mathrm{D}t}
\end{equation}
where $\mathrm{D}/\mathrm{D}t = \partial/\partial t + \bu\boldsymbol{\cdot}\bn$ is the material or total derivative. TB90 is steady-state and 1-dimensional; thus, \ref{eqn:35} becomes 
\begin{equation} \label{eqn:36}
F_{\scriptscriptstyle I}=\rhoL \uf \frac{\partial\uf}{\partial\chi}\text{.}
\end{equation}
Frictional force per unit volume is given by 
\begin{equation} \label{eqn:37}
F_{\scriptscriptstyle F}=\frac{\partial\tauf}{\partial y}
\end{equation}
where $\tauf$ is film shear stress and 
\begin{equation}\label{eqn:38}
y=D-\lambda
\end{equation}
is the spatial inverse of $\lambda$, as already defined. Assuming a Newtonian liquid, \ref{eqn:37} is rewritten using the standard law of viscosity as 
\begin{equation} \label{eqn:39}
F_{\scriptscriptstyle F} = \frac{\partial}{\partial y}\left(\muL\frac{\partial\uf}{\partial y}\right)=\muL\frac{\partial^2\uf}{\partial y^2}\text{.}
\end{equation}
Scaling $y\sim\delta$, inertial and frictional forces are equated $F_{\scriptscriptstyle I}=F_{\scriptscriptstyle F}$ using \ref{eqn:36} and \ref{eqn:39} to yield
\begin{equation}\label{eqn:40}
\delta\sim\sqrt{\frac{\muL\chi}{\rhoL\uf}}\text{.}
\end{equation}
Defining $\delta=\delta_{99}$ where $\delta_{99}$ is BL height at which local velocity reaches $99\%$ of free-stream velocity, proportionality for \ref{eqn:40} is determined to be
\begin{equation}\label{eqn:41}
\delta(\chi)=5\sqrt{\frac{\muL\chi}{\rhoL\uf}}
\end{equation}
based on the original \citet{Blasius1908} solution.\footnote{As reported by SG00 (p.31).} Also of interest is displacement thickness $\delta_1$\textemdash conceptualized as the vertical shift imparted on the outer flow by the BL relative to the inviscid baseline, given here by 
\begin{equation}\label{eqn:42}
\delta_1=0.34\delta\text{.}
\end{equation}

\begin{figure}[!t] 
\centering
 \includegraphics[width=\textwidth]{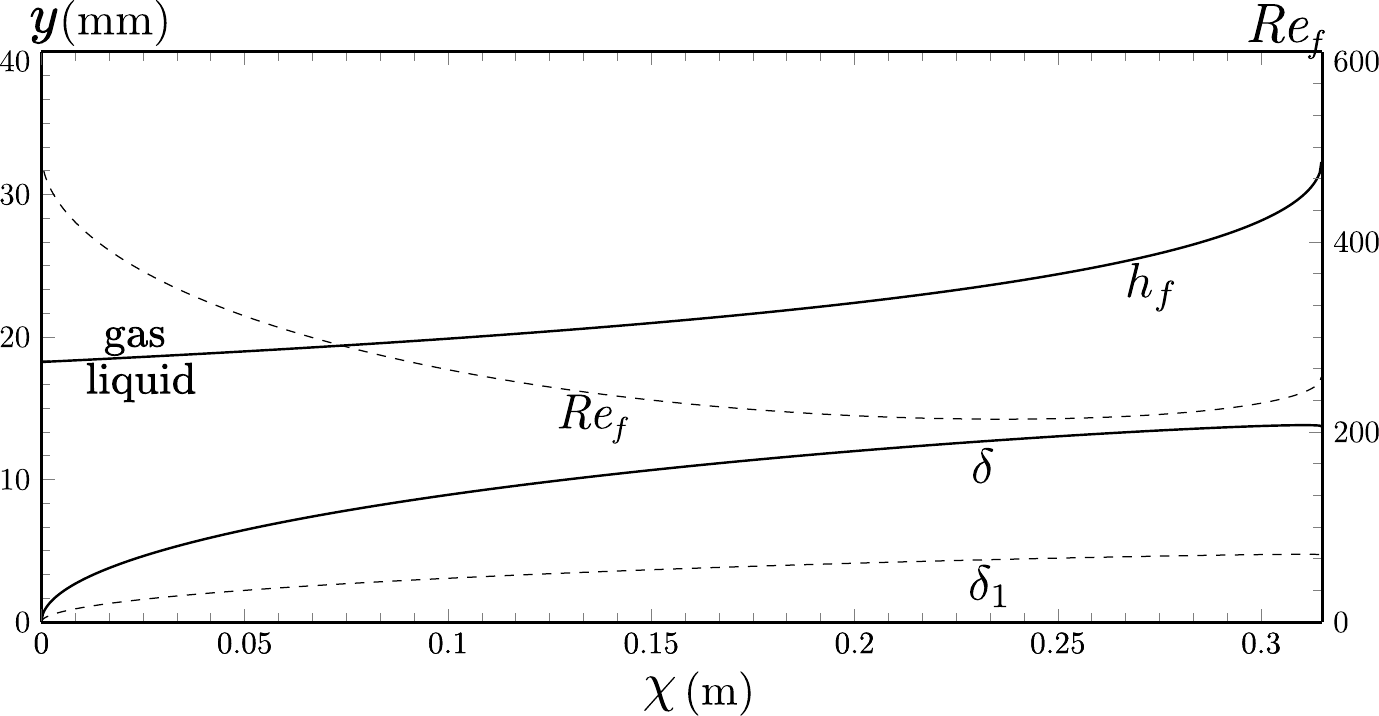}
\caption{Slug flow boundary layer theory applied to HVL case from \citet{KimKim23}: $D=40\unit{mm}$; $\muL=37\unit{mPa.s}$; $\rhoL=878\unit{kg/m^3}$; $(\uGS,\uLS)=(0.6,1.2)\unit{m/s}$. Full-form \citet{TaitelBarnea90} (TB90) model used with $\ls$ data and appropriate closure models to calculate $\hf$ and $\uf$ profiles; $\lB$ data used to validate calibration. Equations \ref{eqn:41}, \ref{eqn:42} and \ref{eqn:28} used to output $\delta$, $\delta_1$ and $\Reyf$, respectively. Flow scenario corresponds to $\chi=\zeta$.}
\label{fig:19}
\end{figure} 

Results of the slug flow BL micro-study\textemdash including film region variables $\hf$, $\Reyf$, $\delta$ and $\delta_1$ for the selected flow-case from \citet{KimKim23}\textemdash are shown in figure \ref{fig:19} wherein the $y$-axis span represents inner pipe diameter. Observed from the $\Reyf$ profile (obtained using equation \ref{eqn:28}) is an unusual trend; namely, $\Reyf$ reaches a minimum ($213.5$) at $\chi=0.75\lB$ rather than at bubble nose-tip\textemdash as would be predicted by \ref{eqn:17}. Furthermore, a maximum value of $494$ is found at long bubble tail, indicating that the entire outer film flow exists in laminarity. Combined with a notable modelled separation between the gas-liquid interface and the BL, this suggests a dynamical environment facilitative of bubble centring realization. Whether this case indeed demonstrates significant detachment is unknown since photographic data are not available; regardless, the above application clearly elucidates the importance of incorporating BL theory into any analytic framework for bubble centring.

\subsection{Wedge theory}\label{sec:4.3}

\begin{figure}[!t] 
\centering
 \includegraphics[width=\textwidth]{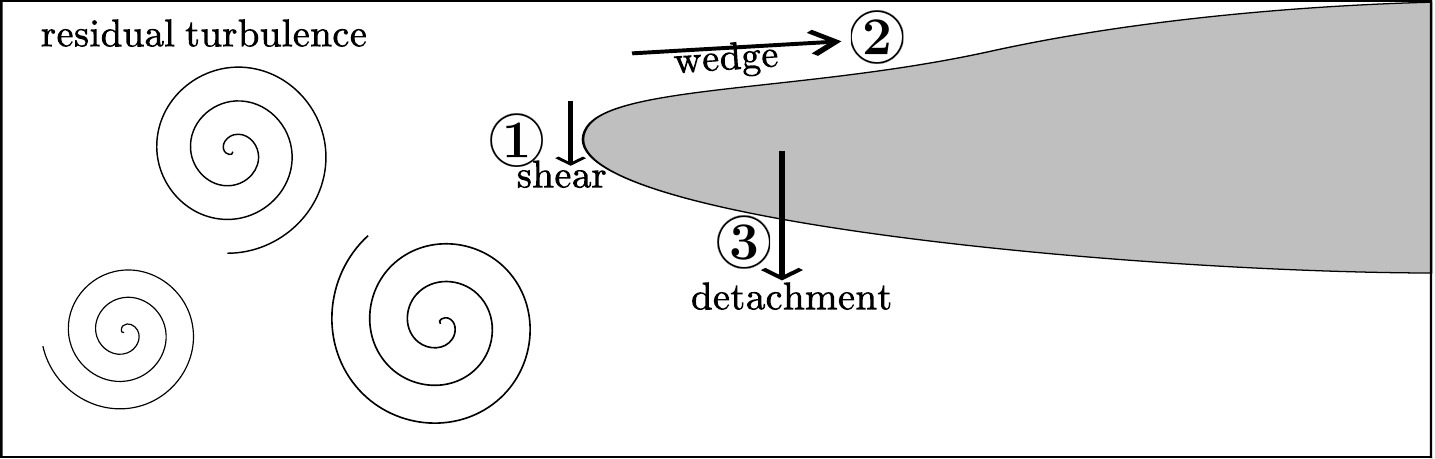}
\caption{Depiction of the wedge theory for partial-centring, comprised of three chronological sub-mechanisms: \textbf{\Circled{1}} residual turbulence in the preceding slug region enacts downward shear on long bubble nose-tip, resulting in slight detachment from upper pipe wall; \textbf{\Circled{2}} relative slug motion causes liquid to enter space created between long bubble and upper pipe wall, driving a wedge-like expansion; and \textbf{\Circled{3}} substantial bubble detachment occurs, manifesting the partial-centring flow configuration.}
\label{fig:20}
\end{figure}

Thus far, film region laminarity has been a postulated prerequisite for the initiation of long bubble centring in horizontal slug flow; however, turbulence is likely in air-water and low-HVL flows. An alternative mechanism\textemdash termed liquid wedging\textemdash is put forth here as a causal element in the realization of partial-centring under certain dynamical conditions. A novel theory, visually represented in figure \ref{fig:20}, is presented as a conglomeration of three distinct flow events. Summarized, they are:

\begin{enumerate}
\item\textbf{Nose shear}: turbulence in the downstream slug region\textemdash stemming residually from mixing region eddies\textemdash contacts long bubble nose due to relative motion, detaching it slightly from the upper pipe wall due to transmission of downward shear;
\item\textbf{Liquid wedging}: liquid enters the created gap with wedge-like shape; and
\item\textbf{Detachment}: since the gas structure is deformable, a segment of the bubble is driven toward pipe centreline, thus manifesting the partial-centring configuration. 
\end{enumerate}

The described process is compounding yet limited; as part of the bubble is centred, the overlying liquid wedge cumulatively accrues volume, thus perpetuating the cycle. However, at some level of detachment, the dual-channel fluid connectivity between wedge and underlying film reaches critical fluency in that proliferation of centring ceases and a localized steady-state emerges. Since the hypothesized chain of wedging mechanisms is functionally dependent on turbulence immediately downstream of long bubble nose, both inertial input $\Fr$ and liquid viscosity $\muL$ are modulators. For similar $\Fr$-characterization, flows with higher $\muL$ incur lower turbulence intensity in the slug mixing region. Further, turbulent kinetic energy dissipation rate $\epsilon$ is positively proportional to $\muL$ \citep{Pope00},
\begin{equation} \label{eqn:43}
\epsilon\sim\muL\text{;}
\end{equation}
therefore, turbulence in the slug mixing zone diffuses faster for high-$\muL$ systems. As such, the likelihood of eddies reaching the long bubble nose with enough energy to catalyze the wedging process is reduced in HVL systems. For this reason, the presented theory is plausibly more applicable to water-based or low-HVL slug flows wherein film region laminarity may not be satisfied. In certain $\muL$-cases, wedging may occur concurrently with the PL20 streamline mechanism; however, cut-off valuation is subject to experimental and further theoretical inquiry. For medium- or high-HVL flow-cases in which film laminarity is expected, the streamline centring model is likely most appropriate for usage. Another aspect of wedge theory to consider is that of eddy rotational orientation near long bubble nose\textemdash within context supplied here, turbulent eddies must rotate such that downward shear is applied to the bubble nose-tip, suggesting a potential element of randomness. 

\subsection{The slug-annular transition}\label{sec:4.4}

Evidently, the bubble centring phenomenon plays a significant role in the microcosm of horizontal gas-HVL slug flow dynamics. Here, it is subsequently hypothesized to host vital functionality within the broader sphere of flow pattern transition theory for HVL-featuring systems. In \S\ref{sec:1} it was shown that, for at least two sets of HVL data with $\muLN\in[1,11000]$, predictive ability of the classical TD76 mechanistic model deteriorates exponentially with increasing $\muL$. For one of the utilized sources, \citet{MatsubaraNaito11}, it was further determined that the proportion of wrongly predicted flow patterns which involve either slug or annular regimes increases with $\muL$.\footnote{This conclusion was found in exclusion of the baseline air-water case. Also, with respect to the data source, the range of investigated $(\uGS,\uLS)$ narrows upon each incremental increase in $\muL$; thus, it is unclear if the trend in question genuinely represents a focused weakening of predictability or, instead, is an artifact of non-constant experimental design.} A different study from \citet{GokcalEA08} (G08)\textemdash which reports HVL flow pattern data for $\muLN\in\{181,587\}$ and $D=50.8\unit{mm}$ overlaid with mechanistic modelling predictions\textemdash is invoked to expand upon the picture illustrated in figure \ref{fig:01}.\footnote{G08: \citet{GokcalEA08} HVL flow pattern study.}

\begin{figure}[!t] 
\centering
 \includegraphics[width=\textwidth]{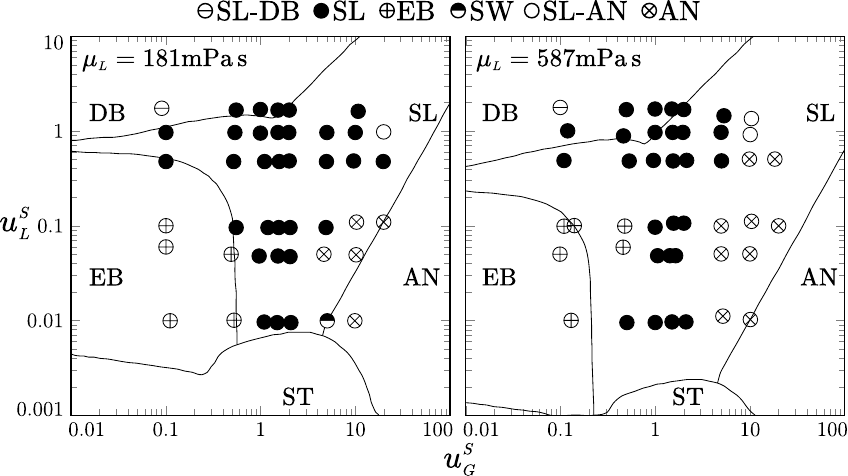}
\caption{HVL flow pattern data from \citet{GokcalEA08} (G08) overlaid with predictions from the \citet{Barnea87} (B87) mechanistic transition model. Above legend corresponds to experimental points; directly labelled regions correspond to model predictions. SL=slug; EB=elongated bubble; DB=dispersed bubble; ST=stratified (smooth or wavy); SW=stratified-wavy. Hyphenated legend entries are transitional. Units for superficial velocity are $\unit{m/s}$. Adapted from figures 7 and 9 of original source.}
\label{fig:21}
\end{figure}

\begin{figure}[!t] 
\centering
 \includegraphics[width=\textwidth]{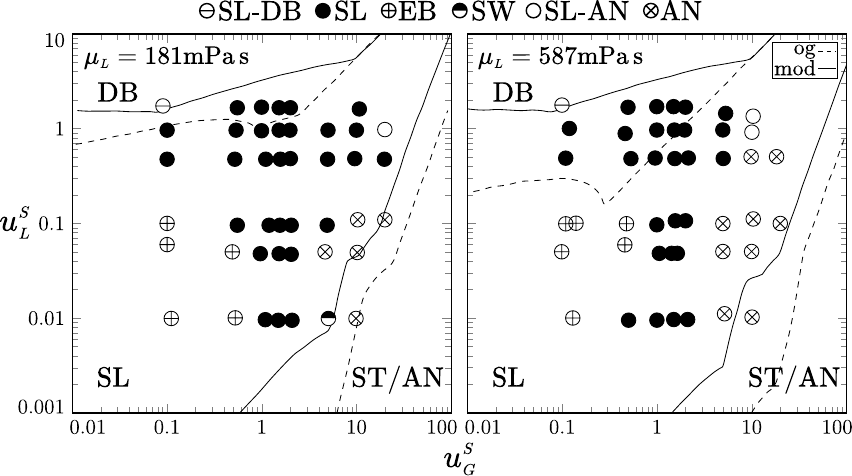}
\caption{HVL flow pattern data from \citet{GokcalEA08} (G08) overlaid with predictions from the \citet{ZhangEA03c} (Z03) mechanistic transition model and its modified form (Z03M) from G08. Dashed lines represent Z03; solid lines represent Z03M. Above legend corresponds to experimental points; directly labelled regions correspond to model predictions. SL=slug, EB=elongated bubble, DB=dispersed bubble, ST=stratified (smooth or wavy), SW=stratified-wavy. Hyphenated legend entries are transitional. Units for superficial velocity are $\unit{m/s}$. Adapted from figures 11 and 12 of original source.}
\label{fig:22}
\end{figure}

G08 utilizes flow pattern transition models from \citet{Barnea87} (B87) and \citet{ZhangEA03c} (Z03), both of which are unified or inclination-flexible.\footnote{B87 and Z03: \citet{Barnea87} and \citet{ZhangEA03c} flow pattern transition models.} When collapsed onto horizontal pipe systems, B87 is a derivative of TD76; Z03, however, is founded on slug flow dynamics and thus is dependent on correlative closure models. Also included in G08 is a modified form of the Z03 model (Z03M) which implements an artificial modulator to correct the ``momentum term for gas entrapment.''\footnote{Z03M: G08-modified form of Z03 model.} Experimental HVL flow pattern data and modelling predictions from B87, Z03 and Z03M\textemdash recreated from G08\textemdash are given in figures \ref{fig:21} and \ref{fig:22}. Overlaid lines and sectional labels represent modelling predictions; discrete points are experimental data. Note that Z03 combines stratified and annular (ST/AN) flows while B87 differentiates between them. The latter also segregates slug ($\HLs\neq 1$) and elongated bubble ($\HLs=1$) flows. 

\begin{figure}[!t] 
\centering
 \includegraphics[width=\textwidth]{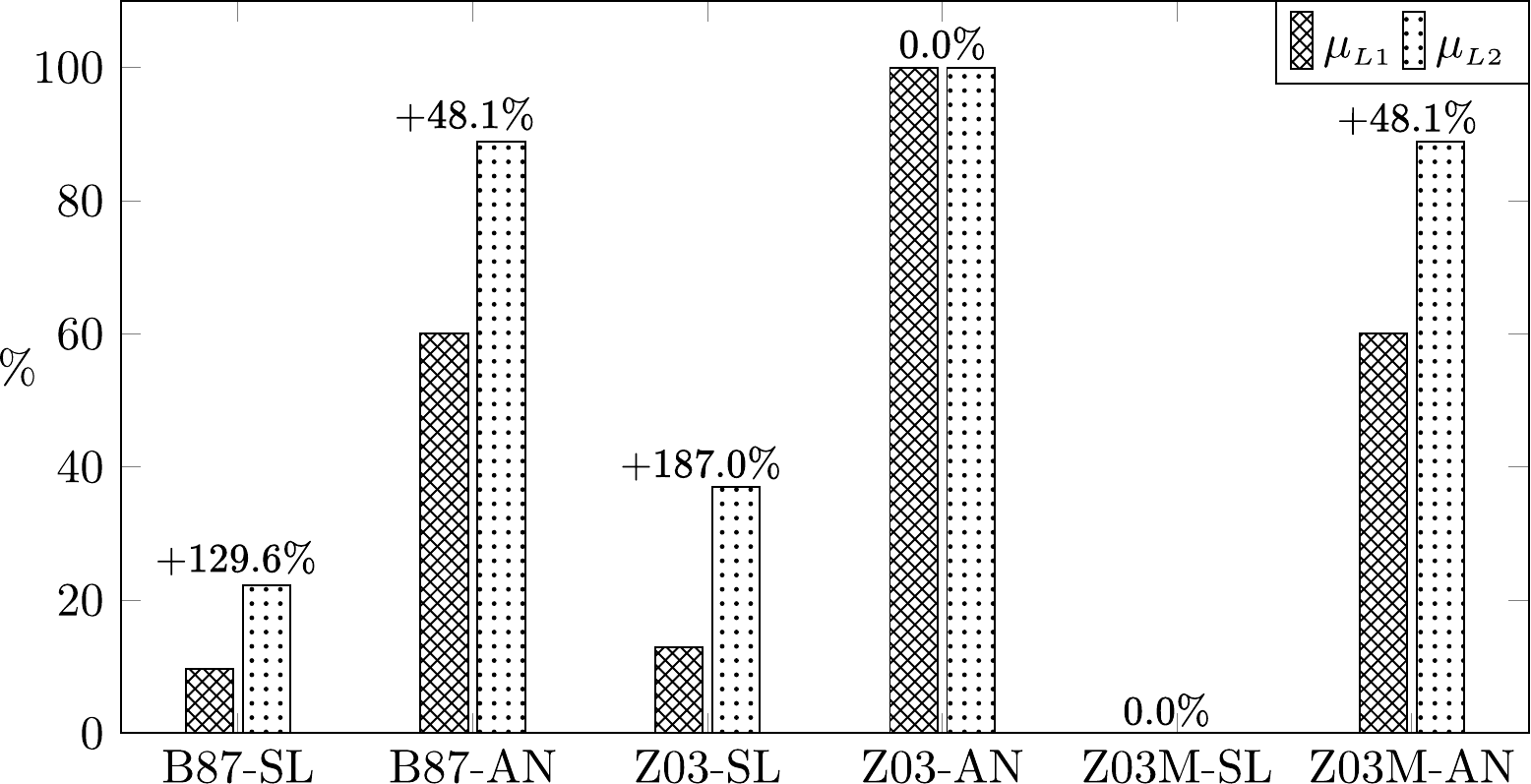}
\caption{Percentage of total experimental slug (SL) and annular (AN) flow-points from \citet{GokcalEA08} (G08) that were wrongly predicted by \citet{Barnea87} (B87), \citet{ZhangEA03c} (Z03) and modified-Z03 (Z03M) mechanistic models where $\mu_{\scriptscriptstyle L1}=181\unit{mPa.s}$ and $\mu_{\scriptscriptstyle L2}=587\unit{mPa.s}$. Overlaid are \%-difference values which represent changes due to $\Delta\muL$.}
\label{fig:23}
\end{figure}

In general, modelling predictability is observed to worsen across the $\muL$-increase of $\Delta\muLN=406$ ($+105.7\%$) for G08 flow pattern data, reiterating the trend explored through figure \ref{fig:01}. Also supported is the notion that the slug-annular boundary is poorly predicted, in particular, for HVL-systems. This is visualized in figure \ref{fig:23} which displays, in bar-chart format, the percentage of total experimental slug or annular flow-points\textemdash for both $(\mu_{\scriptscriptstyle L1}^\circ,\mu_{\scriptscriptstyle L2}^\circ)=(181,587)$\textemdash that were wrongly predicted by each of the tested mechanistic models. For example, B87-SL represents percentage of slug flow data-points wrongly predicted by B87. Z03M, in essence, implements a manual workaround to improve data predictions, suggesting that its improved capacity does not reflect a more robust alignment with underlying mechanics; thus, B87 and Z03 are primarily of interest. Regardless, the takeaway remains unchanged with its inclusion: systemic amplification of $\muL$ in a horizontal gas-liquid pipe flow wreaks havoc on the generalized ability to predict flow pattern occurrence, particularly at the slug-annular boundary. 

Based on G08 modelling endeavours, it appears that increasing $\muL$ alters the range of superficial flow rates for which slug flow is predicted; however, such changes are not reflected in the data, resulting in worsening flow pattern calculation. Specifically, in this case, $\Delta\muL$ expands\textemdash on a flow pattern map\textemdash the outputted slug region at both stratified/non-stratified and slug/annular boundaries while simultaneously constricting it at the slug/dispersed bubble transition, producing a net accumulation. The data, however, show a lesser change. For reference, across $\Delta\muL$, G08 data show a decrease in measured slug (or elongated bubble) flow points from $36$ to $33$ ($82\%$ to $73\%$ of total points) and an increase in annular points from $5$ to $9$ ($11\%$ to $20\%$ of total points). Another relevant inference drawn from G08 stems from their empirical definition of a slug-annular transitional flow pattern (labelled SL-AN; registered once and twice for $\mu_{\scriptscriptstyle L1}$ and $\mu_{\scriptscriptstyle L2}$, respectively). A qualitative description is not given; intuitively, the superposition of slug and annular regimes will, however, exhibit both intermittency and\textemdash at least somewhere within the PCV\textemdash a competent liquid film with non-negligible thickness enclosing a gas core. Such a depiction clearly coincides with slug flow featuring fully-centred long bubbles in that both periodicity and annularity are present. Also, all noted SL-AN designations are encountered at the largest ($\uGS,\uLS$) pairs of their respective flow pattern maps\textemdash logically corresponding to the most significant degree of incurred centring.\footnote{That SL-AN points are observed sparsely, relative to other flow patterns, could be interpreted as a lack of observed centring in the majority of slug flow data; however, centring occurs on a wide spectrum and perfect-centring (radial symmetry\textemdash as is likely found approximately in SL-AN data) is not a prerequisite for centring classification. Further, detection can be tricky\textemdash if flow pattern identification is completed visually, compared to using, for example, advanced image processing techniques, lesser degrees of centring may not be recognized. For relatively small separation distances (i.e., $\yk<0.1D$ yet non-zero) most researchers would likely observe slug rather than transitional flow, especially considering that the phenomenological study of centring is still developing.} The above observations plausibly insinuate a critical role of bubble centring in the slug-annular transition for HVL systems\textemdash one that seemingly lacks field-scale recognition. 

Priorly shown analyses firmly demonstrate the need for fresh perspective in the realm of HVL flow pattern transition theory. Combined with the empiric transitory flow pattern noted above, two elements of reasoning suggest that a novel paradigm for the slug-annular boundary can be construed through incorporation of the bubble centring phenomenon: 1) deterioration of existing mechanistic models' predictive capacity when applied to the slug-annular boundary, positively proportional to $\muL$, and 2) an empirically verifiable connection between full-centring and increasing-$\muL$. A unique theoretical framework\textemdash designed particularly for HVL-containing systems\textemdash is derived here qualitatively. Two key mechanisms, both related to elongated bubble dynamics, are invoked:

\begin{enumerate}
\item[]\textbf{Detachment:} initiation of long bubble centring
\item[]\textbf{Coalescence:} establishment of bulk gas continuity
\end{enumerate}

Figure \ref{fig:24} illustrates the theory using four snapshots of an identical pipe and fluid configuration wherein liquid input $\uLS$ is fixed and gas supply $\uGS$ increases incrementally starting from \Circled{1}. All four $\uGS$-cases exist under temporal development with stratified flow near the PCV inlet (not shown). At some arbitrary distance beyond the inlet, an interfacial instability occurs which, in all instances, triggers the manifestation of non-centred slug flow for differing lengths of pipe. As such, the left-most elongated bubble in \Circled{1} to \Circled{4} represents the first spatial appearance of slug flow. Evolution is as follows:

\begin{figure}[!t] 
\centering
 \includegraphics[width=0.93\textwidth]{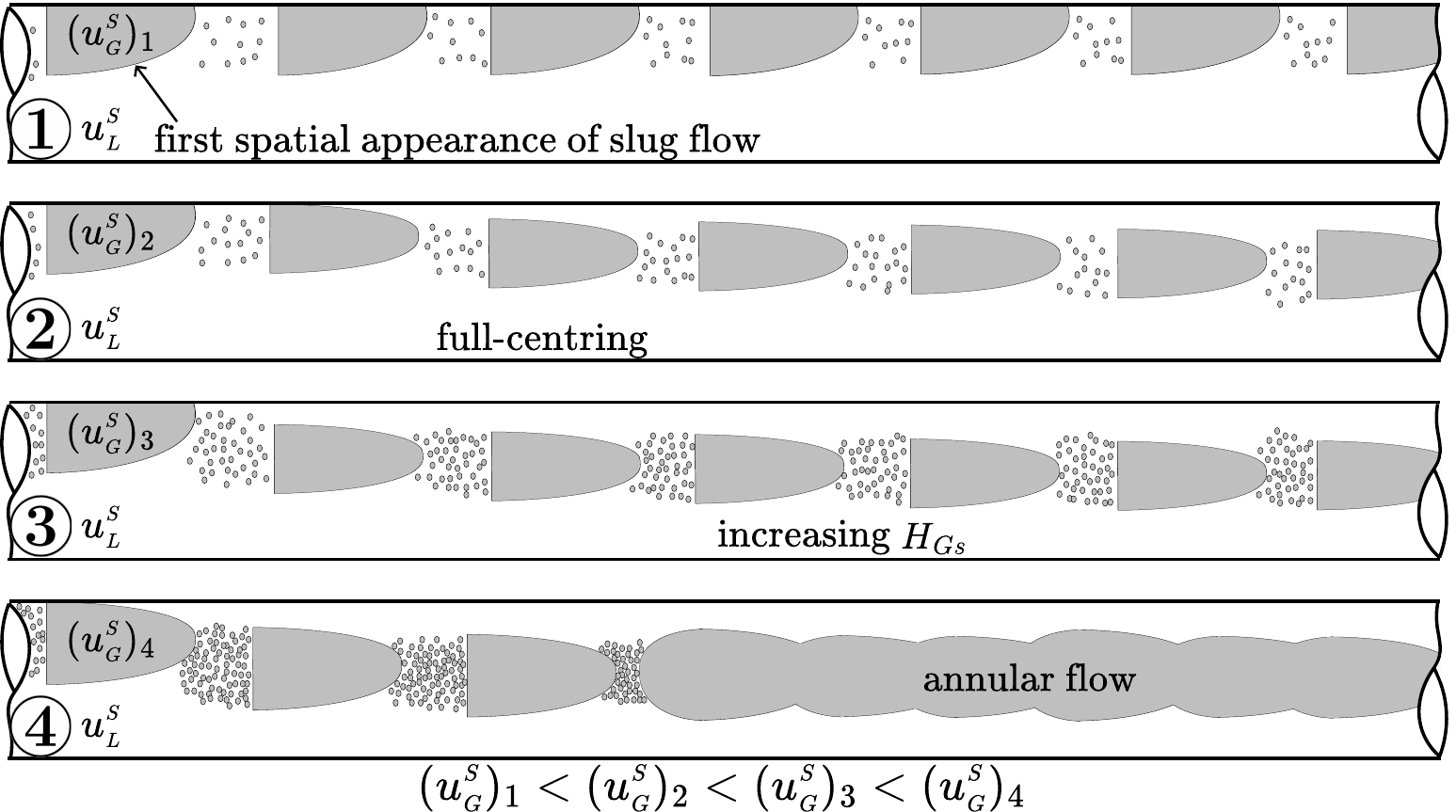}
\caption{Illustration of novel mechanistic framework for the slug-annular transition, designed for gas-HVL flow-systems using four snapshots at incrementally larger superficial gas rates and a fixed liquid rate: \textbf{\Circled{1}} traditional slug flow; \textbf{\Circled{2}} full-centring beyond interfacial instability; \textbf{\Circled{3}} increases in slug gas holdup and degree of centring; and \textbf{\Circled{4}} coalescence occurring after centring, resulting in the annular flow pattern.}
\label{fig:24}
\end{figure}

\begin{description}
\item \textbf{Case \Circled{1}} Typical manifestation of intermittency\textemdash termed here traditional slug flow\textemdash with minimal long bubble centring. 
\item \textbf{Case \Circled{2}} Traditional slug flow exists beyond the interfacial instability; however, an increase in relative motion caused by $\Delta_{12} = (\uGS)_2-(\uGS)_1$, combined with laminarity in the film, results in long bubble detachment and full-centring prevails for the remainder of the PCV. A slight increase in slug gas holdup $\HGs$ is incurred.
\item \textbf{Case \Circled{3}} Traditional slug flow exists momentarily before long bubbles detach and nearly-perfect, radially-symmetric centring is sustained. The further increase in gas inertia $\Delta_{23} = (\uGS)_3-(\uGS)_2$ causes another notable increase in $\HGs$ such that it approaches yet falls below a critical limit for coalescence $\HGsc$. 
\item \textbf{Case \Circled{4}} Traditional slug flow emerges for an instant, likely with some degree of partial-centring. Beyond that, owing to maximal relative gas motion induced by $\Delta_{34} = (\uGS)_4-(\uGS)_3$, long bubbles are immediately centred; however, significant gas volume is dispersed in the liquid slugs with $\HGs>\HGsc$ and thus gas coalesces into a continuous bulk structure. Because centring occurs upstream of coalescence, the gas core flows with a film surrounding its radial entirety; as such, annular flow is maintained for the majority of the PCV. Due to the means of coalescence\textemdash periodic consolidation of curved (nose) and flat (tail) topologies\textemdash the gas-liquid interface is probably wavy as depicted in figure \ref{fig:24}. 
\end{description}

One potentially contradictory aspect of this theory is the existence of both slug and annular flow patterns in the same flowing pipe, as in \Circled{4}, which defies an axiom put forth in TD76; namely, equilibrium stratified liquid height $\hL$\textemdash compared to a critical value\textemdash alone differentiates the two emerging configurations. If $\hL>\hLc$, enough liquid is said to exist at a Kelvin-Helmholtz instability for a stable liquid bridge or slug to form; if $\hL<\hLc$, the liquid is said to be swept around the inner pipe wall to form an annular film.\footnote{For reference, TD76 suggests $\hLc=0.5D$; however, a subsequent study from \citet{BarneaEA82a} concludes that $\hLc=0.35D$ is appropriate, at least for air-water flow-systems. For HVL flows, a recent study by \citet{AlSafranAlQenae18} reports, based on empiricism, that $\hLc=0.45D$ is functional. That modern research is still utilizing this simplistic approach is telling in that a more sophisticated model is lacking.} Such contradiction can be reconciled, however, in realizing that\textemdash in the novel theory\textemdash slug flow exists only briefly before morphing into annularity. Further, because liquid supply remains constant in all four flow-cases, one could argue that if a competent reservoir exists in case \Circled{1}, it can hypothetically exist in case \Circled{4}. Also, to the author's best knowledge, there exist no visual data within the literature that depict and thus confirm the phenomenological distinction put forth in TD76. The principle utilized here is divergent by design since, for novel paradigms to become solidified, accepted fundamentals must be challenged.  

Regarding calculatory procedure, the presented slug-annular framework can be formulated as a combination of bubble centring and coalescence models specifically calibrated and validated for HVL systems. For the former, PL20 is a logical starting point; however, upscaling is required for HVL inclusion, as described in \S\ref{sec:4.2}. For the latter, the B87 model\textemdash utilized in G08\textemdash offers useful parallels; namely, in assuming annular flow to exist, a critical liquid holdup (film plus entrained droplets) is invoked to determine whether a conversion to stable intermittency will occur \citep{Barnea86,BarneaEA82b,BarneaBrauner85}.\footnote{B87 includes an additional mechanism for the slug-annular boundary; that is, a dynamical instability within the film which causes flow reversal and blockage due to buildup. However, in collapsing the unified model onto the case of a horizontal pipe, the critical holdup mechanism is deemed solely relevant.} Based on rudimentary packing theory, B87 posits that $\HLAc=0.24$ represents the slug-annular boundary; however, this criterion is unsubstantiated for HVL systems (as demonstrated priorly). Considering the dynamically unique flow scenarios depicted here, it is plausible that other mechanisms may hold an active role; thus, further research into long bubble coalescence is necessary.

\section{Applicability}
\label{sec:5}
It is worthwhile to discuss the applicability of work outlined in this paper. This will be done in a twofold manner; to follow, applicability will be conveyed for 1) drawn empirical conclusions and 2) derived modelling architecture. 

First, the concept of a ``steady-state'' as it applies to horizontal slug flow must be understood. In reality, slug flow does not attain a true steady-state (i.e., time-independent) configuration; instead, it features an ongoing level of stochasticity which manifests as, for example, variable bubble lengths, morphing interfaces and decaying or growing slugs. Thus, in referring to ``steady-state'' slug flow in the context of laboratory or field (i.e., non-theoretical) environments, two critical, dynamical elements must be instilled: 1) spatial evolution\textemdash a sufficient length of pipe existing between PCV inlet and observational window (to avoid inlet effects); and 2) temporal evolution\textemdash an adequate period of time allotted to the flowing phases prior to the moment of observation (to avoid startup effects). Conversely, then, ``transient'' slug flow would refer to flow conditions whereby spatial and/or temporal development is not reached (e.g., near-inlet or startup slug flow dynamics). Dropping scare quotes for simplicity, types of slug flow found in practice are summarized as:

\begin{enumerate}
\item[]\textbf{Steady-state slug flow:} fully developed in both spatial (sufficient pipe length) and temporal (adequate flow time) dimensions
\item[]\textbf{Transient slug flow:} lacking adequate spatial or temporal development\textemdash encapsulates startup, near-inlet and other unsteady conditions
\end{enumerate}

Based on the circumscribed observational locations within their respective PCV-systems, it appears that adequate spatial evolution was employed in the flow-experiments which outputted pictorial datasets from N23, S24 and K20 (however, this was explicitly confirmed only in S24). Their publications do not mention temporal development or generalized experimental procedure; regardless, it is reasonable to assume that the researchers indeed allowed an adequate passage of time prior to photographic measurement. In light of these considerations, it is inferred that N23, S24 and K20 datasets were derived out of steady-state conditions. 

As such, correlations and relationality deduced from centring measurements extracted for the present study apply, in a strict sense, only to steady-state slug flow systems, as defined above. However, it is important to recognize that centring mechanisms\textemdash as presented originally in PL20\textemdash are not fundamentally unique to such conditions. Realization of the dynamical event which causes transmission of a downward vertical force\textemdash that is, an axial, Bernoulli pressure differential in the presence of coherent streamlines immediately underneath the long bubble and a thin film above it\textemdash is not contingent upon sustaining adequate spatial and temporal development. Therefore, the broad-scale conclusions obtained here based on extracted centring data from N23, S24 and K20 can be plausibly extrapolated to generic horizontal slug flow systems; namely, that the extent of centring is positively correlated with dynamic liquid viscosity $\muL$, regardless of underlying flow-conditions. With that said, to properly validate the posited applicability would require dedicated empirical inquiry which visualizes long bubble behaviour in transient (e.g., startup or inlet) environments. 

Another crucial delineation is as follows: the commonly employed unit-cell assumption is distinctly separate from the steady-state slug flow configuration defined priorly. The former (discussed also in \S\ref{sec:4}) is a modelling device which asserts that interconnected slug units\textemdash consisting each of a film and slug region\textemdash with equivalent velocity, pressure and geometry profiles occupy the entire PCV (or, at least, its majority\textemdash a near-inlet stratified flow could be presumed).\footnote{In essence, the unit-cell assumption models slug flow as if it were a truly steady-state configuration.} The latter is a real-life flow-state featuring sufficient spatial and temporal development. As such, to claim that steady-state conditions exist in an empirical slug flow does not mean that a unit-cell configuration persists; in fact, the unit-cell assumption is merely an approximation of actual flow dynamics that is never truly obtained. Thus, relationality garnered from centring data here is not constrained to unit-cell methodology. 

With respect to mechanistic/phenomenological models of centring (as given in, for example, \S\ref{sec:4} here and PL20), proper care must be taken to differentiate components which pertain solely to the unit-cell assumption and those which may be employed generally (i.e., to both steady-state and transient slug flows, as defined). As alluded to, fundamental centring mechanisms apply to any horizontal (or, to some extent, inclined) slug flow; however, the PL20 predictive model\textemdash which outputs operational conditions at which centring initiates\textemdash is limited, under its original formulation, to the unit-cell assumption owing to its inclusion of the TB90 unit-cell slug flow model.\footnote{The PL20 model could, however, be straightforwardly transposed into non-unit-cell methodology; for instance, the utilized approach to momentum (force) conservation could be implemented with non-constant variables and an alternative underlying slug flow model.}

Theory presented here is constructed using a different approach in contrast with that utilized in PL20; that is, rather than seeking to compose a tangible predictive model, various novel phenomenological aspects of centring are explored in a largely qualitative sense. Since these are untrodden theoretical realms, the baseline (i.e., unit-cell) configuration of slug flow was naturally selected for use as an explanatory artifice. However, the means by which the derived modelling framework can be upscaled to non-unit-cell flow-states should be readily apparent (albeit not overtly simple); further, some elements apply broadly by virtue of their assembly. 

Consider the theoretical investigation into boundary layer (BL) evolution for horizontal slug flow (\S\ref{sec:4.2}). Although the unit-cell assumption is used for clarity, the phenomenological claims are plainly applied to non-unit-cell (i.e., real-life) flows. For example, the film-Reynolds number (equation \ref{eqn:28}) and required BL conditions for full- and partial-centring realization (equations \ref{eqn:29} and \ref{eqn:30}) apply conceptually to any slug flow; however, to implement them into a non-unit-cell modelling framework, $\lB$, $\lD$, $\vf$, $\hf$ and $\delta$ would all require transformation into time-dependent or statistical functions. Further, the outlined application of flat-plate theory (or, for that matter, a hypothetical BL model designed for pipe geometry) is not dependent on the unit-cell assumption (beyond the usage of TB90, which could be replaced).\footnote{One aspect of BL theory given in \S\ref{sec:4.2} whose implementation may potentially be tricky with respect to non-unit-cell methodology is the assertion that subsequent slug mixing regions act to segregate BL development for a given slug unit. Although this is logically coherent, its validity is argued based on its ability to preserve the integrity of the unit-cell assumption. Without needing to conform to such a configuration, a unique, more sophisticated approach may instead be utilized.}

Discussed to follow, in brevity, are projected integrations of bubble centring phenomenology into various non-unit-cell slug flow methodologies; namely, slug-tracking and slug-capturing models, both of which are capable of handling transient phenomena. These deliberations are preliminary, in that further research is required to formalize such novelties. 

An exemplar of the slug-tracking approach is found in \citet{TaitelBarnea98} which outlines an analytical model that follows individual slugs (i.e., using a Lagrangian frame-of-reference) such that ``growth, generation and dissipation'' mechanisms may be simulated. Slug flow is assumed to exist a priori and simplistic conservation equations are applied at either end of the slug \citep{FerrariEA17}. In reviewing the utilized geometric schema (see figure 1 of their paper), a substantial problem becomes evident; namely, a constant liquid level (defined as the equilibrium height $h_{\scriptscriptstyle E}$) is assumed for each film/long bubble region within the PCV. In the PL20 centring model, changes in relative velocity and pressure beneath the long bubble are used to determine whether an adequate downward force is catalyzed to induce detachment. Without modelling the full-form liquid height profile in the film region, centring cannot be characterized or predicted. Thus, the slug-tracking model, as given in the aforementioned source, is rendered unable to integrate the centring phenomenon without considerable theoretical reworking. 

One potential solution would be to use, in tandem, the TB90 unit-cell model to approximate the $\hf$-profile for each isolated long bubble within the PCV of interest. The unit-cell assumption would not be envisaged for the entire PCV; instead, every long bubble would be treated as a unique dynamical system requiring its own set of reasonably accurate closure models (e.g., $\HLs$, $\ut$, $\ls$ and $f_i/f_{\scriptscriptstyle G}$). Plausibly, this would necessitate an interconnected approach to closure modelling; that is, input velocities for an individual long bubble may depend on the liquid shedding rate of the downstream slug. Subsequently, the PL20 centring algorithm could be incorporated to determine which flowing bubbles are subject to some degree of centring, thus allowing resultant pressure loss calculations to gain an order of precision. Clearly, this approach is inherently complex, both from theoretical and computational vantages. Another possible means of centring-inclusion in slug-tracking models would be to directly implement the full-form liquid height profiles, rather than assuming a constant level for each bubble. 

The so-called slug-capturing methodology\textemdash as elucidated in, for example, \citet{FerrariEA17}\textemdash is similar to the slug-tracking approach yet features some key distinctions; namely, it makes use of rigorous conservations equations (i.e., multiphase Navier-Stokes and continuity equations or the ``two-fluid'' model) and, by proxy, is capable of predicting both slug and stratified regimes. The above-mentioned source utilizes a hyperbolic, 1-dimensional system of five mass or momentum equations which are solved numerically to predict intricate slug flow dynamics. Although this model outputs full-form film-liquid profiles, it is unclear as to whether it can accurately characterize the centring phenomenon as it occurs empirically.\footnote{For example, figure 13 of their paper displays dynamic film-region profiles with no degree of centring despite stemming from an air-water system with operational conditions $(\Fr,\gamma)=(4.0,1.3)$. As noted earlier in this study, water-based systems typically feature centring for $\Fr\geq 3.5$.} Rationale for this suspicion is derived from the model's fundamental assumptions; that is, in using 1D (axial) flow with cross-sectional averaging, complex 3D phenomena\textemdash which are important in centring mechanics, particularly at the bubble nose\textemdash will be precluded. Specifically, without a radially-oriented numerical grid with sufficiently small spacing, the existence of a thin film situated above the bubble (a necessary prerequisite for centring) will not be captured; thus, elongated bubbles will always remain attached to the upper pipe wall in modelling predictions\textemdash as confirmed by figure 13 in \citet{FerrariEA17}. Overall, the slug-capturing model, as it presently stands, would require significant complexification in order to account for centring manifestation.

Lastly, the applicability of centring phenomenology in the regime known as severe slugging\textemdash as described in, for example, part III of TB90 and \citet{Taitel86}\textemdash is considered. Severe slugging, also referred to as terrain-induced slugging, is a type of unsteady slug flow wherein liquid accumulation in pipeline valleys results in long slugs that are forced downstream when the preceding long bubble reaches a critical pressure \citep{TaitelBarnea90}. The typical example used to depict severe slugging is a flow-system consisting of a pipeline riser connected to a separator. 

Speculatively, if the long bubble upstream of the vertical pipe section were to be fully-centred, its efficacy in displacing the stagnant liquid may be reduced in comparison with the case of a non-centred bubble. That is, owing to the centred bubble's topology, it may, potentially, be able to force its way through the mid-region of the liquid without fully displacing it, causing a gaseous ``blow-through'' effect. Given that the long bubble remains momentarily static, however, it is unclear whether, upon finally moving, it would reach a sufficient level of inertia such that centring could manifest to a meaningful extent. Liquid viscosity $\muL$ would, of course, play a central role in this dynamical scenario, particularly in light of empirical conclusions found in this study. Whether this hypothetical occurrence is desirable from a functional standpoint is yet another uncertainty\textemdash it could avoid severe slugging to some degree, but the liquid accumulation would eventually reach a critical impasse, beyond which other undesirable or unexpected operational events may unfold. Additionally, beyond manipulating fluid properties and flow rates to enhance the flow event's probabilistic likelihood, the task of ensuring the existence of centred bubbles in a particular PCV location poses a unique operational difficulty. 

\section{Finalities}
\label{sec:6}
Based on empirical and theoretical analyses provided throughout this paper, the following conclusions and suggestions for further work are drawn with respect to horizontal gas-liquid pipe flows:

\subsection{Conclusions}\label{sec:6.1}

\begin{enumerate}
\item Elongated bubble centring is strongly correlated with liquid viscosity $\muL$; specifically, for constant mixture inertial input and pipe size, the degree of incurred centring is positively proportional to $\muL$ as evidenced by thorough analyses of datasets extracted from \citet{NaidekEA23} (N23; $\muL\in[1,30.4]\unit{mPa.s}$), \citet{ShinEA24} (S24; $\muL\in[37.7,352]\unit{mPa.s}$) and \citet{KimEA20} (K20; $\muL\in[510,960]\unit{mPa.s}$). 
\item In this paper, since pipe diameter $D$ is constant within utilized datasets, changing or fixed $\Fr$ represents changing or fixed mixture velocity $\um$ and the influence of pipe size on centring extent is not explicitly studied. Another crucial consideration\textemdash which is only partially explored here owing to limited data availability\textemdash is the impact of individually varying gas or liquid rates, characterized by $\gamma=\uGS/\uLS$. For a given pipe, a slug flow's centring tendency is fully characterized by $\Fr$, $\gamma$ and $\muL$. 
\item Full-centring\textemdash defined as non-negligible separation between upper pipe wall and elongated bubble realized throughout the entire film region in slug flow\textemdash is a highly-probable occurrence for systems with $\muL\geq352\unit{mPa.s}$ in that it prevails even in low-inertial conditions $\Fr\geq0.57$. This is contradictory to air-water dynamics under which full-centring is observed in high-$\Fr$ flows only. 
\item Despite predominance in past literature, long bubble nose-tip position $\yN$ is deemed a non-ideal measure of centring extent due to an inherent level of high-variability; instead, separation distances 1- and 2-diameters upstream of nose-tip $\yOD$ and $\yTD$ along with body and tail separation $\yB$ and $\yT$ are recommended for determination of incurred degree of long bubble centring.
\item The largest extent of bubble centring measured in this study is found in a flow-case with both high-$\Fr$ and high-$\muL$; namely, case $\mathrm{B}.2$ from the S24 dataset with $(\Fr,\muL)=(2.94,352\unit{mPa.s})$ exhibits nearly perfect-centring (radial symmetry), particularly within the nose region. 
\item Elaborating on the phenomenological framework put forth in \citet{PerkinsLi20} (PL20), causality in the measurable interconnection between bubble centring and $\muL$ is investigated; namely, a foundational assumption utilized in the PL20 model\textemdash that coherent streamlines of relative motion immediately underneath the long bubble allow for generation of a downward force necessary in the initiation of centring\textemdash is intrinsically applicable to HVL systems because laminarity persists, in general, for low-$\Rey$ and thus high-$\muL$ flows. 
\item Considering that shear in a wall-bounded flow stems from absolute motion, liquid in the film region of slug flow is conceptualized as a dual-layer configuration: 1) a near-wall boundary layer (BL) governed by absolute velocity $\uf$ and 2) a near-bubble outer flow governed by relative velocity $\vf=\ut-\uf$. For the PL20 streamline mechanism to unfold, the outer flow must be unimpeded by the BL, at least momentarily; and, film laminarity can be assessed using a Reynolds number with characteristic length $\hf-\delta$. 
\item A novel theory for BL development in horizontal slug flow\textemdash consistent with unit-cell methodology yet straightforwardly upscaled to real flows\textemdash is postulated by assuming that the slug's turbulent mixing region acts to deconstruct and recycle near-wall structures, thus segregating individual unit-cell BL growth. As the long bubble translates downstream, the BL is hypothesized to continually grow whilst being simultaneously destroyed at the upstream mixing zone\textemdash eventually restricting the outer flow entirely at a localized steady-state limit.
\item The posited BL theory is tested using a calibrated case of HVL slug flow. Film height and velocity profiles are calculated using the \citet{TaitelBarnea90} (TB90) unit-cell model with appropriate closure inputs while BL thickness is approximated using flat-plate theory. Results demonstrate a significant differentiation between inner and outer film flow regions which\textemdash along with predicted film laminarity\textemdash suggests conditions favourable for manifest bubble detachment. 
\item Two potential mechanisms are explored in connection to partial-centring realization: 1) an axial transition from laminarity into turbulence within the film outer flow at a critical $\vf$ or $\Reyf$ which negates relative motion streamlines and 2) residual slug region turbulence causing downward shear on the long bubble nose-tip, driving it toward the pipe centreline while liquid flows like a wedge into the created gap\textemdash an effect that is compounding yet limited. Film laminarity modulation is predominantly applicable to HVL flows; liquid wedging can be active even for a turbulent film, making it relevant for water-based or low-HVL flows. 
\item Demonstrated using HVL flow pattern data from \citet{MatsubaraNaito11}, \citet{ZhaoEA13} and \citet{GokcalEA08} is a deterioration of mechanistic models' predictive capacity, positively proportional, in general, to $\muL$ as evidenced using outputs from \citet{TaitelDukler76} (TD76), \citet{Barnea87} (B87) and \citet{ZhangEA03c} (Z03). In particular, the slug-annular boundary is inaccurately calculated for HVL systems\textemdash a consequence that amplifies with increasing $\muL$.
\item The mechanistic approach is, intrinsically, to capture underlying phenomena; thus, a knowledge gap in the realm of gas-HVL flow dynamics is elucidated through non-successful application of commonly utilized models. As such, a novel theoretical framework for the slug-annular transition in HVL systems is derived, qualitatively, as a function of two unique long bubble mechanisms: centring and coalescence. In this paradigm, annular flow emerges from slug flow when full-centring occurs prior to coalescence in a flowing pipe system.
\end{enumerate}

\subsection{Future work}\label{sec:6.2}

\begin{enumerate}
\item The field of multiphase pipe flow research will benefit significantly from a purposeful influx of purely theoretical works. Undoubtedly, powerhouse investigators of decades past (e.g., Barnea, Taitel, Dukler, Brauner and others) crafted an exceptional foundation of phenomenological  understanding, the importance of which cannot be overstated; however, there still exist mechanistic gaps as indicated by a demonstrable lack of accuracy in modelling horizontal gas-HVL flow pattern data. Presently, the subfield is characterized by abundant publication of experimental studies utilizing a wide variety of operational conditions, fluid profiles and pipes; necessary, going forward, is a reestablishment of balance through innovations in mechanistic inquiry, capitalizing on the large girth of available exotic data, analogous to 1970-90\textemdash a remarkable era in the development of air-water theory. 
\item A rigorous empirical study focused on the bubble centring phenomenon would bring tremendous value to the field. Variables of interest include mixture flow rate, superficial rate ratio, liquid viscosity, operating pressure, rheological profile, pipe size and inlet configuration. Novel illumination and signal processing techniques are needed to visually capture and measure centring metrics at bubble nose, body and tail and 3-dimensional images can be obtained using methods such as those described in \citet{JamariEA08}. Different regions of the flowing pipe system\textemdash such as near-inlet and far-downstream\textemdash and temporal effects should be considered to better understand centring development. Ultra high-$\muL$ fluids ought to be included to enhance servicing of the heavy oil industry; for example, $\muL\in[1000,50000]\unit{mPa.s}$. 
\item Using appropriate alterations suggested in \S\ref{sec:4.2}, the baseline PL20 centring model must be scaled and tested for HVL system viability. Partial-centring theory put forth here and in PL20 must be validated and expanded upon, considering the phenomenological roles of thin upper film extent, film laminarity and BL growth. Robust distinction between a thin upper film and bubble centring should be clarified, particularly for HVL slug flow wherein detachment may be observed in low-$\Fr$ conditions. Generally speaking, centring theory must be continually developed such that the phenomenon gains widespread recognition within relevant academic and technical communities.  
\item Further work is required to understand BL generation in horizontal slug flow, particularly as it relates to bubble centring mechanics. Dedicated experimental and simulation studies would promote enhancement of theoretical discussion offered here. Particle image velocimetry or dye injection could be used to visualize near-wall and outer flow regions within the film. A novel camera system which translates in tandem with the moving bubble structure could be utilized. Temporal and unit-cell effects in slug flow BL evolution should be investigated; for example, development of free-stream velocity for fixed pipe locations. A BL growth model must be derived specifically for open-channel geometry and 3-dimensional effects require further research attention. 
\item The novel slug-annular transitional framework posited here requires subsequent devoted inquiry, both qualitative and quantitative. The dual-mechanism methodology can be investigated using high-resolution filmography focused on upstream pipe segments under operational conditions straddling the slug-annular boundary at a wide range of $\muL$-values. If deemed credible, clarification on $\muL$-range applicability is desirable and intensive mathematical derivation is needed to yield practical numerical models. 
\item To better understand the role and extent of bubble centring in transient slug flows (e.g., startup conditions, inlet effects or the severe slugging regime), additional theoretical and experimental research is necessary. Considering empirical conclusions uncovered here, future laboratory works should incorporate a wide range of $\muL$-values and operational rates into various spatially- or temporally-unsteady flow scenarios. The potential impact of a centred long bubble situated upstream of the pipeline riser in severe slugging should be investigated; that is, the operational consequences and feasibility of such a configuration warrant mechanistic inquiry. Further, existing non-unit-cell slug flow modelling methodologies (namely, slug-tracking and -capturing) should be modified and expanded such that they properly integrate, model and predict the centring phenomenon as it occurs in practice. 
\end{enumerate}

\newpage
\appendix

\section{Extracted data}
\label{sec:A1}
\begin{table}[h!]
\centering
\resizebox{1.1\columnwidth}{!}{%
\hskip-1cm\begin{tabular}{cccccccccccccc} 
 Case & $\muL$ (\unit{mPa.s}) & $\Fr$ & $\gamma$ & D (\unit{mm}) & $\oyOD$ & $\oyTD$ & $\oyN$ & $\oyB$ & $\oyT$ & INT & BUB & TYPE \\  \toprule 
1.1  & 1.0 & 1.00 & 1.00 & 26 & 1.54 & 1.54 & 14.45 & - & 0.00 & S & N & NC   \\
1.2  & 5.5 & 1.00 & 1.00 & 26 & 2.06 & 2.21 & 19.64 & - & 0.00 & S & N & NC   \\
1.3  & 10.3 & 1.00 & 1.00 & 26 & 6.17 & 2.83 & 34.04 & - & 0.00 & S & N & PC   \\
1.4  & 15.4 & 1.00 & 1.00 & 26 & 6.68 & 3.50 & 36.81 & - & 0.00 & S & N* & PC   \\
1.5  & 20.3 & 1.00 & 1.00 & 26 & 6.53 & 3.34 & 33.62 & - & 0.00 & S & N & PC   \\
1.6  & 30.4 & 1.00 & 1.00 & 26 & 6.99 & 4.22 & 33.57 & - & 1.61 & S & N & PC   \\
2.1  & 1.0 & 1.50 & - & 26 & 2.57 & 1.44 & 28.38 & - & - & S & N & -   \\
2.2  & 5.5 & 1.50 & - & 26 & 3.96 & 2.31 & 22.57 & - & - & S & N* & -   \\
2.3  & 10.3 & 1.50 & - & 26 & 8.28 & 3.75 & 31.05 & - & - & S & N* & -   \\
2.4  & 15.4 & 1.50 & - & 26 & 6.02 & 3.44 & 29.67 & - & - & S & N & -   \\
2.5  & 20.3 & 1.50 & - & 26 & 8.79 & 4.68 & 35.17 & - & - & S & N & -   \\
2.6  & 30.4 & 1.50 & - & 26 & 10.33 & 5.60 & 36.50 & - & - & S & N & -   \\
3.1  & 1.0 & 2.00 & - & 26 & 4.06 & 1.59 & 34.91 & - & 0.00 & T & N* & PC   \\
3.2  & 5.5 & 2.00 & - & 26 & 6.27 & 2.88 & 24.99 & - & 0.00 & T & N* & PC   \\
3.3  & 10.3 & 2.00 & - & 26 & 5.91 & 2.88 & 24.94 & - & 3.35 & S & N* & FC   \\
3.4  & 15.4 & 2.00 & - & 26 & 10.69 & 5.91 & 40.26 & - & 0.00 & S & N* & PC   \\
3.5  & 20.3 & 2.00 & - & 26 & 9.56 & 6.53 & 32.85 & - & 0.00 & S & N* & PC   \\
3.6  & 30.4 & 2.00 & - & 26 & 14.70 & 9.77 & 39.95 & - & 2.69 & S & N* & PC   \\
4.1  & 1.0 & 3.00 & - & 26 & 2.67 & 1.29 & 34.04 & - & 0.00 & W & Y & PC   \\
4.2  & 5.5 & 3.00 & - & 26 & 9.61 & 7.71 & 37.94 & - & 0.00 & W & Y & PC   \\
4.3  & 10.3 & 3.00 & - & 26 & 9.82 & 5.55 & 44.58 & - & 0.00 & W & Y & PC   \\
4.4  & 15.4 & 3.00 & - & 26 & 14.14 & 11.05 & 41.03 & - & 0.00 & S & Y & PC   \\
4.5  & 20.3 & 3.00 & - & 26 & 12.24 & 7.30 & 41.70 & - & 0.00 & S & Y & PC   \\
4.6  & 30.4 & 3.00 & - & 26 & 16.14 & 14.09 & 42.21 & - & 0.00 & S & Y & PC   \\
5.1  & 1.0 & 4.00 & - & 26 & 7.10 & 2.42 & 36.50 & - & 0.00 & W & Y & PC   \\
5.2  & 5.5 & 4.00 & - & 26 & 7.76 & 7.51 & 41.34 & - & 5.95 & W & Y & FC   \\
5.3  & 10.3 & 4.00 & - & 26 & 7.46 & 8.33 & 35.84 & - & 3.49 & W & Y & FC   \\
5.4  & 15.4 & 4.00 & - & 26 & 11.36 & 7.10 & 42.52 & - & 3.41 & W & Y & FC   \\
5.5  & 20.3 & 4.00 & - & 26 & 16.71 & 6.38 & 57.02 & - & 4.58 & W & Y & FC   \\
5.6  & 30.4 & 4.00 & - & 26 & 19.85 & 11.72 & 46.63 & - & 2.64 & S & Y & FC   \\
A.1  & 37.7 & 1.92 & 3.25 & 20 & 13.69 & 9.41 & 40.61 & 1.17 & 2.61 & S & N* & PC   \\
A.2  & 37.7 & 3.23 & 0.78 & 20 & 20.50 &  13.41 & 52.46 & 7.88 & 6.44 & S & Y & FC   \\
B.1  & 352 & 1.81 & 3.00 & 20 & 17.33 &  14.58 & 43.66 & 10.80 & 10.86 & S & Y & FC   \\
B.2  & 352 & 2.94 & 0.59 & 20 & 22.01 & 17.61 & 50.38 & 9.51 & 11.03 & S & Y & FC   \\
C.1  & 352 & 0.68 & 0.50 & 20 & 11.91 & - & 43.85 & - & 11.79 & S & Y & FC   \\ 
$\Omega$.1  & 510 & 0.57 & 1.00 & 50.8 & 10.33 & - & 45.68 & - & 4.68 & S & Y & FC   \\ 
$\Omega$.2  & 680 & 0.57 & 1.00 & 50.8 & 13.10 & - & 46.04 & - & 7.59 & S & Y & FC   \\
$\Omega$.3  & 960 & 0.57 & 1.00 & 50.8 & 14.76 & - & 47.50 & - & 8.31 & S & Y & FC   \\ \bottomrule
\end{tabular}
}
\caption{All data extracted from N23 ($\mathrm{i.j}$), S24 ($\mathrm{m.k}$) and K20 ($\Omega.\mathrm{q}$). $\oy$-values in \%.}
\label{tab:6}
\end{table}

\begin{figure}[h!]
\begin{enumerate}
\item[]\textbf{INT}: tortuosity of (lower) gas-liquid interface
\begin{enumerate}
\item[]S$\ \rightarrow\ $smooth 
\item[]T$\ \rightarrow\ $transitional
\item[]W$\ \rightarrow\ $wavy
\end{enumerate}
\item[]\textbf{BUB}: presence-level of dispersed bubbles in film region
\begin{enumerate}
\item[]N$\ \rightarrow\ $none
\item[]N*$\ \rightarrow\ $near-negligible
\item[]Y$\ \rightarrow\ $significant
\end{enumerate}
\item[]\textbf{TYPE}: type of bubble centring (see \S\ref{sec:2.1})
\begin{enumerate}
\item[]NC$\ \rightarrow\ $no-centring
\item[]PC$\ \rightarrow\ $partial-centring
\item[]FC$\ \rightarrow\ $full-centring
\end{enumerate}
\end{enumerate} 
\caption{Partial legend for table \ref{tab:6}.}
\label{fig:25}
\end{figure}

\section{Additional centring plots}
\label{sec:A2}
% line breaks within captions here may need to be removed in journal formatting, depending on where the last lines naturally land without. 
\begin{figure}[!h]
\centering
\includegraphics[width=3in]{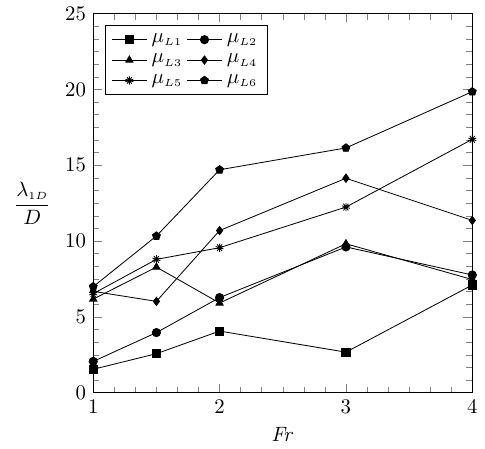}
\caption{Bubble centring data extracted from \citet{NaidekEA23} (N23 dataset): Normalized $\yOD$ (\%) as a function of $\Fr$ ($\um$) for fixed values of $\muL$ (constant $D$).\\ $\mu_{\scriptscriptstyle L1}=1\unit{mPa.s}$; $\mu_{\scriptscriptstyle L2}=5.5\unit{mPa.s}$; $\mu_{\scriptscriptstyle L3}=10.3\unit{mPa.s}$; $\mu_{\scriptscriptstyle L4}=15.4\unit{mPa.s}$; $\mu_{\scriptscriptstyle L5}=20.3\unit{mPa.s}$; $\mu_{\scriptscriptstyle L6}=30.4\unit{mPa.s}$.}
\label{fig:26}
\end{figure}

\begin{figure}[!h]
\centering
\includegraphics[width=3in]{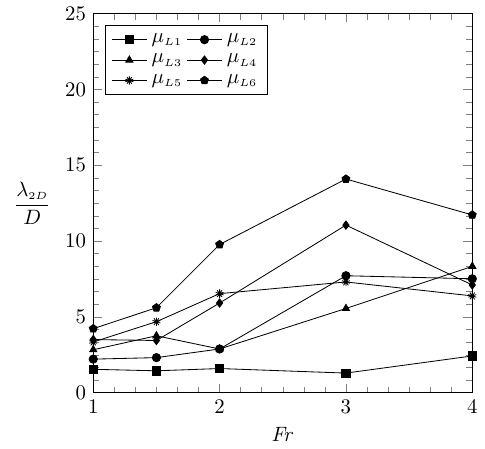}
\caption{Bubble centring data extracted from \citet{NaidekEA23} (N23 dataset): Normalized $\yTD$ (\%) as a function of $\Fr$ ($\um$) for fixed values of $\muL$ (constant $D$).\\ $\mu_{\scriptscriptstyle L1}=1\unit{mPa.s}$; $\mu_{\scriptscriptstyle L2}=5.5\unit{mPa.s}$; $\mu_{\scriptscriptstyle L3}=10.3\unit{mPa.s}$; $\mu_{\scriptscriptstyle L4}=15.4\unit{mPa.s}$; $\mu_{\scriptscriptstyle L5}=20.3\unit{mPa.s}$; $\mu_{\scriptscriptstyle L6}=30.4\unit{mPa.s}$.}
\label{fig:27}
\end{figure}

\begin{figure}[!h]
\centering
\includegraphics[width=3in]{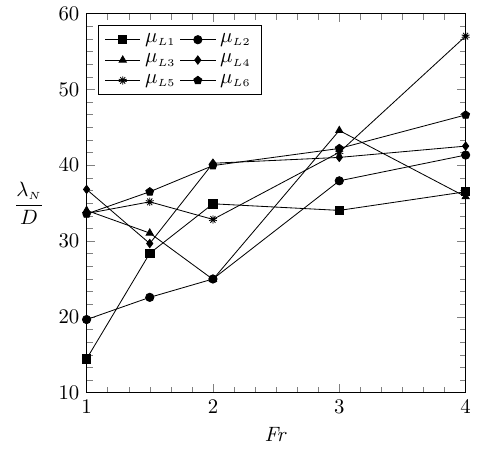}
\caption{Bubble centring data extracted from \citet{NaidekEA23} (N23 dataset): Normalized $\yN$ (\%) as a function of $\Fr$ ($\um$) for fixed values of $\muL$ (constant $D$).\\ $\mu_{\scriptscriptstyle L1}=1\unit{mPa.s}$; $\mu_{\scriptscriptstyle L2}=5.5\unit{mPa.s}$; $\mu_{\scriptscriptstyle L3}=10.3\unit{mPa.s}$; $\mu_{\scriptscriptstyle L4}=15.4\unit{mPa.s}$; $\mu_{\scriptscriptstyle L5}=20.3\unit{mPa.s}$; $\mu_{\scriptscriptstyle L6}=30.4\unit{mPa.s}$.}
\label{fig:28}
\end{figure}

\clearpage
\section{List of acronyms}
\label{sec:A3}
\begin{acronym}
%\acro{xxx}{xxx}
\acro{B87}{\citet{Barnea87} mechanistic flow pattern transition model/study}
\acro{BL}{Boundary layer}
\acro{G08}{\citet{GokcalEA08} experimental HVL flow pattern study}
\acro{HVL}{High-viscosity liquid $\muL>1\unit{mPa.s}$ ($\muL\ge 5.5\unit{mPa.s}$ studied here)}
\acro{K20}{\citet{KimEA20} HVL bubble centring dataset}
\acro{N23}{\citet{NaidekEA23} HVL bubble centring dataset}
\acro{PCV}{Pipe control volume (from phasic inlet to outlet)}
\acro{PL20}{\citet{PerkinsLi20} elongated bubble centring study/model}
\acro{S24}{\citet{ShinEA24} HVL bubble centring dataset}
\acro{SG00}{\citet{Schlichting00} textbook on boundary layer theory}
\acro{TB90}{\citet{TaitelBarnea90} mechanistic unit-cell slug model}
\acro{TD76}{\citet{TaitelDukler76} mechanistic flow pattern transition model/study}
\acro{Z03}{\citet{ZhangEA03c} mechanistic flow pattern transition model/study}
\acro{Z03M}{Modified version of Z03 flow pattern transition model (from G08)}
\end{acronym}

\vspace{-0.5cm}
\section{Elementary set notation}
\label{sec:A4}
\vspace{-0.75cm}
\begin{flalign*}
\{ \mathrm{A}, \mathrm{B}, \mathrm{C}\}  \hspace{1cm} &  \text{set containing A, B and C} \\
\forall x \hspace{1cm} &  \text{for all values of $x$} \\
\{\mathrm{A.k} \mid \forall \mathrm{k}\} \hspace{1cm} & \text{set of A.k cases for all defined values of k}\\
\{\mathrm{A}\cup\mathrm{B}\}\hspace{1cm} & \text{the union of sets A and B: all elements of the combined set }\mathrm{A}+\mathrm{B}\\
\{\Omega.\mathrm{q}\backslash \Omega.3\}\hspace{1cm} & \text{the set }\Omega.\mathrm{q} \text{ minus the element }\Omega.3\\
z\in[a,b] \hspace{1cm} & \text{z-values in } [a,b] \hspace{0.5em} (a\leq z\leq b)\\
z\in(a,b) \hspace{1cm} & \text{z-values in } (a,b) \hspace{0.5em} (a<z<b)\\
\Fr \in \{1,2,3\} \hspace{1cm} & \text{discrete }\Fr\text{-values in the set containing 1, 2 and 3}\\
\forall z\in [0,\lB]\colon x<y  \hspace{1cm} & \text{requirement that, for all values of }z\text{ in }[0,\lB]\text{, }x<y\\
\mathbb{Z}\hspace{1cm} & \text{the set of all integers }\{\mydots -2,-1,0,1,2\mydots\}
\end{flalign*}

% declarations and acknowledgements
\section*{Declaration of competing interests}

\noindent There are no known conflicts of interest associated with this publication and no financial support that could have influenced its outcome. 

\section*{Acknowledgements}

\noindent The author expresses gratitude for partial financial support, during the creation of this publication, stemming from a Discovery Grant from the Natural Sciences and Engineering Research Council of Canada (NSERC) (Grant No. NSERC RGPIN-2020–04571). I further thank my doctoral supervisor Huazhou Li for offering me support and full reign over my research endeavours, along with those who have never stopped believing in me.

%\clearpage
\bibliographystyle{elsarticle-harv} 
%\vspace{1cm}
\bibliography{paper2_references}

\section*{Permissions}
\noindent \textbf{Figure \ref{fig:04}}: Reprinted from \textit{Experimental Thermal and Fluid Science}, Vol. 141, Naidek, B. P., Conte, M. G., Cozin, C., dos Santos, E. N., Rodrigues, H. T., da Fonseca Jr., R., da Silva, M. J. and Morales, R. E. M., \textit{Experimental study of influence of liquid viscosity in horizontal slug flow}, pp. 1-11, copyright 2023, with permission from Elsevier.\\

\noindent \textbf{Figure \ref{fig:05}}: Reprinted from \textit{International Journal of Heat and Mass Transfer}, Vol. 226, Shin, H.-C., Kim, S.-H., Shah, Y. and Kim, S.-M., \textit{An experimental study on air-oil flow patterns in horizontal pipes using two synthetic oils}, pp. 1-19, copyright 2024, with permission from Elsevier. \\

\noindent \textbf{Figure \ref{fig:07}}: Reprinted from \textit{Journal of Petroleum Science and Engineering}, Vol. 191, Kim, T.-W., Al-Safran, E., Pereyra, E. and Sarica, C., \textit{Experimental study using advanced diagnostics to investigate slug aeration and bubble behavior in high liquid viscosity horizontal slug flow}, pp. 1-18, copyright 2020, with permission from Elsevier.

% format
% Reprinted from Publication title, Vol/edition number, Author(s), Title of article / title of chapter, pages no., copyright (year), with permission from Elsevier [or applicable society copyright owner]

\end{document}